\documentclass[a4paper,11pt]{article}

\pdfoutput=1 
\usepackage{jheppub} 

\usepackage[T1]{fontenc} 

\usepackage{bm}         
\usepackage{array}
\usepackage{multirow}   
\usepackage{booktabs}   
\usepackage[normalem]{ulem}
\usepackage{makecell}
\usepackage{multicol}
\usepackage{mleftright}
\usepackage[utf8]{inputenc}
\usepackage{slashed}
\usepackage{abraces}
\usepackage{mathtools}
\usepackage{amssymb}
\usepackage{amsmath}
\usepackage{bbold}
\usepackage{slashed}
\usepackage{cancel}
\usepackage{color}
\usepackage[dvipsnames]{xcolor}
\usepackage{epsfig}

\usepackage{tikz}
\usepackage{pgfplots}
\pgfplotsset{compat=1.15}
\usepgfplotslibrary{polar}
\usepgflibrary{shapes.geometric}
\usetikzlibrary{calc}
\graphicspath{{figures/}}

\usepackage{float} 
\usepackage{feynmp-auto} 
\usepackage{tensor} 
\usepackage{physics} 
\usepackage{subfig}
\usepackage{listings}
\usepackage{url}

\renewcommand{\i}{\mathrm i}
\renewcommand{\(}{\left(}

\renewcommand{\[}{\left[}

\newcommand{\SU}[2]{\mathrm{SU}(#1)_{\mathrm{#2}}}		
\newcommand{\E}[1]{\mathrm{E}_{#1}}	

\definecolor{bostonuniversityred}{rgb}{0.8, 0.0, 0.0}

\usepackage{pifont}
\title{Phenomenology of vector-like leptons with Deep Learning at the Large Hadron Collider}

\author[a]{Felipe F.~Freitas,} 
\emailAdd{felipefreitas@ua.pt}

\author[a]{João Gon\c calves,} 
\emailAdd{jpedropino@ua.pt}

\author[a]{Ant\'onio P. Morais,} 
\emailAdd{aapmorais@ua.pt}

\author[b]{Roman Pasechnik} 
\emailAdd{Roman.Pasechnik@thep.lu.se}

\affiliation[a]{Departamento de F\'\i sica da Universidade 
de Aveiro and \\
Centre  for  Research  and  Development  in  Mathematics  and  Applications  (CIDMA)\\
Campus de Santiago, 3810-183 Aveiro, Portugal}
	
\affiliation[b]{Department of Astronomy and Theoretical Physics,
Lund University, \\ SE 223-62 Lund, Sweden}


\abstract{
In this paper, a model inspired by Grand Unification principles featuring three generations of vector-like fermions, new Higgs doublets and a rich neutrino sector at the low scale is presented. Using the state-of-the-art Deep Learning techniques we perform the first phenomenological analysis of this model focusing on the study of new charged vector-like leptons (VLLs) and their possible signatures at CERN's Large Hadron Collider (LHC). In our numerical analysis we consider signal events for vector-boson fusion and VLL pair production topologies, both involving a final state containing a pair of charged leptons of different flavor and two sterile neutrinos that provide a missing energy. We also consider the case of VLL single production where, in addition to a pair of sterile neutrinos, the final state contains only one charged lepton. All calculated observables are provided as data sets for Deep Learning analysis, where a neural network is constructed, based on results obtained via an evolutive algorithm, whose objective is to maximise either the accuracy metric or the Asimov significance for different masses of the VLL. Taking into account the effect of the three analysed topologies, we have found that the combined significance for the observation of new VLLs at the high-luminosity LHC can range from $5.7\sigma$, for a mass of $1.25~\mathrm{TeV}$, all the way up to $28\sigma$ if the VLL mass is $200~\mathrm{GeV}$. We have also shown that by the end of the LHC Run-III a $200~\mathrm{GeV}$ VLL can be excluded with a confidence of $8.8$ standard deviations. The results obtained show that our model can be probed well before the end of the LHC operations and, in particular, providing important phenomenological information to constrain the energy scale at which new gauge symmetries emergent from the considered Grand Unification picture can be manifest.}

\begin{document}

\maketitle

\section{Introduction}\label{sec:Introduction}
The ultimate goal of any scientific endeavour is to uncover the mysteries of the universe and the world around us and, so far, the best model that we devised to describe all the matter that surrounds us at the most fundamental level is modestly called the Standard Model (SM). 
The SM is a particle physics model based upon modern quantum field theory (QFT) framework whose predictions and results have matched the stringiest of tests and predictions \cite{Chatrchyan:2012ufa,Arnison:1983rp,Hasert:1973ff,Hasert:1973cr,Abe:1995hr,Parker:2018vye,Hanneke:2010au}. However, there are clear indications that something is missing, from the fact that neutrinos have mass, as confirmed by the neutrino oscillation phenomena \cite{Fukuda:1998mi}, and that it does not take into account the existence of dark matter (DM) \cite{Bertone:2004pz}. Besides such experimental evidences, there are also theoretical motivations, as, e.g.~the origin of the family replication found in nature, the fermion masses and mixing hierarchies and the origin of the SM gauge structure, where a consensual understanding is still lacking.

So, we can notice that there are certain deficiencies in our understanding of fundamental particle physics
which leave us with the obvious question: what is missing and how to fix it? Well, so far, the most exotic theories have been put into the forefront, ranging from models where extra spacetime dimensions exist \cite{Tong:2009np} to models with a new symmetry between bosons and fermions known as supersymmetry (SUSY) \cite{Martin:1997ns}. While somewhat separate, these theories have a common underlying idea. The SM is an effective description of a more fundamental theory and is only valid up to a certain energy scale beyond which New Physics (NP) is needed. Therefore, the problems of the SM all result from our lack of understanding of what such theory really is and at which energy scale it becomes manifest.

High-scale theories like the string theory and SUSY,
despite their mathematical complexity, provide a solid theoretical framework from which one can build upon in order to e.g.~obtain NP models well motivated by the first principles. However, the amount of new states and model parameters emerging from such scenarios can be overwhelming. One possibility is to use a brute force method to analyse each combination of parameters and select the most promising ones or, alternatively, follow a smarter approach based upon Deep Learning (DL) techniques and optimization algorithms to find the best parameter space. Furthermore, one might have to overcome the typical challenges inherent to collider phenomenology, where the impact of background events can easily bury possible signal events preventing potential NP signatures from becoming observable. A better approach to handle such problem is the use of multivariate analysis to identify possible deviations from expected events which can be caused by NP. These deviations can be further amplified by combining multiple distributions into multidimensional distributions \cite{Ferreira:2017ymn}. This state of affairs, the need to quickly identify subtle effects in multidimensional distributions of information, clearly calls for artificial intelligence methods. Particularly the use of Machine Learning and DL techniques \cite{Alves:2019ppy}. 

With this being said, in this paper we revisit the key properties of a Grand Unified model recently introduced in \cite{Camargo-Molina:2016yqm,Camargo-Molina:2017kxd,Morais:2020odg,Morais:2020ypd} which attempts to unify all matter and fundamental interactions in a framework inspired by the $\mathrm{E}_8$ symmetry. Among the key features one can highlight a possible explanation for the fermion mass and mixing hierarchies observed in nature, as well as predicting NP states such as vector-like fermions, additional scalars doublets and a rich neutrino sector, well motivated by the model structure and the unification picture. 

The goal of this article is to construct and study the low-energy limit of such a framework which offers interesting phenomenological implications for future explorations at particle colliders. In particular, we focus on the phenomenological study of vector-like leptons (VLLs) and propose potential smoking-gun signatures as direct search channels to probe our model both at the LHC Run-III as well as its high-luminosity upgrade. Furthermore, the techniques that we develop are rather generic and can be used well beyond the scope of the model under consideration. The numerical analysis will be performed using standard Monte Carlo tools, where the final step of our analysis consists in applying the DL techniques for statistical significance studies.

This paper is organized as follows. First, in Sec.~\ref{sec:3HDM}, we discuss the model structure. Here, we briefly review its basic properties both at the unification scale as well as its low energy (SM-like) limit, motivating the parameter choices used in the numerical analysis. The latter represents the main focus of this work which is performed in Sec.~\ref{section:Numerics} where a detailed description of the methods employed in our analysis and the results obtained is given. In Sec.~\ref{sec:Conclusions} we conclude and discuss future work and research directions.

\section{Theoretical background}\label{sec:3HDM}

In this section, we introduce the model that is being explored in this article. We divide it into three main parts. In Sec.~\ref{subsec:Motivation} we motivate and introduce the concept of Grand Unified Theories (GUTs) our model is based upon. In particular, we consider an attractive low-scale picture where the presence of new vector-like fermions is well motivated alongside with three Higgs doublets. Then, in Sec.~\ref{subsec:High-energy} we make a short overview of the key properties of the unified framework. While the main purpose of this article is to study the phenomenological implications at the LHC and in particular the potential observability of VLLs, it is important to explain the origin of such NP and which parameter choices are relevant and well motivated. We then finalise in Sec.~\ref{subsec:Low-energy} with an effective low-energy  description by providing the full Lagrangian density, the particle masses as well as give a brief discussion of the benchmark scenarios that we use in our numerical analysis.

\subsection{A Motivation}\label{subsec:Motivation}

One of the most attractive features of SUSY is an elegant solution to the well known hierarchy problem. Among the key predictions, every known particle in nature should have a SUSY partner with the same mass. However, none of the current or previous experiments have ever observed the existence of such particles. This means that SUSY cannot be an exact symmetry, at least, at phenomenologically relevant scales and should be broken in such a way to generate a larger mass contributions to the superpartners of the SM particles. The actual mass scale of such particles is not known, but the current lack of observation at the LHC \cite{ATLAS:2019vcq,Aad:2019pfy,Aaboud:2018htj,Sirunyan:2018vjp,Sirunyan:2017qaj,Sirunyan:2017lae} indicates that SUSY breaking should occur well 
above the electroweak (EW) scale. However, this by no means excludes SUSY as a well motivated formalism to describe realistic theories. This is the case of the model designed in \cite{Camargo-Molina:2016yqm,Camargo-Molina:2017kxd,Morais:2020odg,Morais:2020ypd} that we analyse in this article. While SUSY does not necessarily manifest at low scale and the effective theory can be treated as a standard non-SUSY model, its high scale limit is indeed supersymmetric with remarkable implications.

As we will see, such model belongs to a class of GUTs that can potentially emerge from a single gauge $\text{E}_8$ group, the unifying force. One of the key properties of this framework is that flavour is promoted to a gauge symmetry that is part of $\text{E}_8$ and treated in the same footing as conventional gauge interactions. The model aims at addressing various issues of the SM delving into fundamental questions such as the origin of gauge interactions and the origin of mass hierarchies for the different matter particles, which is typically known as the flavour problem. As a byproduct of this unification picture NP in the form of vector-like fermions (VLFs) may be manifest at the TeV scale. The emergence of light VLFs from other GUT models had previously been proposed in \cite{Dorsner:2014wva,Raby:2017igl} where one of the key advantages of the presence of their leptonic counterparts is the possibility for explaining the muon anomalies \cite{Raby:2017igl,Poh:2017tfo}. In this article we will pay special attention to this sector since a potential discovery of VLLs at the LHC can offer relevant phenomenological probes of the high-energy theory and hints of the unification picture.

\subsection{High-Energy limit}\label{subsec:High-energy}

Here, we present the high-energy scale formulation of the model under consideration with focus on the main properties needed for a basic understanding relevant for our numerical analysis. A more detailed description can be found in \cite{Camargo-Molina:2016bwm,Camargo-Molina:2016yqm,Camargo-Molina:2017kxd,Morais:2020odg,Morais:2020ypd}, where we highlight \cite{Morais:2020ypd} as the most recent and complete reference.

The main idea of a GUT model is to embed all SM-gauge interactions, i.e.~$\text{SU(3)}_\text{C} \times \text{SU(2)}_\text{L} \times \text{U(1)}_\text{Y}$, into a larger group. As already stated, an interesting possibility resides on the $\E{8}$ symmetry. It has been presented as a GUT candidate in various superstring theories \cite{Green:1987sp,Achiman:1978vg} and is, in fact, a motivation inherent to our model\footnote{Note that in \cite{Morais:2020ypd} and previous publications the exact connection of $\mathrm{E}_8$ to the unification of all interactions has not yet been fully established. However, the main low-scale properties have been thoroughly described and serve as a motivation to the current work.}.
\begin{figure}[t]
	\centering
	\includegraphics[width=0.55\textwidth]{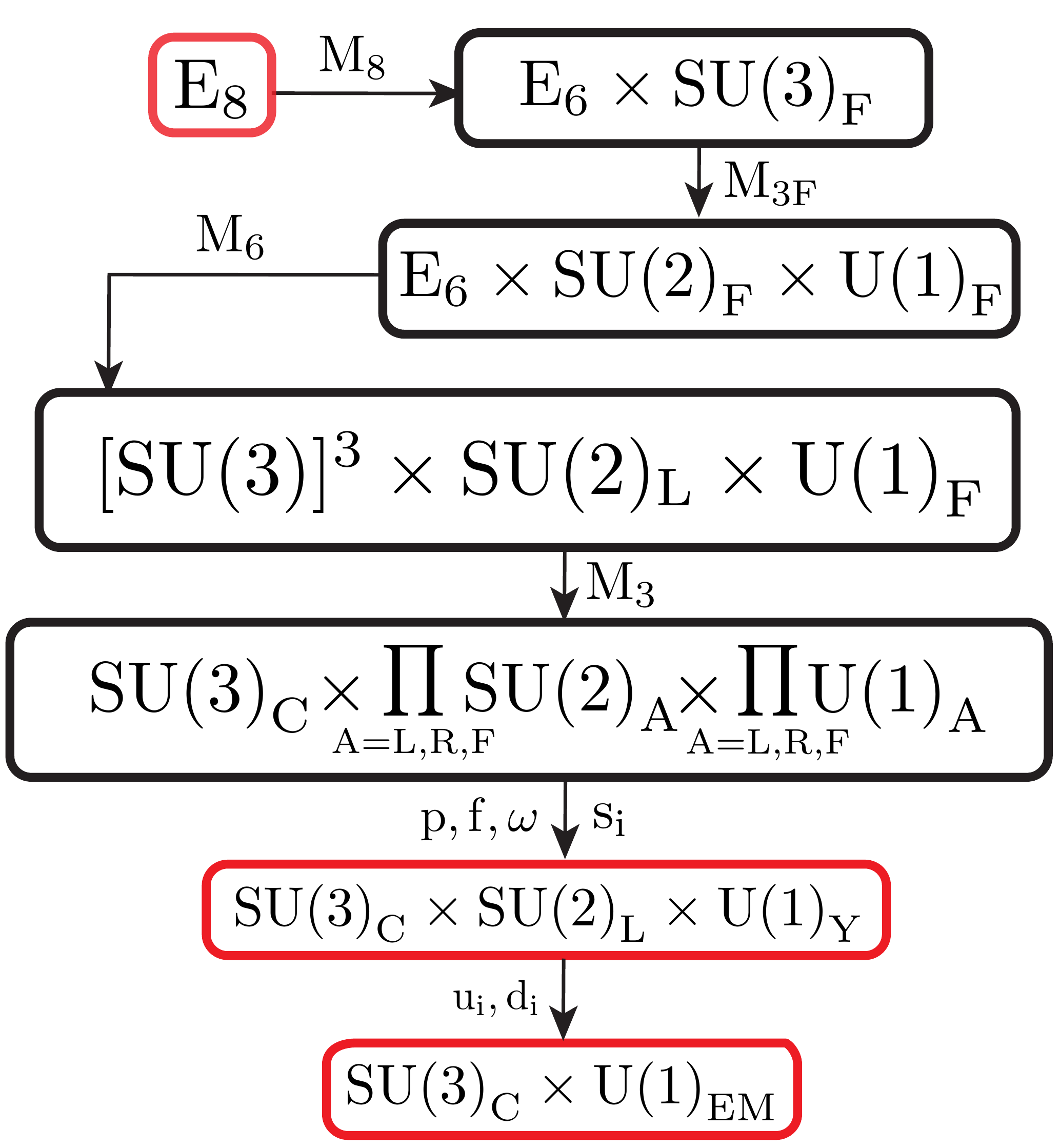}
	\caption[]{Symmetry breaking scheme from the original $\text{E}_8$ gauge symmetry down to the strong and electromagnetic gauge groups ($\text{SU}(3)_\text{C} \times \text{U}(1)_{\text{EM}}$). The various terms between the different boxes ($M_6$, $M_3$, etc) represent the distinct mass scales involved in this scheme according to the discussion in \cite{Morais:2020ypd}. $M_8$ encodes details inspired by theories with extra compact dimensions.}
	\label{fig:Symmetry-breaking}
\end{figure}

Our model was engineered to address some of the main concerns one encounters in the SM. It proposes a first principles explanation for a common origin of the strong and EW interactions as well as the flavour structure observed in nature. The Higgs and matter fields are unified into a single superfield equipping both the scalar and the fermion sectors with the same flavour structures. This results in a rather reduced freedom in the Yukawa interactions allowing only for two free parameters, $\mathcal{Y}_1$ and $\mathcal{Y}_2$, which will provide the dominant contributions to three generations of exotic vector-like quarks (VLQs) as well as the third and second generation SM-like quark masses. All remaining fermions, including the first-generation quarks and charged leptons, have their masses radiatively generated making them naturally lighter. The CKM mixing is also emergent in this framework provided that there are, at least, three Higgs doublets developing vacuum expectation values (VEVs). Given its rather unique properties, this model has been named as SUSY Higgs-matter Unified Trinification or SHUT for short. Note that the trinification group emerges as a subgroup of $\mathrm{E}_8$ which we consider below the $M_6$ scale in Fig.~\ref{fig:Symmetry-breaking}.

As stated before, the starting point is the $\text{E}_8$ gauge symmetry, and the first symmetry breaking step reads
\begin{equation}\label{eq:First_reduction_E8}
\text{E}_8 \rightarrow \text{E}_6 \times \text{SU(3)}_\text{F} \,,
\end{equation}
where the subscript F denotes the family symmetry. From this point on, the sequence of steps by which we obtain the SM gauge group is schematically illustrated in Fig.~\ref{fig:Symmetry-breaking}. The SM particle content and all NP emergent at low-energy scales correspond to the states that remain light after the various breaking stages. Only second- and third-generation SM-like quarks and all three VLQ masses are tree-level generated, with their relative sizes controlled by the only two Yukawa couplings in the theory, which are of SUSY origin. To see this, let us consider the theory after the breaking step denoted by $M_3$ in Fig.~\ref{fig:Symmetry-breaking} whose superpotential reads \cite{Morais:2020ypd} 
\begin{equation}\label{eq:Superpotential}
\begin{aligned}
W = &\mathcal{Y}_1 \varepsilon_{ij}\qty(\bm{\chi}^i  \bm{q}_\mathrm{L}^3 \bm{q}_\mathrm{R}^j + \bm{\ell}^i_\mathrm{R}  \bm{D}^3_\mathrm{L}  \bm{q}_\mathrm{R}^j + \bm{\ell}_\mathrm{L}^i \bm{q}_\mathrm{L}^3 \bm{D}^j_\mathrm{R} + \bm{\phi}^i\bm{D}_\mathrm{L}^3\bm{D}^j_\mathrm{R}) \\ -& \mathcal{Y}_2\varepsilon_{ij} \qty(\bm{\chi}^i\bm{q}_\mathrm{L}^j\bm{q}_\mathrm{R}^3 + \bm{\ell}^i_\mathrm{R} \bm{D}_\mathrm{L}^j\bm{q}_\mathrm{R}^3 + \bm{\ell}^i_\mathrm{L}\bm{q}_\mathrm{L}^j\bm{D}^3_\mathrm{R} + \bm{\phi}^i\bm{D}_\mathrm{L}^j\bm{D}^3_\mathrm{R}) \\ +&\mathcal{Y}_2 \varepsilon_{ij}\qty(\bm{\chi}^3\bm{q}_\mathrm{L}^i\bm{q}_\mathrm{R}^j + \bm{\ell}^3_\mathrm{R} \bm{D}^i_\mathrm{L}\bm{q}_\mathrm{R}^j + \bm{\ell}^3_\mathrm{L} \bm{q}_\mathrm{L}^i \bm{D}^j_\mathrm{R} + \bm{\phi}^3\bm{D}^i_\mathrm{L}\bm{D}^j_\mathrm{R}) \,,
\end{aligned}
\end{equation}
where $\varepsilon_{ij}$ is the two-dimensional Levi-Civita symbol in the generations' (or flavour, in what follows) space, $\mathcal{Y}_1$ and $\mathcal{Y}_2$ are the Yukawa couplings, and $\mathrm{L}/\mathrm{R}$ denotes $\text{SU(2)}_\text{L/R}$ doublet superfields. Note that while $\bm{\chi}$ is a $\text{SU(2)}_\text{L} \times \text{SU(2)}_\text{R}$ bi-doublet where the light Higgs sector resides, $\bm{\phi}$ is a singlet carrying only family symmetry charges, typically dubbed as flavon. Despite some allowed mixing after symmetries are sequentially broken, the left- and right-handed leptons are essentially embedded in $\bm{\ell}_\text{L}$ and $\bm{\ell}_\text{R}$ respectively, whereas the SM-like quark sector belongs to both $\bm{q}_\text{L}$ and $\bm{q}_\text{R}$. Note also that this model addresses neutrino masses due to the existence of six right-handed sates, three in $\bm{\ell}_\text{R}$ and three in $\bm{\phi}$. Last but not least, new down-type $\text{SU}(2)_\text{L}$ singlet VLQs and $\text{SU}(2)_\text{L}$ doublet VLLs are predicted in the SHUT model emerging from the fermionic components of $\bm{D}_\text{L,R}$ and $\bm{\chi}$, respectively. In this work, we will study the collider phenomenology of the latter and discuss possible implications for the high scale picture. All exotic scalars are assumed to be decoupled at the soft SUSY breaking scale beyond the reach of the LHC. 

As we have mentioned above, one of the features of the SHUT model is that only second and third generation chiral quarks as well as the three VLQ generations are allowed to obtain masses at tree-level. For a better understanding of this statement let us inspect the superpotential in Eq.~\eqref{eq:Superpotential}. First, after the second last breaking stage in Fig.~\ref{fig:Symmetry-breaking} all six neutral scalars in $\tilde{\bm{\phi}}^i$ and $\tilde{\bm{\ell}}_\text{R}^i$ develop VEVs which are denoted by $p$, $f$, $\omega$ and $s_{i}$ (see \cite{Morais:2020ypd} for details). We immediately see that mass terms for VLQs are generated from $\expval{\tilde{\bm{\phi}}} \bm{D}_L\bm{D}_R$ and $\expval{\tilde{\bm{\ell}_\text{R}}} \bm{D}_L\bm{q}_R$ type of terms resulting in \cite{Morais:2020ypd}
	\begin{equation}
	\begin{aligned}
	& m_\text{D/S}^2\simeq \frac{1}{2} (f^2+p^2)\mathcal{Y}_2^2 \,, \quad
	m_\text{S/D}^2\simeq \frac{\omega^2(f^2+p^2+\omega^2)}{2(f^2+\omega^2)}\mathcal{Y}_2^2 \,, \\
	& m_\text{B}^2\simeq \frac{1}{2} (f^2+\omega^2)\mathcal{Y}_1^2 + \frac{f^2p^2}{2(f^2+\omega^2)}\mathcal{Y}_2^2, \,
	\end{aligned}\label{eq:mVLQ}
	\end{equation}
where, for simplicity, we have ignored the subdominant effect of the $s_i$ VEVs and where we adopt a notation such that the lightest VLQ is the D-quark. Along the same lines, SM-like quark masses are generated from $\expval{\tilde{\bm{\chi}}} \bm{q}_L\bm{q}_R$ type of terms where, even for a generic setting with all six EW doublets in $\tilde{\bm{\chi}}$ developing nonzero VEVs, the up and down-quark masses are always zero. Furthermore, it was shown in \cite{Morais:2020odg,Morais:2020ypd} that a \textit{proto-realistic} description of the CKM matrix requires a minimum of three light Higgs doublet VEVs where the quark masses become 
	\begin{equation}\label{eq:mu}
	m_\text{u} = 0, \qquad m_\text{c}^2 = \tfrac{1}{2} \mathcal{Y}_2^2 \left(u_1^2 + u_2^2\right),  \qquad m_\text{t}^2 = \tfrac{1}{2} \mathcal{Y}_1^2 \left(u_1^2 + u_2^2\right) \,
	\end{equation} 
and
	\begin{equation}\label{eq:md}
	m_\text{d} = 0 \,, \qquad   
	m_\text{s}^2 = \tfrac{1}{2}\mathcal{Y}_2^2 \frac{d_2^2 p^2}{(f^2+p^2+\omega^2)} \,, \qquad
	m_\text{b}^2 = \tfrac{1}{2}\mathcal{Y}_1^2 d_2^2 \,,
	\end{equation}
with $u_i$ and $d_i$ being i-th family EW-symmetry breaking VEVs from Higgs doublets coupling to up-type and down-type quarks, respectively. If we consider, for simplicity, that $p\approx f \approx \omega$, we obtain the following ratios
\begin{equation}
\label{eq:ratio}
\frac{\mathcal{Y}_1}{\mathcal{Y}_2} ~=~ \frac{m_\text{t}}{m_\text{c}} ~\approx~ \frac{m_\text{b}}{m_\text{s}}  ~\approx~ \frac{m_\text{B}}{m_\text{D,S}} ~\sim~ \mathcal{O}(100) \,,
\end{equation}
implying also the presence of up to two generations of VLQs at the reach of the LHC if $\omega$ and $f$ are around $100~\mathrm{TeV}$. This relation fixes the size of $\mathcal{Y}_1$ and $\mathcal{Y}_2$ and implies that mass ratios in the VLQ sector are the same as the ones found among their chiral counterparts. Note that in this work we will only study the VLL sector and leave a detailed numerical study of VLQs for a future work.

For the case of both SM-like leptons as well as VLLs, there are no allowed terms of the form $\expval{\tilde{\bm{\chi}}} \bm{\ell}_L\bm{\ell}_R$ and $\expval{\tilde{\bm{\phi}}} \bm{\chi}\bm{\chi}$, respectively, which means that, at tree-level, their masses are zero just like the first-generation quarks. 
However, below the second to last symmetry breaking stage in Fig.~\ref{fig:Symmetry-breaking}, such type of operators become allowed which means that they can be radiatively generated via loops with internal heavy scalar and fermion propagators.

\subsection{Low-Energy effective limit}\label{subsec:Low-energy}

While a direct probe for the high energy limit of the SHUT model at, or above, the $\omega-f-p$ scales is far beyond the reach of the LHC, exploring the corresponding NP signatures at the TeV-scale can offer us solid indications about the structure of the model at higher scales. Furthermore, such an analysis will provide an important piece of information about the low-scale properties of the model, which, although not explored in this work, can become relevant for matching of the low and the high scale regimes of the theory.

We consider in this section a possible low-energy scale limit of the SHUT model whose gauge symmetry is given in the second to last box of Fig.~\ref{fig:Symmetry-breaking}. All the quantum numbers for the gauge groups are shown in Tabs.~\ref{tab:SM-like},~\ref{tab:BSM} and \ref{tab:Scalar} 
\begin{table}[H]
	\centering
	\begin{tabular}{c|c|c|c|c}
		\textbf{Field} & $\mathbf{SU(3)_\text{C}}$ & $\mathbf{SU(2)_\text{L}}$ & $\mathbf{U(1)_\text{Y}}$ & \textbf{\# of generations} \\ \hline
		$Q_\mathrm{L}$          & \textbf{3}                & \textbf{2}                & $1/3$   & 3                          \\
		$L$            & \textbf{1}                & \textbf{2}                & $-1$  & 3                          \\
		$d_\mathrm{R}$          & \textbf{3}                & \textbf{1}                & $-2/3$  & 3                          \\
		$u_\mathrm{R}$          & \textbf{3}                & \textbf{1}                & $4/3$   & 3                          \\
		$e_\mathrm{R}$          & \textbf{1}                & \textbf{1}                & $-2$              & 3                         
	\end{tabular}
	\caption{\label{tab:SM-like}SM-like sector for the fermions and quarks.}
\end{table}
\begin{table}[H]
	\centering
	\begin{tabular}{c|c|c|c|c}
		\textbf{Field} & $\mathbf{SU(3)_\text{C}}$ & $\mathbf{SU(2)_\text{L}}$ & $\mathbf{U(1)_\text{Y}}$ & \textbf{\# of generations} \\ \hline
		$E_\mathrm{L,R}$          & \textbf{1}                & \textbf{2}                & $-1$   & 3                          \\
		$D_\mathrm{L,R}$          & \textbf{3}                & \textbf{1}                & $-2/3$  & 2                          \\
		$\nu_\mathrm{R}$          & \textbf{1}                & \textbf{1}                & $0$   & 6                          \\
	\end{tabular}
	\caption{\label{tab:BSM}Beyond-the-SM sector for the fermions and quarks.}
\end{table}
\begin{table}[H]
	\centering
	\begin{tabular}{c|c|c|c|c}
		\textbf{Field} & $\mathbf{SU(3)_\text{C}}$ & $\mathbf{SU(2)_\text{L}}$ & $\mathbf{U(1)_\text{Y}}$ & \textbf{\# of generations} \\ \hline
		$\phi$          & $\bm{1}$                & $\bm{2}$                & $1$   & 3                          \\
	\end{tabular}
	\caption{\label{tab:Scalar}Scalar sector.}
\end{table}
where the $\text{SU(2)}_\text{L}$ doublets are defined as follows,
\begin{equation}\label{eq:Doublets}
\begin{aligned}
Q_\mathrm{L}^i= \begin{bmatrix}
u_\mathrm{L}\\ 
d_\mathrm{L}
\end{bmatrix}^i \quad
L^i= \begin{bmatrix}
\nu_{e_\mathrm{L}}\\ 
e_\mathrm{L}
\end{bmatrix}^i \quad 
E_\mathrm{L,R}^i= \begin{bmatrix}
\nu_{e_\mathrm{L,R}}'\\ 
e_\mathrm{L,R}'
\end{bmatrix}^i\,, \quad 
\end{aligned}
\end{equation}
with $Q_\mathrm{L}^i$ denoting the fermionic components of the $\bm{q}_\mathrm{L}^{i=1,2,3}$ superfields, whereas $L^i$ are the lepton doublet components of $\bm{\ell}^i_\mathrm{L}$, and $E^i_\mathrm{L,R}$ belong to the $\bm{\chi}^i$ bi-doublet superfields.

Let us now describe the low-scale version of the SHUT model, step by step. The gauge boson's quantum numbers are not shown\footnote{In fact, the model does allow for extra vector bosons, however those only become relevant at higher energy scales that are not particularly important for our discussion here.} since they are 
identical to the SM. On the other hand, the matter sector can be subdivided into two sub-sectors. The first, shown in Tab.~\ref{tab:SM-like}, represents the SM-like fermions from where ordinary matter emerges. The second sector, shown in Tab.~\ref{tab:BSM}, is where NP appears including three new VLL generations, $E_\mathrm{L,R}$, and two light VLQ generations which we denote as $D_\mathrm{L,R}$. The Beyond-the-SM (BSM) sector also offers a rich neutrino content including six left-handed states originating from the $E_\mathrm{L}$ and $E_\mathrm{R}$ $\text{SU(2)}_\text{L}$ doublets and six right-handed SM-singlet Majorana neutrinos which we denote as $\nu_\mathrm{R}$. Recall that the latter are embedded in three $\bm{\ell}^i_\mathrm{R}$ $\SU{2}{R}$-doublets and three $\bm{\phi}^i$ flavons as stated above. Note that the lightest of the right-handed neutrinos, which we cast as $\nu_\text{BSM}$ in the remainder of this article, can be sterile enough to provide a good DM candidate \cite{Boyarsky:2018tvu}. While we do not perform DM studies in the current work, we will consider this scenario in our numerical analysis by setting its mass in the keV-MeV range and the mixing to the SM-like neutrinos to zero. In such a scenario $\nu_\text{BSM}$ escapes the detector and is treated as missing energy. While the scalar sector also offers NP we will not further study it in this paper leaving any further details for a future work.

We can now introduce the relevant interaction terms for our analysis. We start with the low-scale Yukawa Lagrangian 
that reads as
\begin{equation}\label{eq:YukawaTerms}
\begin{aligned}
\mathcal{L}_{\text{y}} = &\qty(Y^a)_{ij}\qty(\bar{Q}_\mathrm{L})^i\qty(D_\mathrm{R})^j\phi_a + \qty(\Gamma^a)_{ij}\qty(\bar{Q}_\mathrm{L})^i\qty(d_\mathrm{R})^j\phi_a + \qty(\Delta^a)_{ij}\qty(\bar{Q}_\mathrm{L})^i\qty(u_\mathrm{R})^j\Tilde{\phi}_a + \\ & + \qty(\Theta^a)_{ij}\qty(\bar{E}_\mathrm{L})^i\qty(e_\mathrm{R})^j\phi_a + \qty(\Upsilon^a)_{ij}\qty(\bar{E}_\mathrm{L})^i\qty(\nu_\mathrm{R})^j\Tilde{\phi}_a + \qty(\Sigma^a)_{ij}\qty(\bar{L})^i\qty(\nu_\mathrm{R})^j\Tilde{\phi}_a + \\
& + \qty(\Pi^a)_{ij}\qty(\bar{L})^i\qty(e_\mathrm{R})^j\phi_a + \qty(\Omega^a)_{ij}\qty(\bar{E}_\mathrm{R})^i\qty(\nu_\mathrm{R})^j\tilde{\phi}_a + \text{h.c.} \,,
\end{aligned}
\end{equation}
where $\Gamma$, $\Delta$, $\Theta$ and $\Pi$ are the $3\times3$ Yukawa matrices, $\Upsilon$, $\Sigma$,  and $\Omega$ are $3\times6$ matrices whereas $Y$ is a $3\times2$ one. Note that only $Y$, $\Gamma$ and $\Delta$ contain entries whose leading contributions are proportional to $\mathcal{Y}_1$ and $\mathcal{Y}_2$. The remaining ones are purely of a radiative origin.
Unlike what we have in the SM, here, the gauge symmetries allow for explicit construction of invariant bilinear and mass terms 
\begin{equation}\label{eq:Bilinear}
\begin{aligned}
\mathcal{L}_{\text{bil}} = &\qty(M_D)_{ij}\qty(\bar{D}_\mathrm{L})^i\qty(D_\mathrm{R})^j + \qty(M_E)_{ij}\qty(\bar{E}_\mathrm{L})^i\qty(E_\mathrm{R})^j + \frac{1}{2}\qty(M_{\nu_\mathrm{R}})_{ij}\qty(\bar{\nu}_\mathrm{R})^i\qty(\nu_\mathrm{R})^j + \\ & + \qty(M_{LE})_{ij}\qty(\bar{L})^i\qty(E_\mathrm{R})^j+ \qty(\Xi)_{ij}\qty(\bar{D}_\mathrm{L})^i\qty(d_\mathrm{R})^j \,.
\end{aligned}
\end{equation}
These arise from the vector-like nature of the involved fields where $\text{SU}(2)_\text{L}$ transformations do not distinguish between left and right chiralities. All such mass terms in \eqref{eq:Bilinear} are generated at the $\omega$-$f$-$p$ scales, thus larger than the EW scale. Note that the neutrino mass matrix $M_{\nu_R}$ is generated once the $p$, $f$, $\omega$ and $s_i$ VEVs are developed. However, contrary to all remaining bilinear and Yukawa terms in the leptonic sector, its entries are generated by tree-level diagrams once the corresponding operators become allowed (see \cite{Morais:2020ypd} for details). Therefore, small loop factors will not suppress the size of $M_{\nu_\mathrm{R}}$, whose entries can be up to an order of $p$, $f$ and $\omega$ scales. As a byproduct, the neutrino sector automatically contains a seesaw mechanism and hence an explanation for the smallness of SM neutrino masses as we further discuss below. For completeness, we show the remaining Lagrangian terms in appendix~\ref{app:Feynman Rules}.

With the model fully defined, we finalise this section by showing the fermion mass matrices in the gauge eigenbasis that are implemented in our numerical analysis. First, for the quarks, and considering the components of the $Q_\mathrm{L}$ $\text{SU}(2)_\text{L}$ doublets as in \eqref{eq:Doublets}, the new Lagrangian is written as 
\begin{equation}\label{eq:QuarkLag_HM}
\begin{aligned}
&\mathcal{L}_{q,\text{SB}} = \frac{v_a}{\sqrt{2}}\qty(Y^a)_{ij}\qty(\bar{d}_\mathrm{L})^i\qty(\bar{D}_\mathrm{R})^j + \frac{v_a}{\sqrt{2}}\qty(\Gamma^a)_{ij}\qty(\bar{d}_\mathrm{L})^i\qty(\bar{d}_\mathrm{R})^j + \frac{v_a}{\sqrt{2}}\qty(\Delta^a)_{ij}\qty(\bar{u}_\mathrm{L})^i\qty(\bar{u}_\mathrm{R})^j +\\ &+\qty(M_D)_{ij}\qty(\bar{D}_\mathrm{L})^i\qty(\bar{D}_\mathrm{R})^j + \qty(\Xi)_{ij}\qty(\bar{D}_\mathrm{L})^i\qty(\bar{d}_\mathrm{R})^j \,,
\end{aligned}
\end{equation}
with $v_a$ being the VEV of the respective Higgs doublet $\phi_a$. The up-type quark mass matrix written in the basis \{$u_\mathrm{L}^1$,$u_\mathrm{L}^2$,$u_\mathrm{L}^3$\} $\otimes$ \{$u_\mathrm{R}^1$,$u_\mathrm{R}^2$,$u_\mathrm{R}^3$\} 
takes the form
\begin{equation}\label{Mass_up}
[M_u]=\frac{v_a}{\sqrt{2}}
\begin{bmatrix}
\Delta^a_{11}&\Delta^a_{12}  &\Delta^a_{13} \\ 
\Delta^a_{21}&\Delta^a_{22}  &\Delta^a_{23} \\ 
\Delta^a_{31}&\Delta^a_{32}  &\Delta^a_{33}
\end{bmatrix} \,.
\end{equation}
The eigenvalues of $[M_u]$ give the masses of the up-type quarks whose leading contributions are proportional 
to \eqref{eq:mu}. A similar strategy can be now employed for the down quark sector where, in the basis \{$d_\mathrm{L}^1$,$d_\mathrm{L}^2$,$d_\mathrm{L}^3$,$D_\mathrm{L}^1$,$D_\mathrm{L}^2$\} $\otimes$ \{$d_\mathrm{R}^1$,$d_\mathrm{R}^2$,$d_\mathrm{R}^3$,$D_\mathrm{R}^1$,$D_\mathrm{R}^2$\}, we have
\begin{equation}\label{Mass_down}
[M_d] = 
\begin{bmatrix}
\frac{v_a}{\sqrt{2}}\Gamma^a_{11} &\frac{v_a}{\sqrt{2}}\Gamma^a_{12} &\frac{v_a}{\sqrt{2}}\Gamma^a_{13} &\frac{v_a}{\sqrt{2}}Y^a_{11} &\frac{v_a}{\sqrt{2}}Y^a_{12} \\ 
\frac{v_a}{\sqrt{2}}\Gamma^a_{21} &\frac{v_a}{\sqrt{2}}\Gamma^a_{22} &\frac{v_a}{\sqrt{2}}\Gamma^a_{23} &\frac{v_a}{\sqrt{2}}Y^a_{21} &\frac{v_a}{\sqrt{2}}Y^a_{22} \\ 
\frac{v_a}{\sqrt{2}}\Gamma^a_{31} &\frac{v_a}{\sqrt{2}}\Gamma^a_{32} &\frac{v_a}{\sqrt{2}}\Gamma^a_{33} &\frac{v_a}{\sqrt{2}}Y^a_{31} &\frac{v_a}{\sqrt{2}}Y^a_{32} \\ 
\Xi_{11} &\Xi_{21} &\Xi_{31} &\qty(M_D)_{11} &\qty(M_D)_{12} \\ 
\Xi_{12} &\Xi_{22} &\Xi_{32} &\qty(M_D)_{21} &\qty(M_D)_{22} 
\end{bmatrix} \,.
\end{equation}
Unlike the up sector, here we have NP contributions. Besides the down, strange and bottom quarks, we also have two new VLQs which we name as $d_4$ and $d_5$\footnote{This rather simplistic nomenclature is used to facilitate the designation when doing numerical analysis, as this is the name of the particle as defined in the \texttt{UFO} files. The designation in \cite{Morais:2020odg,Morais:2020ypd} and above in Eq.~\eqref{eq:mVLQ} corresponds to $d_4 \equiv D$, $d_5 \equiv S$.} defined in such a way that $m_{d_5} > m_{d_4}$. The leading contributions to the down-type quark masses are proportional to Eqs.~\eqref{eq:mVLQ} and \eqref{eq:md}.

We can now extend this analysis to the lepton sector and write down the mass matrices for the charged leptons and neutrinos. Starting with the charged leptons, in the basis \{$e_\mathrm{L}'^i$,$e_\mathrm{L}^i$\} $\otimes$ \{$e_\mathrm{R}'^j$,$e_\mathrm{R}^j$\} one gets
\begin{equation}\label{eq:Mass_matrix_lepton}
[M_L] = 
\begin{bmatrix}
\begin{bmatrix}\qty(M_E)_{ij}\end{bmatrix}_{3\times3} & \begin{bmatrix}\dfrac{v_a}{\sqrt{2}}\qty(\Theta^a)_{ij}\end{bmatrix}_{3\times3} \\ 
\begin{bmatrix}\qty(M_{LE})_{ij}\end{bmatrix}_{3\times3}& \begin{bmatrix}\dfrac{v_a}{\sqrt{2}}\qty(\Pi^a)_{ij}\end{bmatrix}_{3\times3} 
\end{bmatrix} \,,
\end{equation}
and for the neutrinos, in the basis \{$\nu_{e_\mathrm{L}}^i$,$\nu_{e_\mathrm{L}}'^i$,$\nu_{e_\mathrm{R}}'^i$,$\nu_\mathrm{R}^j$\} $\otimes$ \{$\nu_{e_\mathrm{L}}^i$,$\nu_{e_\mathrm{L}}'^i$,$\nu_{e_\mathrm{R}}'^i$,$\nu_\mathrm{R}^j$\} we arrive at 
\begin{equation}\label{eq:Mass_matrix_neutrino}
[M_\nu] = 
\begin{bmatrix}
\begin{bmatrix}0\end{bmatrix}_{3\times 3} & \begin{bmatrix}0\end{bmatrix}_{3\times 3} & \begin{bmatrix}M_{LE}\end{bmatrix}_{3\times 3} & \begin{bmatrix}\dfrac{v_a\Sigma^a}{\sqrt{2}}\end{bmatrix}_{3\times 6} \\
\begin{bmatrix}0\end{bmatrix}_{3\times 3} & \begin{bmatrix}0\end{bmatrix}_{3\times 3} & \begin{bmatrix}M_{E}\end{bmatrix}_{3\times 3} & \begin{bmatrix}\dfrac{v_a\Upsilon^a}{\sqrt{2}}\end{bmatrix}_{3\times 6} \\
\begin{bmatrix}M_{LE}\end{bmatrix}^\dagger_{3\times 3} & \begin{bmatrix}M_{E}\end{bmatrix}^\dagger_{3\times 3} & \begin{bmatrix}0\end{bmatrix}_{3\times 3} & \begin{bmatrix}\dfrac{v_a\Omega^a}{\sqrt{2}}\end{bmatrix}_{3\times 6} \\
\begin{bmatrix}\dfrac{v_a\Sigma^a}{\sqrt{2}}\end{bmatrix}_{6\times 3}^\dagger & \begin{bmatrix}\dfrac{v_a\Upsilon^a}{\sqrt{2}}\end{bmatrix}_{6\times 3}^\dagger & \begin{bmatrix}\dfrac{v_a\Omega^a}{\sqrt{2}}\end{bmatrix}_{6\times 3}^\dagger & \begin{bmatrix}M_{\nu_R}\end{bmatrix}_{6\times 6}
\end{bmatrix}\,,
\end{equation}
where $i=1,2,3$ as usual and $j=1,\ldots,6$. For charged leptons, besides the SM-like states we also have exotic VLLs which we name as $e_4$, $e_5$ and $e_6$\footnote{Again, in accordance with \cite{Morais:2020odg,Morais:2020ypd}, we have $e_4 \equiv E$, $e_5 \equiv M$, $e_6 \equiv T$.}, defined in such a way that $m_{e_6} > m_{e_5} > m_{e_4}$. The neutrino sector is quite rich in new particles, besides the three SM-like ones, we have a total of twelve new states. The numerical analysis will only consider the three lightest, keV-MeV scale BSM neutrinos which are denoted as $\nu_4 \equiv \nu_\textrm{BSM}$, $\nu_5$ and $\nu_6$. 

\subsubsection{Physically viable benchmark scenarios for masses}\label{subsubsec:Benchmarks_masses}

Before moving to the numerical analysis we present possible benchmark scenarios for couplings and masses in such a way to preserve the key properties emergent from the unification picture as well as complying with the measured phenomenological quantities.

The main focus of this work is the construction of an analysis framework dedicated to the study of VLLs and how important the DL techniques can be. This will enable us to propose robust signal events to be tested via direct searches at the LHC as well as understanding whether the model under consideration can be probed in such a sector. As it was shown in \cite{Morais:2020ypd}, under certain approximations and before EW symmetry breaking (EWSB), the lepton mass matrix is reduced to\footnote{It is important to note that \textbf{this does not} represent a one-to-one correspondence between \eqref{eq:lepton_mass_matrix} and \eqref{eq:Mass_matrix_lepton}. One should interpret \eqref{eq:lepton_mass_matrix} as a matrix one would get by following all symmetry breaking steps as seen in Fig.~\ref{fig:Symmetry-breaking}, while \eqref{eq:Mass_matrix_lepton} corresponds to the stage immediately after the $\omega$, $f$ and $p$ VEVs.}
\begin{equation}\label{eq:lepton_mass_matrix}
[M_L] = \begin{bmatrix}
0 & 0 & 0 & 0 & 0 & 0 \\
0 & 0 & 0 & 0 & \kappa_7\omega & \kappa_5\omega \\
0 & 0 & 0 & 0 & \kappa_6\omega & \kappa_8\omega \\
0 & 0 & 0 & 0 & \kappa_1 p & \kappa_3 f \\
0 & 0 & 0 & \kappa_2 p & 0 & 0 \\
0 & 0 & 0 & \kappa_4 f & 0 & 0 
\end{bmatrix}\,,
\end{equation}
where the various $\kappa_i$ terms are radiatively generated Yukawa couplings, thus expected to be smaller than unity. 
The VLL masses are then
\begin{equation}\label{eq:VLL_mass}
\begin{aligned}
&m_{e_6}^2 = p^2 \kappa_2^2 + f^2\kappa_4^2 \,, \\
&m_{e_5,e_4}^2 = \frac{1}{2}\Bigg(\omega^2\Lambda_1 + p^2\kappa_1^2 + f^2\kappa_3^2 \pm \Big[\qty(\omega^2\Lambda_1 + p^2\kappa_1 + f^2\kappa_3^2)^2 \\ &\quad\quad\quad-4\omega^2 \qty(\omega^2\Lambda_2 - 2fp\Lambda_3 + p^2\Lambda_4 + f^2\Lambda_5)\Big]^{1/2}\Bigg) \,,
\end{aligned}
\end{equation}
where we defined $\Lambda_1 = \kappa_5^2 + \kappa_6^2 + \kappa_7^2 + \kappa_8^2$, $\Lambda_2 = (\kappa_5\kappa_6 - \kappa_7\kappa_6)^2$, $\Lambda_3 = (\kappa_5\kappa_7 + \kappa_6\kappa_8)\kappa_1\kappa_3$, $\Lambda_4 = (\kappa_5^2 + \kappa_8^2)\kappa_1^2$ and $\Lambda_5 = (\kappa_6^2 + \kappa_7^2)\kappa_3^2$. The plus sign in \eqref{eq:VLL_mass} 
corresponds to $e_5$ and the minus sign to $e_4$.

Considering a scenario where $\omega \sim f \ll p$, Taylor expansion of \eqref{eq:VLL_mass} leads to the simplified 
expressions
\begin{equation}\label{eq:VLL_mass_taylor}
\begin{aligned}
&m_{e_6} \approx p \kappa_2\,, \\
&m_{e_5} \approx p\kappa_1\,, \\
&m_{e_4} \approx \omega\sqrt{\kappa_5^2 + \kappa_8^2} \,.
\end{aligned}
\end{equation}
Along the lines of what was discussed in \cite{Morais:2020ypd}, let us consider a set of possible solutions with
	\begin{itemize}
		\item $\kappa_2 \sim \mathcal{O}(10^{-2})$, $\kappa_1 \sim \mathcal{O}(10^{-3.5} - 10^{-2})$ 
		and $\kappa_{5,8} \sim \mathcal{O}(10^{-3} - 10^{-2})$,
		\item $p \sim \mathcal{O}(500 - 1000~\mathrm{TeV})$ and $\omega \sim f \sim \mathcal{O}(100 ~\mathrm{TeV})$.
	\end{itemize}
	This benchmark scenario leads to the following mass ranges
	\begin{itemize}
		\item $m_{e_6} \sim \mathcal{O}(5 - 10 ~\mathrm{TeV})$,
		\item $m_{e_5} \sim \mathcal{O}(0.15 - 10 ~\mathrm{TeV})$,
		\item $m_{e_4} \sim \mathcal{O}(0.1 - 1 ~\mathrm{TeV})$,
	\end{itemize}
	which we will use as a guiding principle for our numerical analysis. In particular, we see that for the model under consideration $e_4$ can be light enough to be probed at the LHC. On another hand, $e_6$ will always be rather heavy and a potential observation at the LHC would likely be very challenging. Regarding $e_5$, we see that it can either be as heavy as $e_6$ or as light as $e_4$ depending on yet unexplored model details. Based on this estimation, we will consider both possibilities in the numerical studies. 
	
To finalise this subsection, let us consider the neutrino sector. Before EWSB, the mass matrix is block diagonal,
\begin{equation}\label{eq:Neutrino_blocs}
M_\nu = \begin{bmatrix}
\bar{M}_{9\times 9} & 0 \\
0 & M_{6\times 6}
\end{bmatrix}\,,
\end{equation}
where $\bar{M}$ represents neutral components belonging to $\SU{2}{L}$ doublets while $M$ denotes SM singlets corresponding to $\nu_R$ in Tab.~\ref{tab:BSM}. Starting with the $M_{6\times6}$ block, which corresponds to $M_{\nu_R}$ in \eqref{eq:Bilinear}, its components offer the larger contributions to the neutrino mass matrix. In this sector, hierarchies among gauge eigenstates result from the relative sizes of the EW-preserving VEVs. On the other hand, the $\bar{M}$ components are radiatively generated and share the same properties as the VLLs. Thus, after the $p$, $f$ and $\omega$ VEVs one can write
\begin{equation}\label{eq:M_9times9}
\bar{M} = \begin{bmatrix}
0 & 0 & 0 & 0 & 0 & 0 & 0 & 0 & 0 \\
0 & 0 & 0 & 0 & 0 & 0 & 0 & \kappa_7\omega & \kappa_5\omega \\
0 & 0 & 0 & 0 & 0 & 0 & 0 & \kappa_6\omega & \kappa_8\omega \\
0 & 0 & 0 & 0 & 0 & 0 & 0 & \kappa_1p & \kappa_3f \\
0 & 0 & 0 & 0 & 0 & 0 & \kappa_2p & 0 & 0 \\
0 & 0 & 0 & 0 & 0 & 0 & \kappa_4f & 0 & 0 \\
0 & 0 & 0 & 0 & \kappa_2p & \kappa_4f & 0 & 0 & 0 \\
0 & \kappa_7\omega & \kappa_6\omega & \kappa_1p & 0 & 0 & 0 & 0 & 0 \\
0 & \kappa_5\omega & \kappa_8\omega & \kappa_3f & 0 & 0 & 0 & 0 & 0 \\
\end{bmatrix}
\end{equation}
with eigenvalues,
\begin{equation}\label{eq:Eigen_neutrinos}
\begin{aligned}
m^2_{\nu_{1,2,3}} = 0\,,  \quad m^2_{\nu_{4,5}} = m^2_{e_6}\,, \quad 
m^2_{\nu_{6,7}} = m^2_{e_5}\,, \quad m^2_{\nu_{8,9}} = m^2_{e_4}\,,
\end{aligned}
\end{equation}
such that, the left-handed neutrino components, at this stage, share the same masses as their charged lepton partners. In total, and before EWSB, we have three massless, and twelce massive neutrinos (six from the doublets and six from singlets). In the corresponding mass basis, if we identify the massive states as $\mu_i$ ($i=1,\dots,12$), we can recast the neutrino mass matrix in a condensed notation as
\begin{align}
\label{eq:mn-seesaw}
m_\nu & ~=~  \left(
\begin{array}{cc}
 \bm{0}_{3 \times 3}  \; &\;
\dfrac{v_{\text{EW}}}{\sqrt{2}}\left(\bm{y}_\nu\right)_{3 \times 12} \\
 \dfrac{v_{\text{EW}}}{\sqrt{2}}\left(\bm{y}_\nu^\top\right)_{12 \times 3}  
 \; & \; \left(\bm{\mu}_{N}\right)_{12 \times 12}  
\end{array}
\right)\,,
\end{align}
where the contribution of EWSB VEVs was already included. Note that $\bm{y}_\nu$ are the $3\times12$ Yukawa matrices whose entries are all radiatively generated. While a more dedicated analysis is beyond the scope of this work, this structure can potentially offer three sub-eV states as well as light keV-MeV order sterile neutrinos as we will assume in our numerical studies. 

\subsubsection{Physically viable benchmark scenarios for couplings}\label{subsubsec:Benchmarks_couplings}

In our numerical analysis we will be using \texttt{MadGraph5} \cite{Alwall:2014hca} which is a tool that requires a theory written in the mass basis. Therefore, not only masses but also couplings need to be rotated to such a basis. To this end we use \texttt{SARAH} \cite{Staub:2013tta}, which also offers a complete set of Feynman rules with physical fields. All relevant diagrams for our studies are shown in appendix \ref{app:Feynman Rules}. Note that all signal and background processes that we will consider involve only triple gauge self-interactions as well as fermion-fermion-gauge vertices. While the gauge sector is purely SM-like with well-known parameters, the Feynman rules involving fermion vertices will be sensitive to elements of the mixing matrices in the charged lepton (including VLLs) and neutrino sectors, defined by the bi-unitary transformations
\begin{equation}\label{eq:Mixing_matrices}
\begin{aligned}
& U^e_L\cdot M_L\cdot {U^e_R}^\dagger = m_e^{\text{diag}}\,, \\
& U_\nu \cdot m_\nu\cdot  U_\nu^\dagger = m_\nu^{\text{diag}} \,.
\end{aligned}
\end{equation}

Let us now discuss which phenomenological constraints are applied to these matrices. First, for the charged leptons, we consider the limit where the SM-like sector is flavour-diagonal. Therefore, in $U^e_L$ and $U^e_R$, we add a $3\times3$ identity block and consider a limiting scenario where there is no mixing with VLLs. While this may not be the case in general, a realistic scenario cannot strongly deviate from the flavour alignment limit that we impose. A complete study with flavour mixing is beyond the scope of this work. For the VLL block, we consider a generic mixing with the only restriction being that both $U^e_L$ and $U^e_R$ are unitary. To summarize, the lepton mixing matrices used in the numerical analysis are given by
\begin{equation}\label{eq:lepton-mixing}
U^e_{L,R} = \begin{bmatrix}
\mathbb{1}_{3\times 3} & 0_{3\times 3} \\[0.15cm]
0_{3\times 3} & (U^{\text{VLL}}_{L,R})_{3\times 3}
\end{bmatrix}\,,
\end{equation}
where $U^{\text{VLL}}_{L,R}\cdot{U^{\text{VLL}}_{L,R}}^\dagger = \mathbb{1}_{3\times 3}$.

For the neutrino sector, we also consider a limiting scenario where, for simplicity, the mixing between the three light active neutrinos and the remaining twelve BSM states is zero. Once again, a more generic case with flavour mixing is beyond the scope of our analysis and does not significantly affect our main conclusions. Note, however, that mixing among light neutrinos is allowed and fixed by the PMNS matrix. For the remaining BSM $12\times12$ block we recall that the mixing among right-handed and left-handed components is radiatively generated and is thus likely small. Here, we consider that those elements are always smaller than $10^{-3}$. Having said this, the full neutrino mixing matrix reads
\begin{equation}\label{eq:neutrino-mixing}
U_\nu = \begin{bmatrix}
\mathbb({U}_{\text{PMNS}})_{3\times 3} & 0_{3\times 12} \\[0.15cm]
0_{12\times 3} & (U^{\text{BSM}}_{\nu})_{12\times 12}
\end{bmatrix}\,,
\end{equation}
with,
\begin{equation}\label{eq:neutri-BSM-mixing}
U_\nu^{\text{BSM}} = \begin{bmatrix}
\mathbb(U_1)_{6\times 6} & (D_1)_{6\times 6} \\[0.15cm]
(D_1)^\dagger_{6\times 6} & (U_2)_{6\times 6}
\end{bmatrix}\,.
\end{equation}
We set the matrix elements in $D_1$ to be of order $\mathcal{O}(10^{-3}-10^{-8})$ while in $U_{1,2}$ they are randomly generated in a way that preserves unitarity and guarantees that $e_4$ couples democratically to the sterile neutrinos. With the above ingredients we have defined a possible benchmark scenario to start our collider phenomenology studies while preserving the essential features of the model under consideration.

\section{Searching for vector-like leptons at the LHC}\label{section:Numerics}

As a first step, we create the necessary \texttt{UFO} \cite{Degrande:2011ua} files using \verb|SARAH|. These are later used by \verb|MadGraph5| (MG5) to generate the signal and background Monte-Carlo events. Hard-scattering events are generated with \verb|Pythia8| \cite{Sjostrand:2014zea}, and then \verb|Delphes| \cite{deFavereau:2013fsa} is used to include hadronization and detector effects. 

\begin{figure}[ht!]
	\centering
	\includegraphics[width=0.40\textwidth]{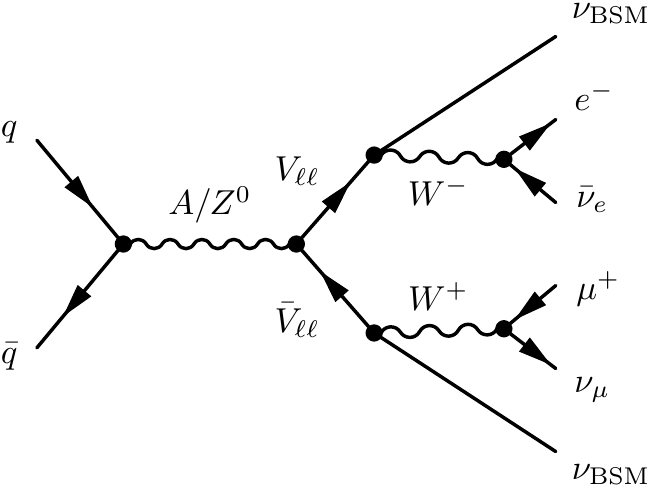}
	\caption{Leading-order Feynman diagram for the ZA topologies. Here, $q$ and $\bar{q}$ correspond to quarks originating from the colliding protons, $V_{\ell\ell}$ represents VLLs and $\nu_{\text{BSM}}$ denotes the lightest BSM neutrino. There are two purely SM leptonic final channels identified with $\ell$ and $\nu_\ell$.}\label{fig:ZA-events}
\end{figure}

All hard-scattering events are generated using $pp$ collisions at 14 TeV center-of-mass energy, with the parton distribution function \textit{nn23lo1} and with the strong coupling constant $\alpha_{s}$ fixed automatically by MG5. We have generated a total of 250000 events. The background channels up to two extra jets are generated with the MLM matching scheme \cite{Hoche:2006ph}. While substantial theoretical work has already been done over the last decades \cite{Bell:2019mbn,Falkowski:2013jya,Garcia:2015sfa,Kumar:2015tna,Dermisek:2013gta,Dermisek:2014qca,Holdom:2014rsa,Ellis:2014dza,Dorsner:2014wva,Raby:2017igl,Poh:2017tfo,Bhattacherjee:2017cxh,Kawamura:2019hxp,Bhattiprolu:2019vdu,Fujikawa:1994we}, only recently, the searches for exotic charged leptons have started. The most recent analysis was done by the CMS collaboration at the LHC in 2019 \cite{Sirunyan:2019ofn}, where a search for VLLs coupling to taus was performed. In fact, one of the three topologies that we propose in our analysis is very similar to the one in Fig.~1 of \cite{Sirunyan:2019ofn}, and more in line to what we see, for example, in Fig.~30 of \cite{Bhattacharya:2018fus}. In the context of our model, such topology can be seen in Fig.~\ref{fig:ZA-events} which, in what follows, will be referred to as ``ZA''.

Recall that we are treating the lightest, sterile, BSM neutrinos as missing energy as long as their mass is in the keV-MeV mass range. This means that possible decays are kinematically forbidden and therefore, such neutrinos escape the detector. Of course, it would be interesting to study a scenario where we have SM neutrinos instead of the BSM ones provided that the final state would be purely the SM one. However, due to the structure of the mixing matrix used in this study \eqref{eq:lepton-mixing}, such a mixing is non-existing. We are leaving for a future work the scenario where inclusion of a non-zero mixing between SM leptons and VLLs, in consistency with both the flavour observables constraints and predictions from the high-scale theory, is implemented.

\begin{figure}[t]
	\centering
	\subfloat{	\includegraphics[width=0.35\textwidth]{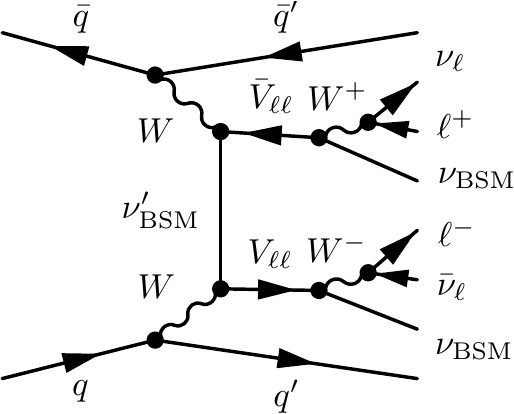}} 
	\subfloat{	\includegraphics[width=0.35\textwidth]{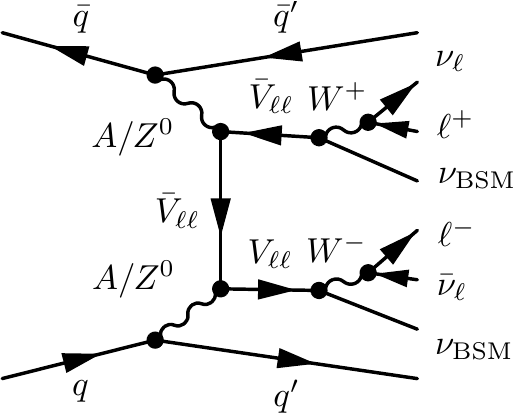}}
	\caption{Leading-order Feynman diagrams for the VBF topologies. The same nomenclature as seen in Fig.~\ref{fig:ZA-events} applies here. $\nu'_{\text{BSM}}$ correspond to any BSM neutrino.}
	\label{fig:VBF-events}
\end{figure}

\begin{figure}[h!]
	\centering
    \includegraphics[width=0.40\textwidth]{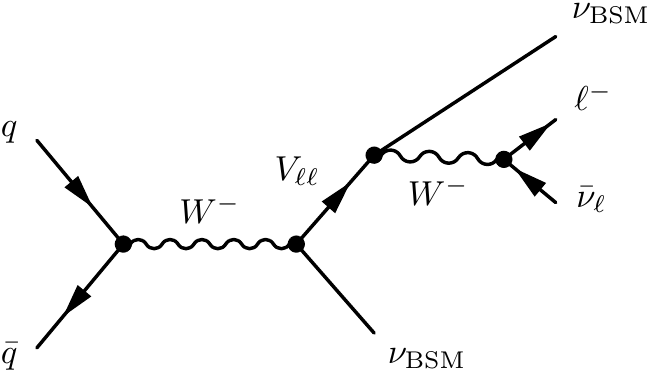}
	\caption{Leading-order Feynman diagrams for the VLBSM topologies. The same nomenclature as seen in Fig.~\ref{fig:ZA-events} applies here.}\label{fig:VLBSM-events}
\end{figure}

We also consider vector-boson fusion events whose topology is shown in Fig.~\ref{fig:VBF-events} and that we shall refer to as ``VBF''. While the latter is expected to have a smaller cross section, the presence of two well-defined forward jets enables us to tag such events using the high transverse mass of the forward jets. The signal channels, we propose here, provide a good starting point for our analysis. However, due to the expected low cross section for the signal events in comparison with the overwhelming cross-section of the irreducible background, searching for such particles at the LHC solely considering these two processes can become rather challenging. A third channel, denoted as ``VLBSM'', with only four internal vertices (VBF diagrams contain eight vertices while ZA ones -- six) is then considered and can be seen in Fig.~\ref{fig:VLBSM-events}. Furthermore, we use DL techniques inspired from previous works \cite{Cogollo:2020afo,Freitas:2019hbk,Alves:2019ppy} and tailored for our analysis, in order to efficiently discriminate signal from background.

Another possible signal topology would be to consider diagrams similar to Fig.~\ref{fig:VLBSM-events} but with a neutral boson $Z^0$/$A$ decaying directly into a pair $\bar{e}_4e_4$, that is, $pp \rightarrow Z^0/A \rightarrow \bar{e}_4e_4$. This would provide us a sizeable cross section, a clean signal and could appear as charged tracks in the detector potentially offering a good smoking gun for our model. However, preliminary numerical calculations showed that $e_4$ decays too quickly and does not reach the detector track chamber such that one would have to reconstruct $e_4$ from their main product decays. Therefore, we will only consider the ZA, VBF and VLBSM topologies in our analysis and leave other possibilities for a future work.

\begin{figure}[t!]
	\centering
	\subfloat{\includegraphics[width=0.31\textwidth]{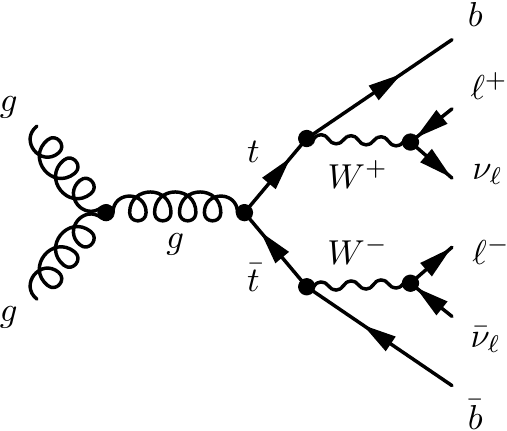}} 
	\hspace{1cm}
	\subfloat{\includegraphics[width=0.31\textwidth]{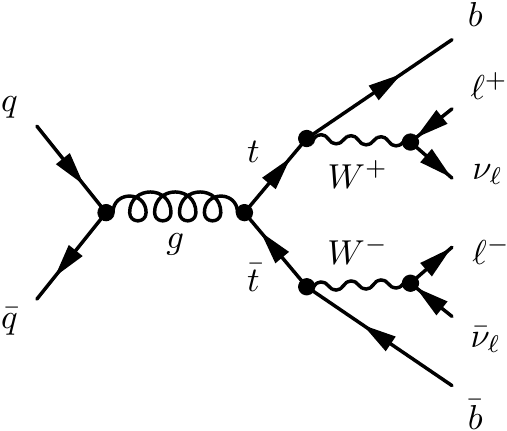}}\\
	\subfloat{\includegraphics[width=0.31\textwidth]{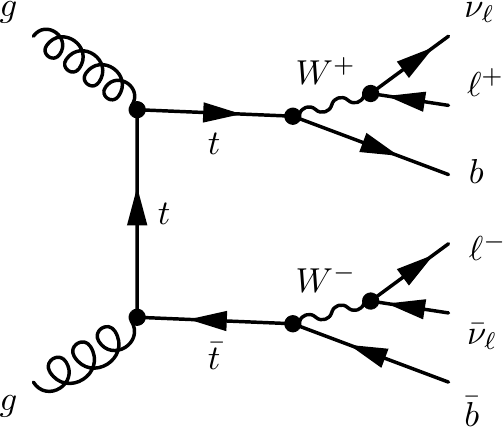}}
	\caption{$t\bar{t}$ background topologies with fully leptonic final states. All particles shown here are purely SM ones. The final leptonic channels in both the signal events and the $t\bar{t}$ background processes shown here are chosen to be identical.}
	\label{fig:ttbackground}
\end{figure}

\begin{figure}[t!]
	\centering
	\subfloat{\includegraphics[width=0.31\textwidth]{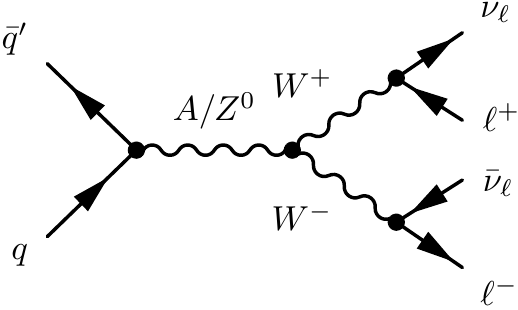}} 
	\hspace{1cm}
	\subfloat{\includegraphics[width=0.31\textwidth]{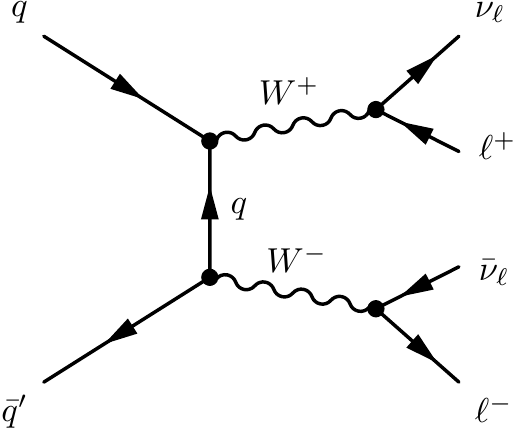}}
	\caption{$W^+ W^-$ background topologies. All particles shown here are purely SM ones. The final leptonic channels in both 
	the signal events and the $W^+ W^-$ background processes shown here are chosen to be identical.}
	\label{fig:WWbackground}
\end{figure}

The main irreducible background for each signal channel is chosen as follows:
\begin{itemize}
    \item For ZA topologies, we consider $t\bar{t}$ and $W^+W^-$ both with fully leptonic final states, $t\bar{t} +Z^0$, with $Z^0$ decaying into lepton/anti-lepton pair and $t\bar{t}$ fully leptonic decay and finally $t\bar{t} +Z^0$ with $Z^0$ decaying into neutrinos.
    \item For VBF topologies, we consider $W^+W^-$ with fully leptonic final states, $t\bar{t} + (j,jj)$, where tops decay into 
    leptons accompanied with either one or two light jets.
    \item For VLBSM topologies, we consider the single lepton production $pp\rightarrow \ell\nu_\ell$ with zero, one and two light jets.
\end{itemize}
Both the background and signal leptonic final states are chosen to be identical. The Feynman diagrams for $W^+W^-$ and $t\bar{t}$ with fully leptonic final states are displayed in Figs.~\ref{fig:ttbackground} and \ref{fig:WWbackground}.

To facilitate the signal detection and reduce the background contamination we consider final-state leptons of different flavour. In particular, we choose $W^{+}$ decaying to $\mu^+$ and $\nu_\mu$, while for $W^-$ we consider the $e^- + \bar{\nu}_e$ channel. We also choose the following simple kinematic cuts as event selection criteria:
\begin{enumerate}
\label{selc_cut}
\item Charged leptons ($e^-$ and $\mu^+$) are required to have a transverse momentum $p_T > 25$ GeV and $\abs{\eta} \leq 5$ and
\item Missing transverse energy $\slashed{\it{E}}_{T} > 15$ GeV. 
\item For events with jets, we use the Cambridge/Aachen jet clustering algorithm with  $\Delta R = 1.0$, transverse momentum $p_T > 35$ GeV and pseudo-rapidity $\abs{\eta}\leq 5$. 
\end{enumerate}	

At this point we are able to reconstruct all particles up to the VLLs with relative precision using the information from the final state visible particles, tracks and calorimetric towers provided by the \verb|Delphes| output. All chosen observables in our studies are detailed in Sec.~\ref{subsec:Results}. We compute the observables both in the lab frame, as well as the $W$ frame. We also emphasize that all Monte Carlo simulations and posterior data processing (\texttt{PYTHIA8}, \texttt{MadGraph}, \texttt{Delphes}) are performed in the \texttt{blafis}\footnote{Technical details can be found at the Gr@v's website \cite{blafis}.} and \texttt{ARGUS} computer clusters as part of the overall computing infrastructure at the University of Aveiro.

\subsection{Methodology: Deep Learning models and dataset}\label{DLmodels}

In this section, we describe the construction of our neural network (NN) models and what are the best architectures we found to accurately separate and identify signal events from their respective backgrounds. For the unfamiliar reader, NNs, and by extension DL algorithms, are rooted in the \textit{universal approximation theorem}, which essentially states that a given NN with a given number of hidden layers and a finite number of hidden units, a.k.a neurons, can approximate any arbitrary continuous function on compact subsets of $\mathbb{R}^{n}$; this function can describe a hyperdimensional plane which separate samples from distinct classes (classification problems) or predict new samples based om previous one (regression problems). However, the universal approximation theorem does not tells us how deep the NN, neither the number of hidden units needed to better approximate the desired arbitrary function. This is a problem that one faces when identifying clever solutions and finding better designs for NNs.

The main goal of our NN model is to classify the signal channels defined in the previous section over each respective background. This procedure is often denoted as classification. For a better performance one must determine what kind of architecture should be employed, the number layers, how many neurons each layer needs, etc. An appropriate choice of such parameters can lead to models which are capable of giving very accurate predictions, which in the context of high energy physics can potentially lead to discoveries using the available data. 

The problem of selecting the correct parameters is refereed to as hyperparameter optimization or tuning. There are many methods to search for the best combinations of parameters including a brute force approach by testing each possible combination until the optimal NN model is recovered. However, as one might expect such a method is very time consuming. A more efficient procedure consists in using an evolutionary algorithm search \cite{Freitas:2019hbk,Alves:2019ppy}. We define for our analysis the following set of hyperparameters:
\begin{itemize}
    \item number of hidden layers: 1 to 5
    \item number of neurons in each layer: 256, 512, 1024 or 2048
    \item kernel initializer: 'normal','he normal','he uniform'
    \item L2 regularization penalty: 1e-3, 1e-5, 1e-7
    \item activation function: 'relu', 'elu', 'tanh', 'sigmoid'
    \item optimizer: 'adam', 'sgd', 'adamax', 'nadam'
\end{itemize}
\noindent Our evolutionary algorithm is initialized by building a set of ten NNs using \texttt{Keras} \cite{chollet2015keras} with \texttt{TensorFlow} \cite{Abadi:2016kic} as back-end. The hyperparameters are then randomly chosen from the ones presented in the list above. Each NN is trained for 200 epochs and once the training phase is complete we select the top five NNs that have shown better performances in order to ``breed'' new NNs for the next iteration. Such NNs are then initialized with the hyperparameters of the selected ones and treated as ``parent traits'' while randomly including new ones as mutations. We have set a 20\% probability of a random mutation to occur. We then construct a new population set and repeat the training/evaluation process for five times, five generations until finally retrieving the best NN architecture. 

Another important aspect of the evolutionary algorithm is the fitness function. This helps the algorithm to select the best architectures based on a pre-defined metric. In our case, we set two fitness functions, one where the best models are ranked according to their accuracy on the test set, and a second one that ranks the models according to their Asimov significance defined as:
\begin{equation}\label{eq:Asimov_sig}
\mathcal{Z}_A = \Bigg[2\qty((s+b)\ln(\frac{(s+b)(b+\sigma_b^2)}{b^2 + (s+b)\sigma_b^2}) - \frac{b^2}{\sigma_b^2}\ln(1 + \frac{\sigma_b^2 s}{b(b+\sigma_b^2)}))\Bigg]^{1/2} \,,
\end{equation}
with $s$ and $b$ being the number of signal and background events, respectively, and $\sigma_b^2$ is the variance of background events. We include a similar procedure to that described in \cite{Elwood:2018qsr} and the necessary modifications for the training methodology into our evolutionary algorithm.

Although, the main ``body'' of our NN is built using the principles of natural selection \cite{bhl135954}, some characteristics of the model construction are universal to all NN models that we summarize as follows:
\begin{itemize}
	\item As input data, the NN receives a standard normalised vector, i.e.~such a vector has mean 0 and standard deviation 1, and data sets from all observables are extracted from the \texttt{ROOT} detector output. This data set is then reshuffled and divided into a training set (80 \% of data) 
	and a test set (20 \% of data). To avoid overfitting, we use the cross-validation with a five-fold scheme during the training of the NN.
	\item We employ a cyclic learning rate during the training phase with 0.01 initial value and maximal value of 0.1.
    \item In the output layer, the data is transported to a vector, with entries between 0 and 1 (which correspond to probabilities), in the format $(S,B_j)$, where $S$ is the signal and $B_j$ correspond to different backgrounds. The index $j$ runs over the number of backgrounds chosen for a given signal. As an example, in ZA we consider 4 distinct backgrounds ($j = 4$). So, such a vector would correspond to $(S,B_1,B_2,B_3,B_4)$. 
	\item Batch size of 32768 entries. 
	\item We impose a total limit of 200 epochs with a patience of 5 epochs and a validation loss monitor, i.e.~if the loss value on the test/validation set did not change for 5 epochs the training is resumed and all metrics are computed and stored to be passed to the evolutionary algorithm.
	\item To select the NN models with better accuracy we use the binary cross entropy (BCE) loss, while for the selection of those models that maximize the Asimov significance we use Eq.~\eqref{eq:Asimov_sig} as a loss function during the evolutionary scan of the best hyperparameters.
\end{itemize}

Another important aspect to mention is the fact that our data is unbalanced, i.e., we have more data points for some of the classes we are analyzing (in some cases, with a ratio of $s/b \approx 1.83$). This is a result of the event selection criteria that we impose for the signal and background. Unbalanced data can lead to models with lower predictive power for substantially outnumbered classes, which is a serious issue if one wants to search for NP phenomena. To avoid this, we use the Synthetic Minority Oversampling Technique (SMOTE) \cite{2011arXiv1106.1813C} to create synthetic entries for the minority classes in our training dataset. Note that we do not employ any re-sampling technique on the test dataset. This method is faster and more efficient than generating additional Monte-Carlo events and passing them through subsequent hadronization and detector effects.


Our dataset is stored in a table format where each row corresponds to an event entry that has successfully passed the selection cuts described in \ref{selc_cut}, and each column corresponds to the observables described in Tabs.~\ref{tab:vars_Zp} and \ref{tab:vars_Zp_VLBSm}. The dimensions of each training and test datasets are displayed in Tab.~\ref{tab:dataset}.
\begin{table*}[ht!]    
	\centering
\resizebox{\textwidth}{!}{\begin{tabular}{|c|c|c|c|}
		\toprule
		\hline
		& Dimension-full & \multicolumn{2}{|c|}{Dimensionless} \\
		\hline
		\hline
		\midrule
		\makecell{Lab. \\
			frame}  & \makecell{$p_T(e^-)$, $p_T(\mu^+)$,$p_T(e_4)$ \\
			$p_{T}(\bar{e}_4)$, $m_{e_4}$, $m_{\bar{e}_4}$ \\
			$M_T(W^-)$, $M_T(W^+)$, MET} & 
		\makecell{
			$\cos(\theta_{\bar{\nu}_e e})$, $\cos(\theta_{\bar{\nu}_\mu \mu^+})$, \\
			$\cos(\theta_{W^- W^+})$, \\ $\cos(\Delta \phi)$,
			$\cos(\Delta \theta)$, \\ $\eta_e$, $\eta_{\mu^+}$, $\eta_{e_4}$, $\eta_{\bar{e}_4}$ } & 
		\makecell{$\Delta R(e, \bar{\nu_e})$, $\Delta R(\mu^+, \nu_{\mu^+})$}\\
		\hline
		\makecell{$W^-$ \\
			frame} & \makecell{$p_T(e^-)$, $p_T(e_4)$}& \makecell{
			$\cos(\theta_{\bar{\nu}_e e})$, \\
			$\eta_e$, $\eta_{e_4}$}  & 
		\makecell{}\\
		\hline
		\makecell{$W^+$ \\
			frame} &\makecell{$p_T(\mu^+)$, $p_T(\bar{e}_4)$} & \makecell{
			$\cos(\theta_{\nu_\mu \mu^+})$, \\
			$\eta_{\mu^+}$, $\eta_{\bar{e}_4}$} & 
		\makecell{}\\           
		\hline
		\makecell{$\ell'\bar{\ell}'$ \\
			frame} &\makecell{} & \makecell{
			$\cos(\Delta \phi)$,
			$\cos(\Delta \theta)$} & 
		\makecell{}\\
		\hline
		\hline
	\end{tabular}}
	\caption{Kinematic (dimension-full) and angular (dimensionless) observables selected to study the ZA and VBF channels. We include observables in four different frames of reference: laboratory frame (top row), $W^-$ rest frame (second row), $W^+$ rest frame (third row) and $\ell'\bar{\ell}'$ frame. $\theta_{i,j}$ denotes the angle between the respective particles from either the final state or reconstructed objects. }
	\label{tab:vars_Zp}
\end{table*}

\begin{table*}[ht!]    
	\centering
    \begin{tabular}{|c|c|c|}
		\toprule
		\hline
		& Dimension-full & {Dimensionless} \\
		\hline
		\hline
		\midrule
		\makecell{Lab. \\
			frame}  & \makecell{$p_T(\mu^+)$,$M_T(W)$, \\
			  $p_T(W)$, MET} & 
		\makecell{
			$\cos(\theta_{\mu^+})$, $\cos(\theta_{\bar{\nu}_\mu \mu^+})$, \\
			$\cos(\theta_{W})$, $\eta_{\mu^+}$, $\eta_{W}$, $\phi_{\mu^{+}}$}  \\
		\hline
		\hline
	\end{tabular}
	\caption{Kinematic (dimension-full) and angular (dimensionless) observables selected to study the VLBSM channel. We compute observables in the laboratory frame where $\theta_{i,j}$ is the angle between the respective particles from either the final state or reconstructed objects.}
	\label{tab:vars_Zp_VLBSm}
\end{table*}

\begin{table*}[ht!]    
	\centering
	\resizebox{\textwidth}{!}{\begin{tabular}{|c|c|c|c|c|c|c|}
		\toprule
		\hline
    		& \multicolumn{2}{|c|}{ZA} & \multicolumn{2}{|c|}{VBF} & \multicolumn{2}{|c|}{VLBSM} \\
		\cline{2-7}
		dataset:\makecell{Original \\ Training (SMOTE) \\ Test}  & \makecell{Signal} & 
		\makecell{Backgrounds} & 
		\makecell{Signal} & 		\makecell{Backgrounds} &
		\makecell{Signal} &
		\makecell{Backgrounds} \\
		\hline
		\makecell{\texttt{$m_{e_4} = 200$ GeV}} & \makecell{ (77405, 36) \\(65983, 36) \\ (15481, 36)}& \makecell{\begin{math}\begin{aligned}
			&t\bar{t}: &&\makecell{(44725,36) \\ (65983, 36) \\ (8945, 36)} \\
			&t\bar{t}, Z^0(l^+l^-) :&& \makecell{(81825,36) \\ (65983, 36) \\ (16365, 36)} \\ 
			&t\bar{t}, Z^0(\nu_\ell \bar{\nu}_\ell) : &&\makecell{(46705,36) \\ (65983, 36) \\ (9341, 36)} \\
			&W^+W^- : &&\makecell{(48475,36) \\ (65983, 36) \\ (9695, 36)}		\end{aligned}\end{math}}  & 
		\makecell{(115330,36) \\ (91444, 36) \\ (23066, 36)} &
		\makecell{\begin{math}\begin{aligned}	&t\bar{t} + j (jj) : &&\makecell{(85870,36) \\ (91444, 36) \\ (17174, 36)} \\
			        &W^+W^- : &&\makecell{(48170,36) \\ (91444, 36) \\ (9634, 36)} \end{aligned}\end{math}} &
		\makecell{(147825,10) \\ (128870, 10) \\ (29565, 10)} &
		\makecell{\begin{math}\begin{aligned} &\ell\nu_\ell : &&\makecell{(84185,10) \\ (128870, 10) \\ (16837, 10)} \\
		&\ell\nu_\ell + j (jj) : &&\makecell{(160630,10) \\ (128870, 10) \\ (32126, 10)}\end{aligned}\end{math}} \\
		\hline
		\makecell{\texttt{$m_{e_4} = 486$ GeV}} & \makecell{(125455,36) \\ (100656, 36) \\ (25091, 36)}& \makecell{\begin{math}\begin{aligned}
			&t\bar{t} : &&\makecell{(45625,36) \\ (100656, 36) \\ (9125, 36)} \\
			&t\bar{t}, Z^0(l^+l^-) : &&\makecell{(81845,36) \\ (100656, 36) \\ (16369, 36)} \\ 
			&t\bar{t}, Z^0(\nu_\ell \bar{\nu}_\ell) : &&\makecell{(46885,36) \\ (100656, 36) \\ (9377, 36)} \\
			&W^+W^- : &&\makecell{(48630,36) \\ (100656, 36) \\ (9726, 36)}\end{aligned}\end{math}}  & 
		\makecell{(143455,36) \\ (114892, 36) \\ (28691, 36)} &
		\makecell{	\begin{math}\begin{aligned}  &t\bar{t} + j (jj) : &&\makecell{(85840,36) \\ (114892, 36) \\ (17168, 36)} \\
			&W^+W^- : &&\makecell{(49145,36) \\ (114892, 36) \\ (9829, 36)}\end{aligned}\end{math}} &
		\makecell{(187530,10) \\ (149901, 10) \\ (37506, 10)} &
		\makecell{\begin{math}\begin{aligned} &\ell\nu_\ell : &&\makecell{(83065,10) \\ (149901, 10) \\ (16613, 10)} \\
		&\ell\nu_\ell + j (jj) : &&\makecell{(161460,10) \\ (149901, 10) \\ (32292, 10)}\end{aligned}\end{math}} \\
		\hline
		\makecell{\texttt{$m_{e_4} = 677$ GeV}} & \makecell{(137310,36) \\ (111257, 36) \\ (27462, 36)}& \makecell{\begin{math}\begin{aligned}
			&t\bar{t} : &&\makecell{(45060,36) \\ (111257, 36) \\ (9012, 36)} \\
			&t\bar{t}, Z^0(l^+l^-) : &&\makecell{(83905,36) \\ (111257, 36) \\ (16781, 36)} \\ 
			&t\bar{t}, Z^0(\nu_\ell \bar{\nu}_\ell) : &&\makecell{(46790,36) \\ (111257, 36) \\ (9358, 36)} \\
			&W^+W^- : &&\makecell{(48345,36) \\ (111257, 36) \\ (9669, 36)}\end{aligned}\end{math}}  & 
		\makecell{(143455,36) \\ (122589, 36) \\ (28691, 36)} &
		\makecell{	\begin{math}\begin{aligned} &t\bar{t} + j (jj) : &&\makecell{(85840,36) \\ (122589, 36) \\ (17168, 36)} \\
		&W^+W^- : &&\makecell{(49145,36) \\ (122589, 36) \\ (9829, 36)}\end{aligned}\end{math}} &
		\makecell{(195230,10) \\ (156047, 10) \\ (39046, 10)} &
		\makecell{\begin{math}\begin{aligned} &\ell\nu_\ell : &&\makecell{(83640,10) \\ (156047, 10) \\ (16728, 10)} \\
		&\ell\nu_\ell + j(jj) : &&\makecell{(160870,10) \\ (156047, 10) \\ (32174, 10)}\end{aligned}\end{math}} \\
		\hline
		\makecell{\texttt{$m_{e_4} = 868$ GeV}} & \makecell{(146085,36) \\ (116302, 36) \\ (29217, 36)}& \makecell{\begin{math}\begin{aligned}
			&t\bar{t} : &&\makecell{(44460,36) \\ (116302, 36) \\ (8892, 36)} \\
			&t\bar{t}, Z^0(l^+l^-) : &&\makecell{(82045,36) \\ (116302, 36) \\ (16409, 36)} \\ 
			&t\bar{t}, Z^0(\nu_\ell \bar{\nu}_\ell) : &&\makecell{(47240,36) \\ (116302, 36) \\ (9448, 36)} \\
			&W^+W^- : &&\makecell{(48380,36) \\ (116302, 36) \\ (9676, 36)}\end{aligned}\end{math}}  & 
		\makecell{(157250,36) \\ (125058, 36) \\ (31450, 36)} &
		\makecell{\begin{math}\begin{aligned}	&t\bar{t} + j (jj) : &&\makecell{(85245,36) \\ (125058, 36) \\ (17049, 36)}\\
			&W^+W^- : &&\makecell{(48870,36) \\ (125058, 36) \\ (9774, 36)}\end{aligned}\end{math}} &
		\makecell{(198290,10) \\ (158405, 10) \\ (39658, 10)} &
		\makecell{\begin{math}\begin{aligned} &\ell\nu_\ell : &&\makecell{(83740,10) \\ (158405, 10) \\ (16748, 10)} \\
		&\ell\nu_\ell + j (jj) : &&\makecell{(160680,10) \\ (158405, 10) \\ (32136, 10)}\end{aligned}\end{math}} \\
		\hline
		\makecell{\texttt{$m_{e_4} = 1250$ GeV}} & \makecell{(151020,36) \\ (120489, 36) \\ (30204, 36)}& \makecell{\begin{math}\begin{aligned}
			&t\bar{t} : &&\makecell{(44370,36) \\ (120489, 36) \\ (8874, 36)} \\
			&t\bar{t}, Z^0(l^+l^-) : &&\makecell{(81935,36) \\ (120489, 36) \\ (16387, 36)} \\ 
			&t\bar{t}, Z^0(\nu_\ell \bar{\nu}_\ell) : &&\makecell{(47425,36) \\ (120489, 36) \\ (9485, 36)} \\
			&W^+W^- : &&\makecell{(48635,36) \\ (120489, 36) \\ (9727, 36)}\end{aligned}\end{math}}  & 
		\makecell{(159445,36) \\ (126781, 36) \\ (31889, 36)} &
		\makecell{\begin{math}\begin{aligned}	&t\bar{t} + j (jj) : &&\makecell{(86005,36) \\ (126781, 36) \\ (17201, 36)} \\
			&W^+W^- : &&\makecell{(48080,36) \\ (126781, 36) \\ (9616, 36)}\end{aligned}\end{math}} &
		\makecell{(196480,10) \\ (156931, 10) \\ (39296, 10)} &
		\makecell{\begin{math}\begin{aligned} &\ell\nu_\ell : &&\makecell{(84140,10) \\ (156931, 10) \\ (16828, 10)} \\
		&\ell\nu_\ell + j (jj) : &&\makecell{(160255,10) \\ (156931, 10) \\ (32051, 10)}\end{aligned}\end{math}} \\
		\hline
		\hline
	\end{tabular}}
	\caption{Dataset dimensions for each value of the lightest VLL mass,  $m_{e_4}$. In the pairings $(X,Y)$, $X$ denotes the number of events (rows) while $Y$ is the number of features in each dataset. The $(X,Y)$ pairs are organized in groups of three. The top ones correspond to the original dataset before splitting (80/20\% ratio) and re-sampling using the SMOTE technique, the middle ones are the training set already balanced while the bottom ones correspond to the remaining test set (20\% of the original one).}
	\label{tab:dataset}
\end{table*}

\clearpage
\subsection{Results}\label{subsec:Results}

We start our discussion by presenting a specific benchmark point whose parametric choice was guided by our discussion in Secs.~\ref{subsubsec:Benchmarks_masses} and \ref{subsubsec:Benchmarks_couplings}. Note that the analysis methodology is independent of such a parametric choice. We will then study events for a light VLL ($e_4$) accompanied by the lightest BSM neutrino ($\nu_4$), which is treated as missing energy, and whose masses read as
\begin{equation}\label{eq:choice}
m_{e_4} = 677 \hphantom{.}\text{GeV} \,, \quad m_{\nu_4} = 216 \hphantom{.}\text{keV} \,.
\end{equation}
The decay width is automatically calculated in \texttt{MadGraph} in the narrow width approximation. While for the ZA and VLBSM events heavier neutrinos are not important, in the VBF case they should be taken into account since they appear as intermediate states. We then fix their masses as
\begin{equation}\label{eq:choice_neutrinos}
\begin{aligned}
&m_{\nu_5} = 0.138 \hphantom{.}\text{GeV}\,, \quad m_{\nu_6} = 36.7 \hphantom{.}\text{GeV}\,, \quad m_{\nu_7} = 2140\hphantom{.}\text{GeV}\,, \quad m_{\nu_8} = 2537\hphantom{.}\text{GeV}\,,\\ &m_{\nu_9} = 3035\hphantom{.}\text{GeV}\,, \quad m_{\nu_{10,11}} = m_{e_4}\,, \quad m_{\nu_{12,13}} = m_{e_5}\,, \quad m_{\nu_{14,15}} = m_{e_6}\,,
\end{aligned}
\end{equation}
with 
\begin{equation}\label{eq:choice_VLLs}
\begin{aligned}
m_{e_5} = 3258\hphantom{.}\text{GeV}\,, \quad m_{e_6} = 4240\hphantom{.}\text{GeV}\,.
\end{aligned}
\end{equation}
The BSM couplings are essentially the mixing matrices seen in Sec.~\ref{subsubsec:Benchmarks_couplings}. Here we adopt,
\begin{equation}\label{lepton_matrices}
\begin{aligned}
&U^{\text{VLL}}_{L} = \begin{bmatrix}
-0.162 - 0.381i & -0.683 - 0.321i & 0.318 + i0.379\\
-0.315-0.089i & -0.341+0.225i & -0.746-0.411i\\
0.844-0.0970i & -0.035+0.498i & -0.105-0.134i
\end{bmatrix}\,,\\
&U^{\text{VLL}}_{R} = \begin{bmatrix}
-0.186-0.490i & -0.266-0.462i & -0.654 + i0.113\\
0.389-0.153i & -0.750-0.366i & 0.306-0.186i\\
0.640-0.375i & 0.035+0.133i & -0.497-0.429i
\end{bmatrix} \,,
\end{aligned}
\end{equation}
for the VLLs, while for the neutrinos we have\footnote{The numerical values for $D_{1}$ are small, $\mathcal{O}(10^{-3}-10^{-8})$, so not to occupy too much space, they are omitted. The dominant contributions all come from $U_1$ and $U_2$.}
\begin{equation}\label{neutrino_matrices}
\begin{aligned}
&U_{1} = \begin{bmatrix}
-0.490 & -0.408 & -0.081 & -0.192 & 0.568 & 0.476 \\
0.491 & -0.745 & 0.436 & -0.063 & 0.014 & -0.10 \\
0.098 & -0.109 & -0.511 & -0.337 & 0.385 & -0.675 \\ 
-0.445 & 0.003 & 0.455 & 0.511 & 0.278 & -0.506 \\
-0.553 & -0.267 & 0.073 & -0.434 & -0.614 & -0.227 \\
0.068 & 0.443 & 0.574 & -0.630 & 0.270 & -0.029
\end{bmatrix}\,, \\
&U_{2} = \begin{bmatrix}
-0.413 & 0.166 & 0.360 & -0.313 & 0.270 & 0.708 \\
0.304 & -0.082 & 0.292 & -0.630 & -0.647 & 0.017 \\
-0.053 & 0.842 & 0.374 & 0.062 & -0.033 & -0.379 \\
-0.080 & 0.145 & -0.040 & 0.597 & -0.643 & 0.449 \\
-0.837 & -0.259 & 0.064 & -0.062 & -0.282 & -0.380 \\
0.163 & -0.411 & 0.800 & 0.376 & 0.121 & -0.095
\end{bmatrix} \,.
\end{aligned}
\end{equation}
Let us stress here that the numerical values above were randomly generated but in consistency with the theory requirements discussed in Secs.~\ref{subsubsec:Benchmarks_masses} and \ref{subsubsec:Benchmarks_couplings}.

The overall cross sections for both signal and background events are estimated by \texttt{MadGraph} and for our benchmark point read:
\begin{equation}\label{eq:cross_section}\nonumber
\begin{aligned}
&\text{ZA:} \quad \sigma  = 4.40 \times 10^{-7} \pm 2.62 \times 10^{-10} \hphantom{.}\text{pb}\,, \\
&\text{VBF:} \quad \sigma = 8.96 \times 10^{-7} \pm 5.88 \times 10^{-10} \hphantom{.}\text{pb}\,,\\
&\text{VLBSM:} \quad \sigma = 7.70 \times 10^{-5} \pm 4.33 \times 10^{-8} \hphantom{.}\text{pb}\,,\\
&t\bar{t}\hphantom{.}\text{:}\quad \sigma = 6.72 \pm 3.01 \times 10^{-3} \hphantom{.}\text{pb}\,,\\
&t\bar{t} + j\hphantom{.}\text{:}\quad \sigma = 7.85 \pm 5.06 \times 10^{-3} \hphantom{.}\text{pb}\,,\\
&t\bar{t} + jj\hphantom{.}\text{:}\quad \sigma = 5.99 \pm 3.70 \times 10^{-3} \hphantom{.}\text{pb}\,,\\
&t\bar{t} + Z^0(\ell^+\ell^-)\hphantom{.}\text{:}\quad \sigma = 5.36 \times 10^{-4} \pm 3.81 \times 10^{-7} \hphantom{.}\text{pb}\,,\\
&t\bar{t} + Z^0(\bar{\nu}_\ell\nu_\ell)\hphantom{.}\text{:}\quad \sigma = 1.06 \times 10^{-3} \pm 6.95 \times 10^{-7} \hphantom{.}\text{pb}\,,\\
&W^+W^-\hphantom{.}\text{:}\quad \sigma = 0.839 \pm 5.45 \times 10^{-4} \hphantom{.}\text{pb}\,,\\
&pp \rightarrow \ell\nu_\ell\hphantom{.}\text{:}\quad \sigma = 10309.1  \pm 5.4\hphantom{.}\text{pb}\,,\\
&pp \rightarrow \ell\nu_\ell + j\hphantom{.}\text{:}\quad \sigma = 2943.6 \pm 2.1\hphantom{.}\text{pb}\,,\\
&pp \rightarrow \ell\nu_\ell + jj\hphantom{.}\text{:}\quad \sigma = 1233.2 \pm 0.7 \hphantom{.}\text{pb}\,.\\
\end{aligned}
\end{equation}
As one can see the main problem we face is the overwhelming background resulting from $t\bar{t}$ events whose cross-section largely overtakes that of ZA and VBF channels, as well as $pp \rightarrow \ell\nu_\ell + (0,1 \ \text{and} \ 2) \text{ jets}$ whose cross-section exceeds that of the VLBSM channel by at least eight orders of magnitude. While each diagram in Fig.~\ref{fig:VBF-events} has a larger suppression factor associated with the presence of more interaction vertices and internal propagators, it ends up generating more contributions to the overall cross section. Indeed, while in Fig.~\ref{fig:ZA-events} we only have $e_4$ as an intermediate state, so less combinations are concerned, in Fig.~\ref{fig:VBF-events} we have all BSM neutrinos contributing to the $\nu'_\mathrm{BSM}$ propagator thus implying a larger number of possible VBF processes. Furthermore, two of the BSM neutrinos in $\nu_\mathrm{BSM}^\prime$ are rather light with masses of the order of $100~\mathrm{keV}$ (for $\nu_4$) and $100~\mathrm{MeV}$ (for $\nu_5$), which, on its own, offers an enhancement factor of at least 6 and 3 orders of magnitude, respectively, in comparison to massive EW-scale (or above) propagators.

\begin{figure}[]
    \centering
    \includegraphics[width=0.65\textwidth]{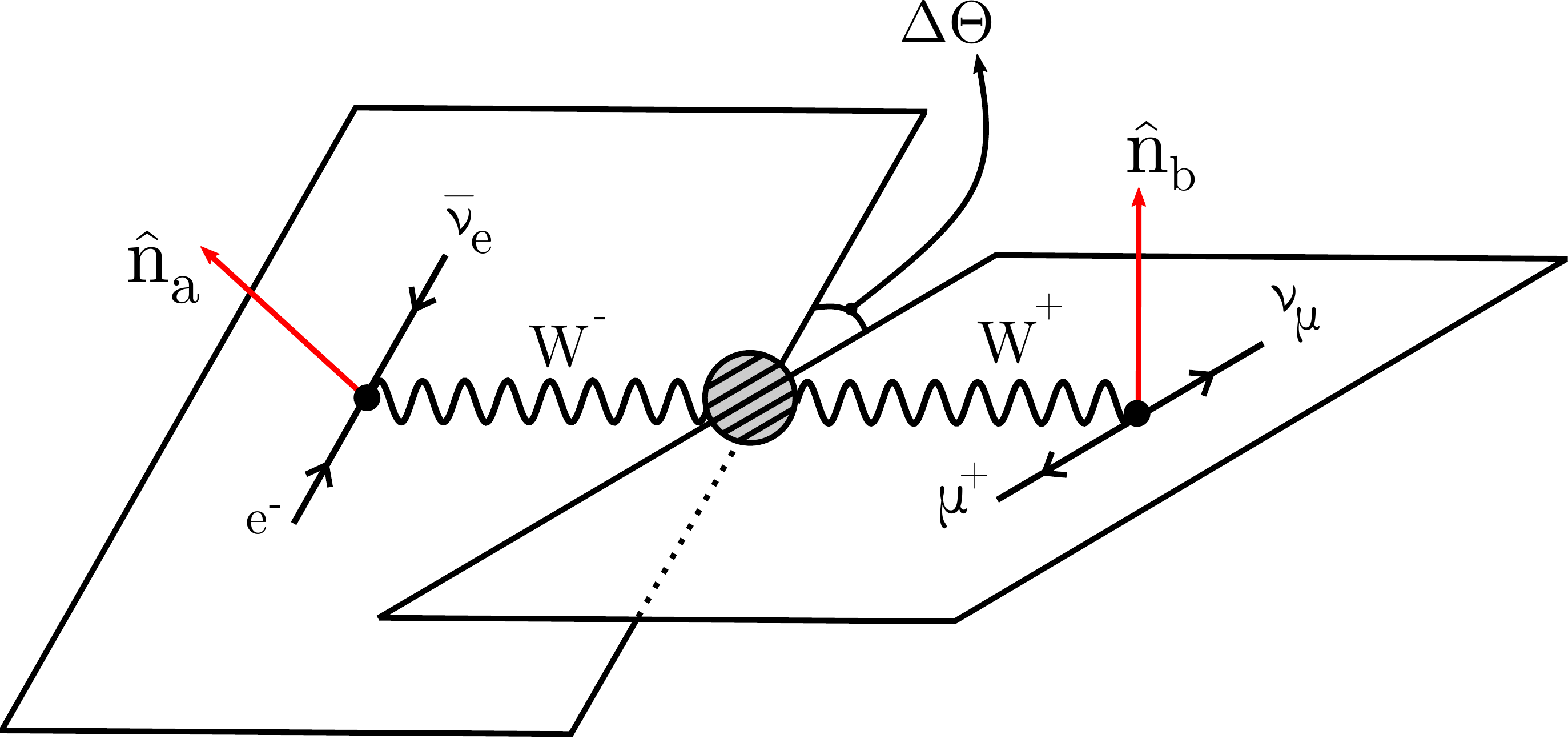} 
	\caption{Schematic representation of the new angles for the variables $\cos(\Delta\theta)$ and $\cos(\Delta\phi)$. The vectors $\hat{n}_a$ and $\hat{n}_b$ are normal to the plane formed by the decay products of $W^-$ and $W^+$. We can use these vectors to determine the angle $\Delta\theta$.
	\label{fig:New_angular_vars}}
\end{figure}

The relevant observables for VLBSM signals are detailed in Tab.~\ref{tab:vars_Zp_VLBSm}, where both angular and kinematic variables are determined in the laboratory frame. The most obvious distinction between this dataset and the previous two is in the number of features. While the VLBSM signal with less internal propagators and only one lepton final state yields cross-sections larger than those of VBF and ZA events, as a drawback, we do not have a wealth of distinct variables to choose from. Not only that, the VLBSM topology and its corresponding backgrounds do not allow for a direct one-to-one correspondence between variables such as VLL invariant mass distributions, as well as the azimuthal and polar angles $\cos(\Delta \phi)$ and $\cos(\Delta \theta)$ are absent in VLBSM events. A schematic representation of these new angular variables can be seen in Fig.~\ref{fig:New_angular_vars}, where $\Delta\theta$ is the angle between the $W^-$ and $W^+$ planes, and $\Delta\phi$ is the azimuthal angle between those two planes. The only relevant distributions, as specified in Tab.~\ref{tab:vars_Zp_VLBSm}, can be seen in Fig.~\ref{fig:VLBSM-vars} of appendix \ref{app:Kin-Ang-vars}.

Starting with the ZA/VBF signal and background topologies one can reconstruct the top, $W$ and $e_4$ masses within the expected range provided that there is a noticeable difference between them. We also observe a sizeable separation between signal events and background, especially for $\Delta  R$ distributions, where, for the former, ${\Delta R}_{e^-\bar{\nu}_e}$ and ${\Delta R}_{\mu^+\nu_\mu}$ showcase a peak near zero, well separated from background events. For pseudo-rapidity distributions the majority of signal events have a peak at around zero indicating particle trajectories perpendicular to the beam axis, which helps to separate events from some of the background channels ($t\bar{t}, W^{+}W^{-}$). 
\noindent While there are indeed sizeable differences in some variables, others are clearly dominated by background, especially for angular variables $\cos(\theta)$. Similar conclusions arise when observing the distributions for the VLBSM channel (see Fig.~\ref{fig:VLBSM-vars}), with pseudo-rapidity distributions providing a good signal-to-background separation, whereas angular distributions suffer from the considered backgrounds.

Generally speaking, for all studied channels, kinematic distributions such as transverse momentum and missing transverse energy can offer some degree of distinction in order to separate signal and background events. In fact, for backgrounds, these distributions tend to accumulate at lower energies when compared to signals. In particular, for transverse momentum distributions we have such an accumulation of events at $p_T < 200~\mathrm{GeV}$ and for missing energy the preferred region is $\text{MET} < 200~\mathrm{GeV}$. On another hand, for signal events, we have a significant accumulation in the high energy region where $p_T/\text{MET} > 300~\mathrm{GeV}$. In fact, due to a rich neutrino sector, missing energy distributions are of particular interest as the signals we are considering here which contain both BSM and SM missing energy.

However, it is important to note that the information available at experiments is limited, typically referred to as low level observable. This operates mostly on counting the number of hits (or events) that were ``observed'' by a given detector. A high level approach would combine the different information from these detectors with further complex and sophisticated observables, such as the ones we explore in this work. The use of such a multitude of observables, including the variables in the $W$ frame, will then serve as an important step in the subsequent analysis. It allows us to build a vast dataset for DL studies, which in turn, allows for a quicker training and a greater overall accuracy despite lower cross-sections.

\begin{figure}[]
    \centering
	\subfloat[ZA topologies]{{\includegraphics[width=\textwidth]{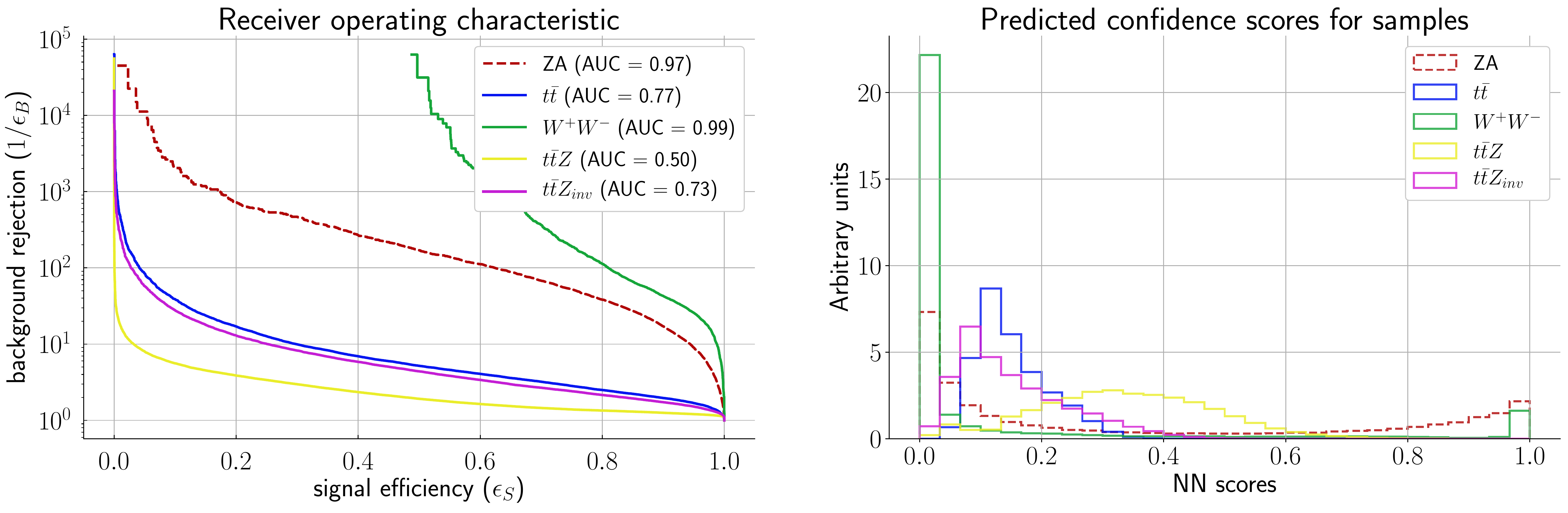} }} \\
	\subfloat[VBF topologies]{{\includegraphics[width=\textwidth]{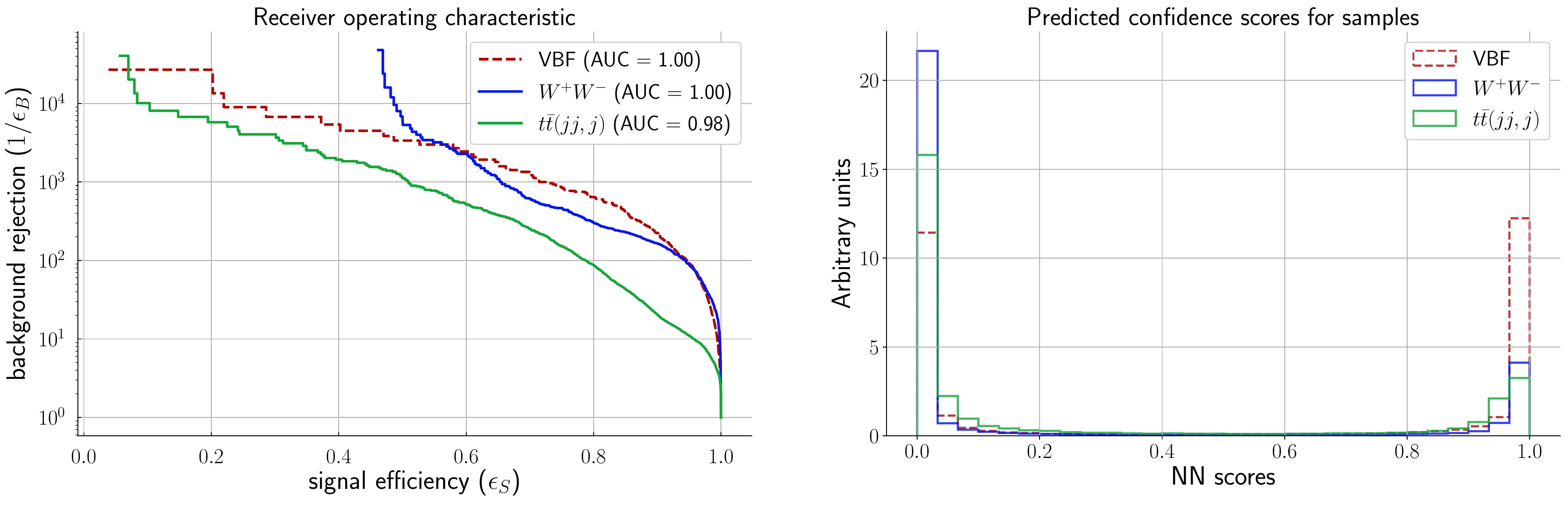} }} \\
	\subfloat[VLBSM topologies]{{\includegraphics[width=\textwidth]{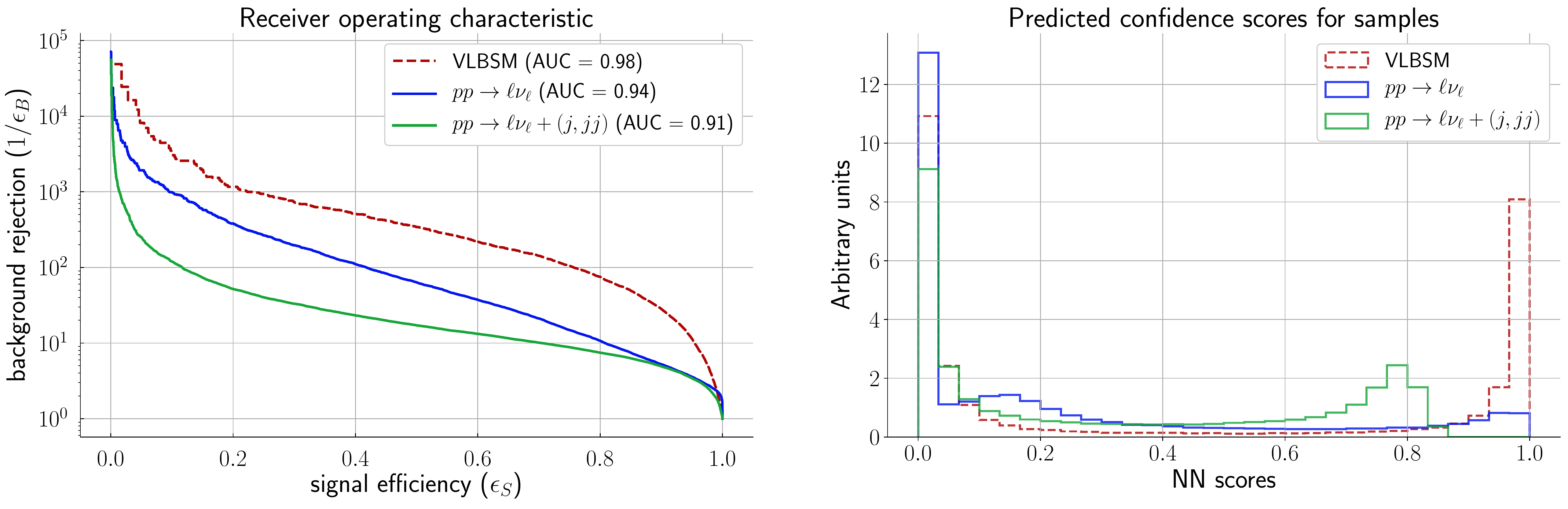} }}
	\caption{ROC and predicted confidence scores for each signal for a light VLL with mass $m_{e_4} = 677~\mathrm{GeV}$ and integrated luminosity $\mathcal{L} = 3000$ $\mathrm{fb}^{-1}$. Signal events are represented by dashed curves in red. AUC denotes accuracy. The distributions are computed following an implementation of an evolutive algorithm that maximizes the accuracy metric.
	\label{fig:ACC-ROC-plots}}
\end{figure}

\begin{figure}[]
    \centering
	\subfloat[ZA topologies]{{\includegraphics[width=\textwidth]{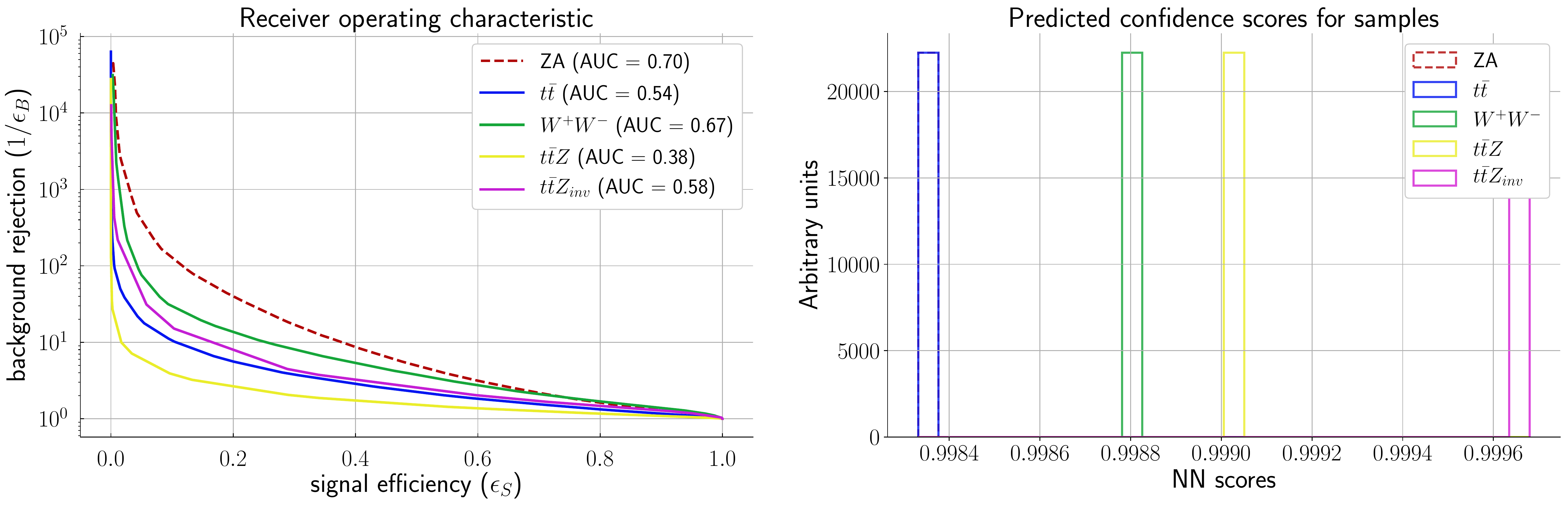} }} \\
	\subfloat[VBF topologies]{{\includegraphics[width=\textwidth]{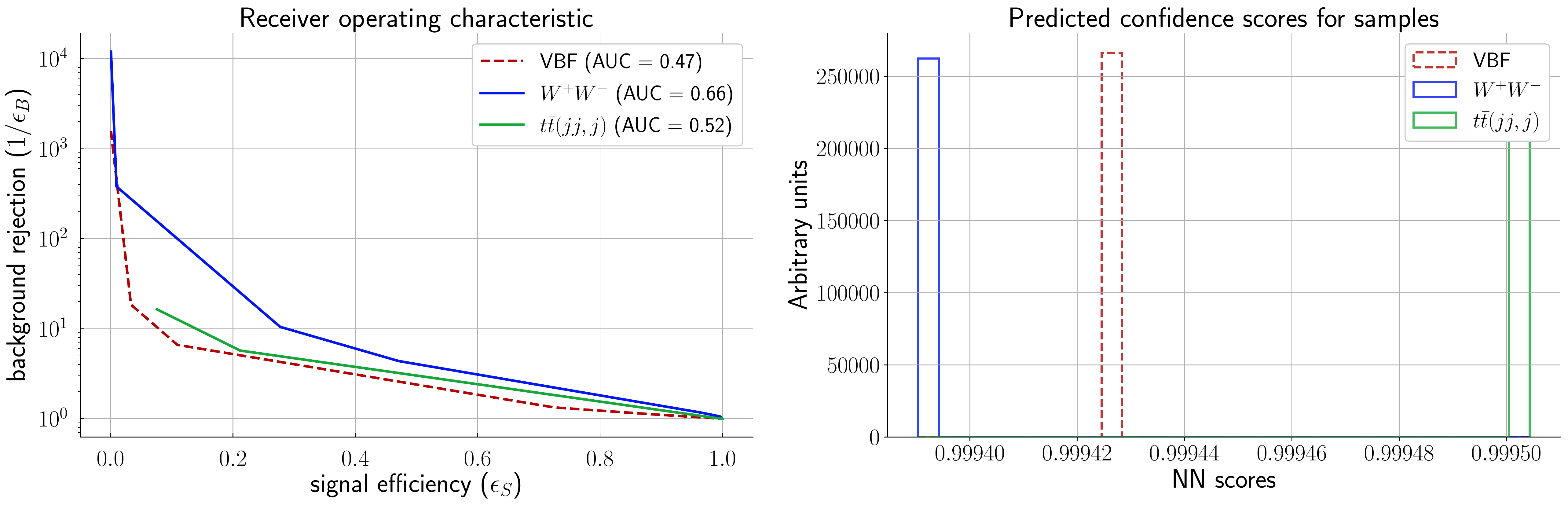} }} \\
	\subfloat[VLBSM topologies]{{\includegraphics[width=\textwidth]{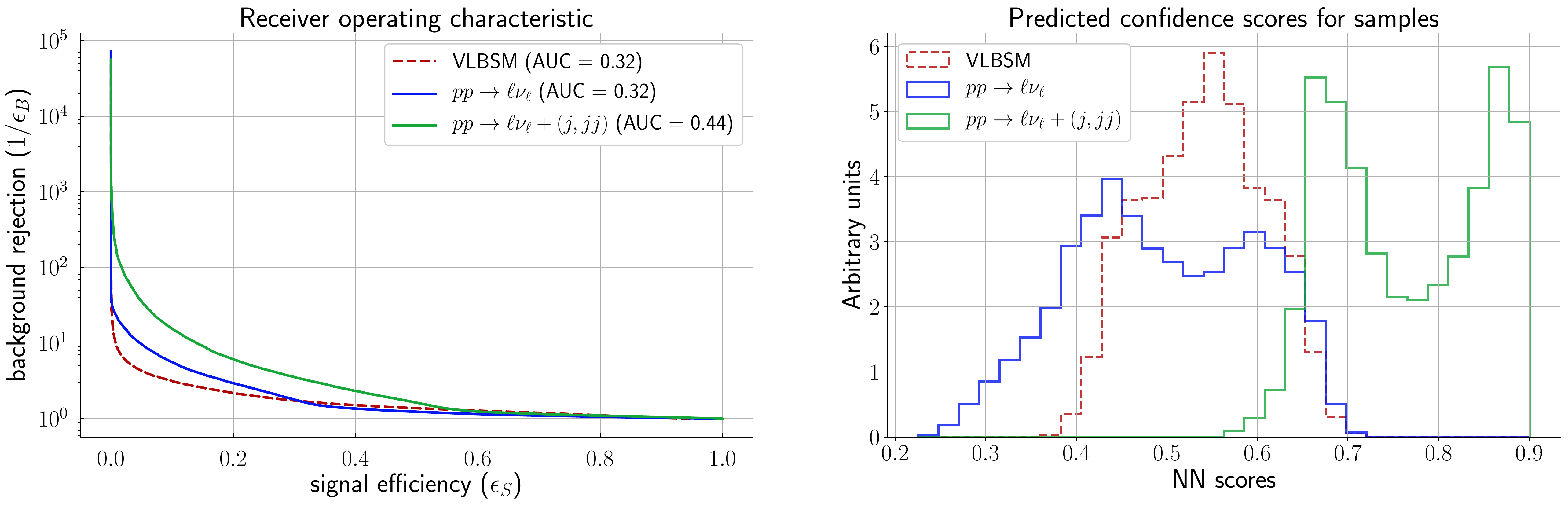} }}
	\caption{The same as in Fig.~\ref{fig:ACC-ROC-plots} but now computed for an evolutive algorithm that maximizes the Asimov significance.
	\label{fig:Asimov-ROC-plots}}
\end{figure}

For the benchmark scenario considered above (a VLL with $m_e = 667~\mathrm{GeV}$ and a BSM neutrino in the keV range), the architecture that maximizes the accuracy can be seen in Tabs.~\ref{tab:table-ZA-EVO}, \ref{tab:table-VBF-EVO}  and \ref{tab:table-VLBSM-EVO}. On the other hand, the architecture that maximizes the Asimov significance is shown in Tabs.~\ref{tab:table-ZA-Asimov-EVO}, \ref{tab:table-VBF-Asimov-EVO} and \ref{tab:table-VLBSM-Asimov-EVO}. A quantitative approach to evaluate our NN models can be done with the help of ROC (Receiver operating characteristic) curves, which represent a measure of how well the NN classification has performed. In Fig.~\ref{fig:ACC-ROC-plots} we show our results for the best accuracy whereas the best Asimov significance can be seen in Fig.~\ref{fig:Asimov-ROC-plots}. As one can observe, for the models which perform with a better accuracy, the selected architectures are capable of separating signal events from background with almost 100 \% efficiency. In particular, we see that signal events are above background events with signal efficiencies of about $\varepsilon_S = 1$, with 97$\%$ accuracy for ZA, 100$\%$ for VBF and 98$\%$ for VLBSM channels. 

However, we note that a large significance does not necessarily imply a good accuracy. In fact, for the NN architectures that maximize the Asimov significance, the accuracy is substantially reduced. A particularly relevant example is that of the VLBSM channel exhibiting the lowest accuracy with a value AUC = 0.32. The predicted confidence scores can be found on the right panels of Fig.~\ref{fig:ACC-ROC-plots}. Note that the NN assigns a different score to each prediction. For example, taking Fig.~\ref{fig:ACC-ROC-plots}(a), the NN score of 1.0 labels an event that is either a signal or a $W^+W^-$ background event.

\begin{figure*}[]
    \hspace{-1.0cm}
	\subfloat[ZA topologies]{{\includegraphics[width=0.36\textwidth]{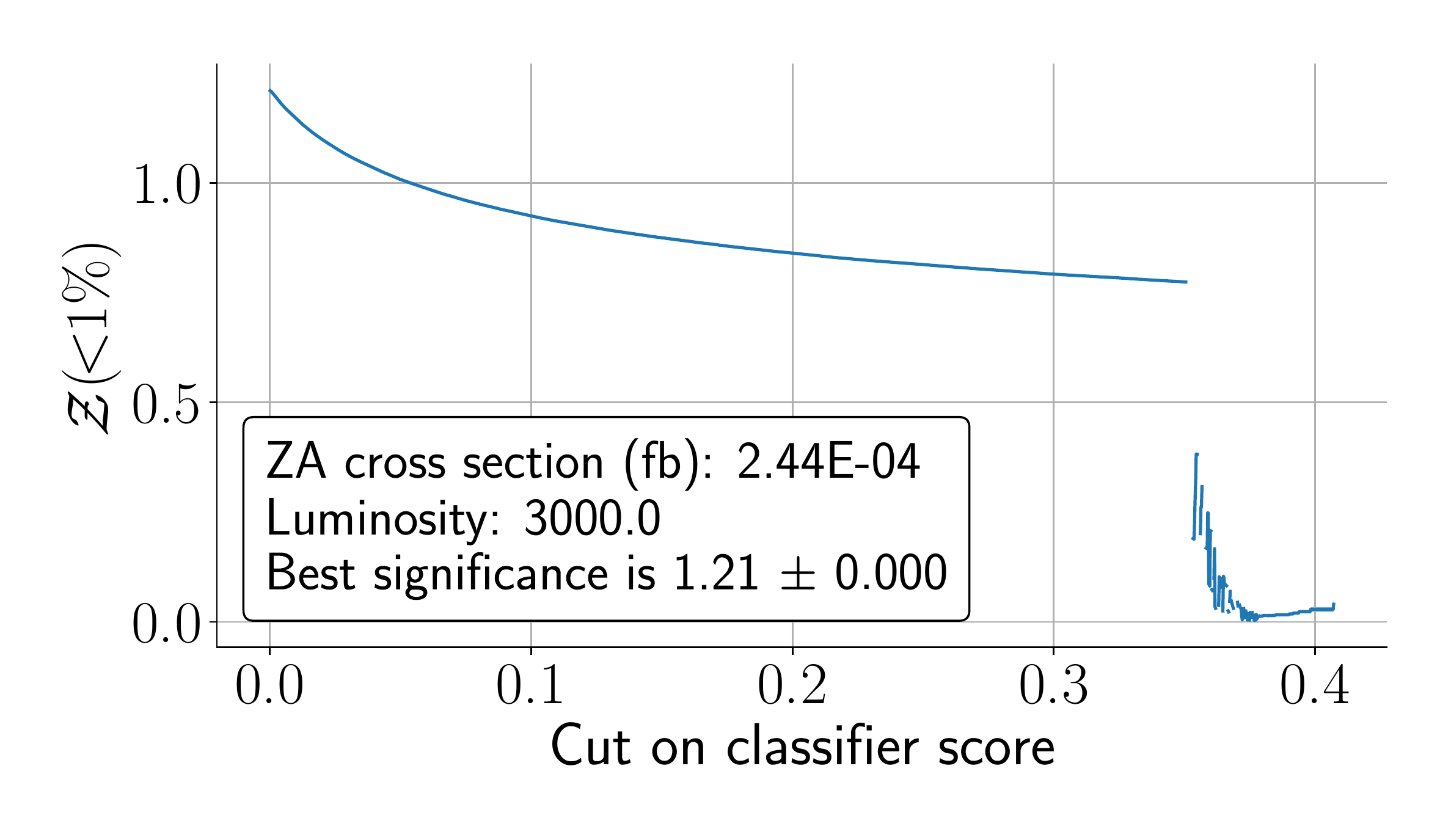} }} 
	\subfloat[ZA topologies]{{\includegraphics[width=0.36\textwidth]{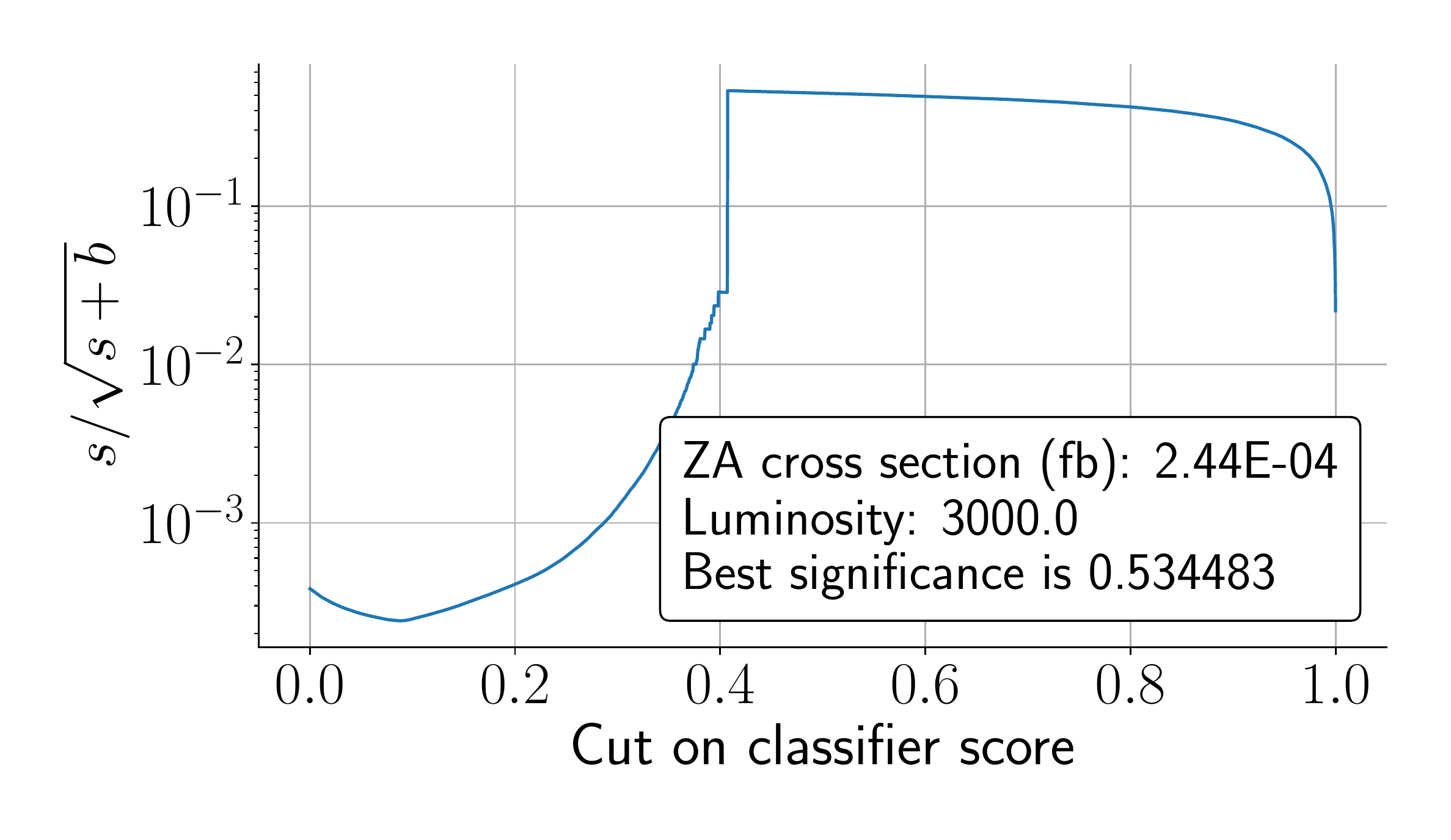} }} 
	\subfloat[ZA topologies]{{\includegraphics[width=0.36\textwidth]{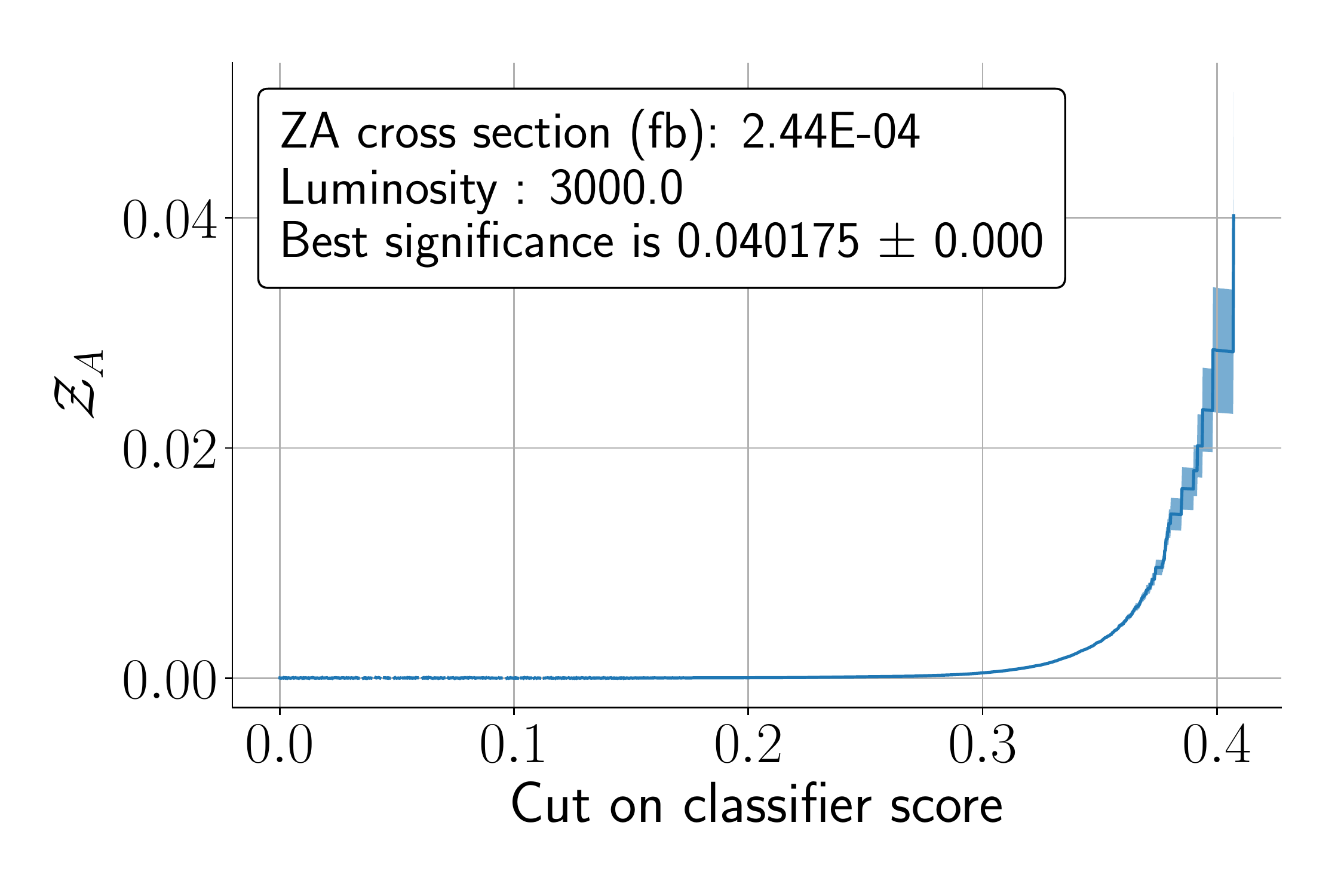} }}\\
    \hspace*{-1.0cm}
	\subfloat[VBF topologies]{{\includegraphics[width=0.36\textwidth]{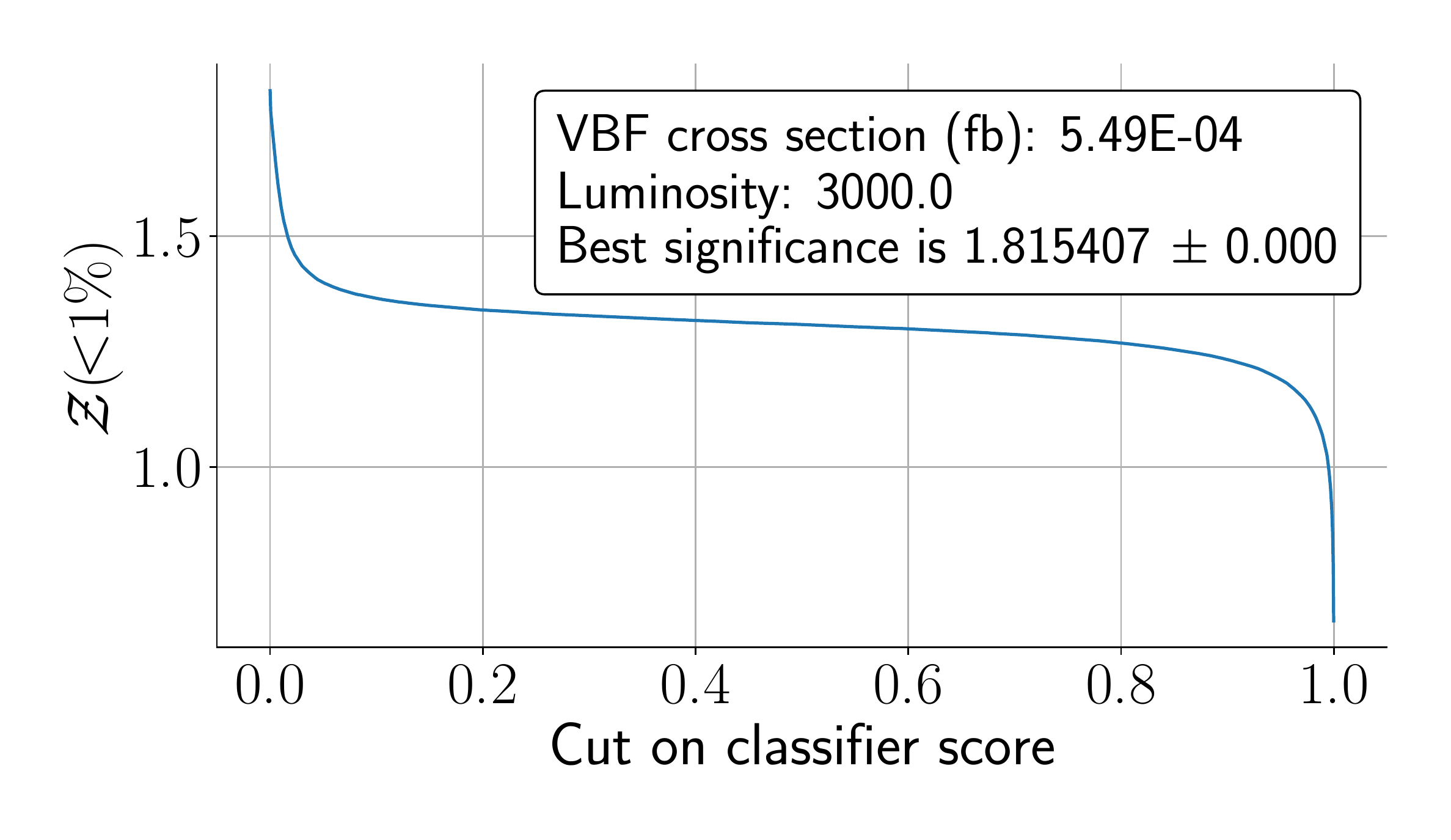} }} 
	\subfloat[VBF topologies]{{\includegraphics[width=0.36\textwidth]{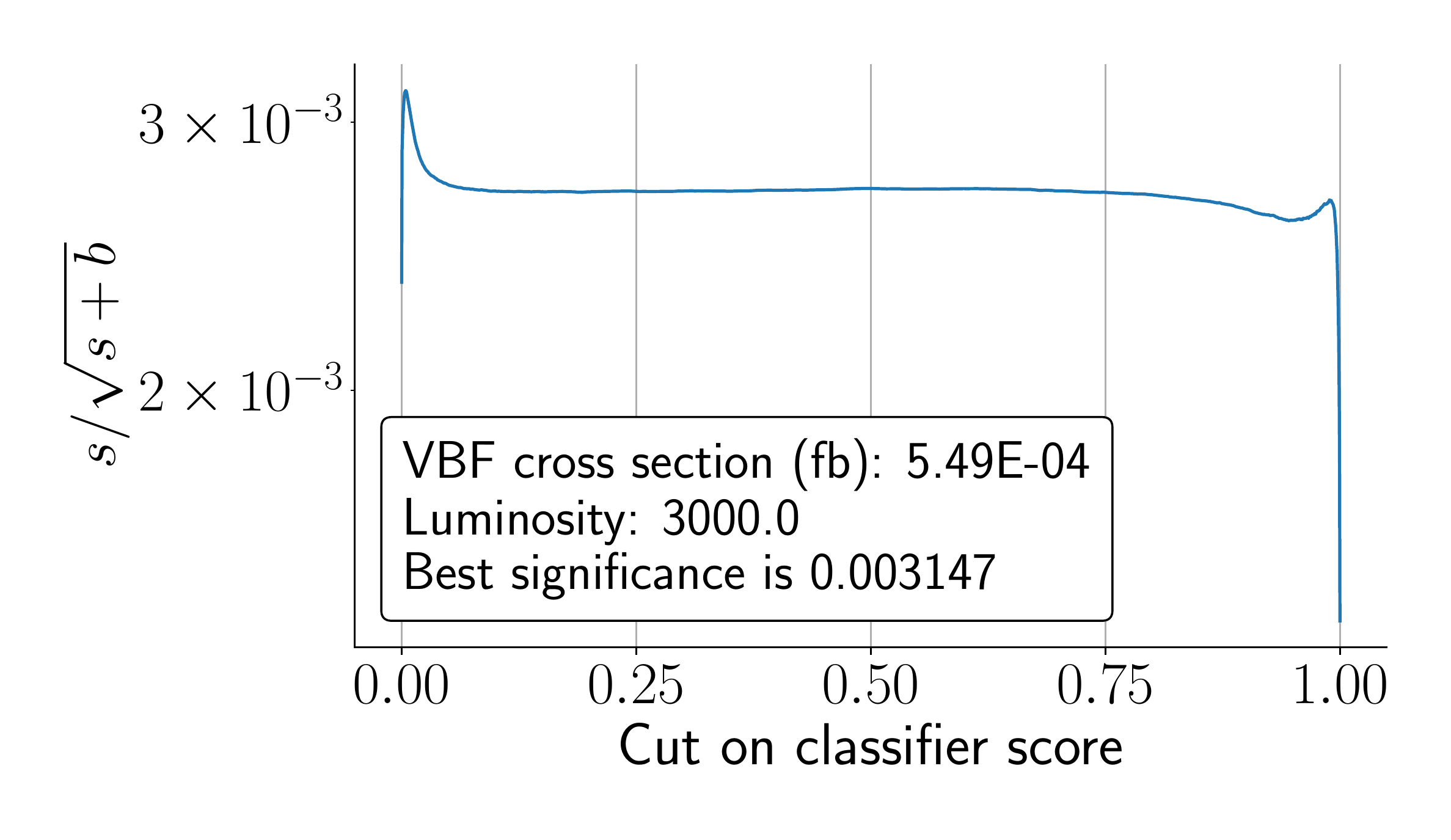} }} 
	\subfloat[VBF topologies]{{\includegraphics[width=0.36\textwidth]{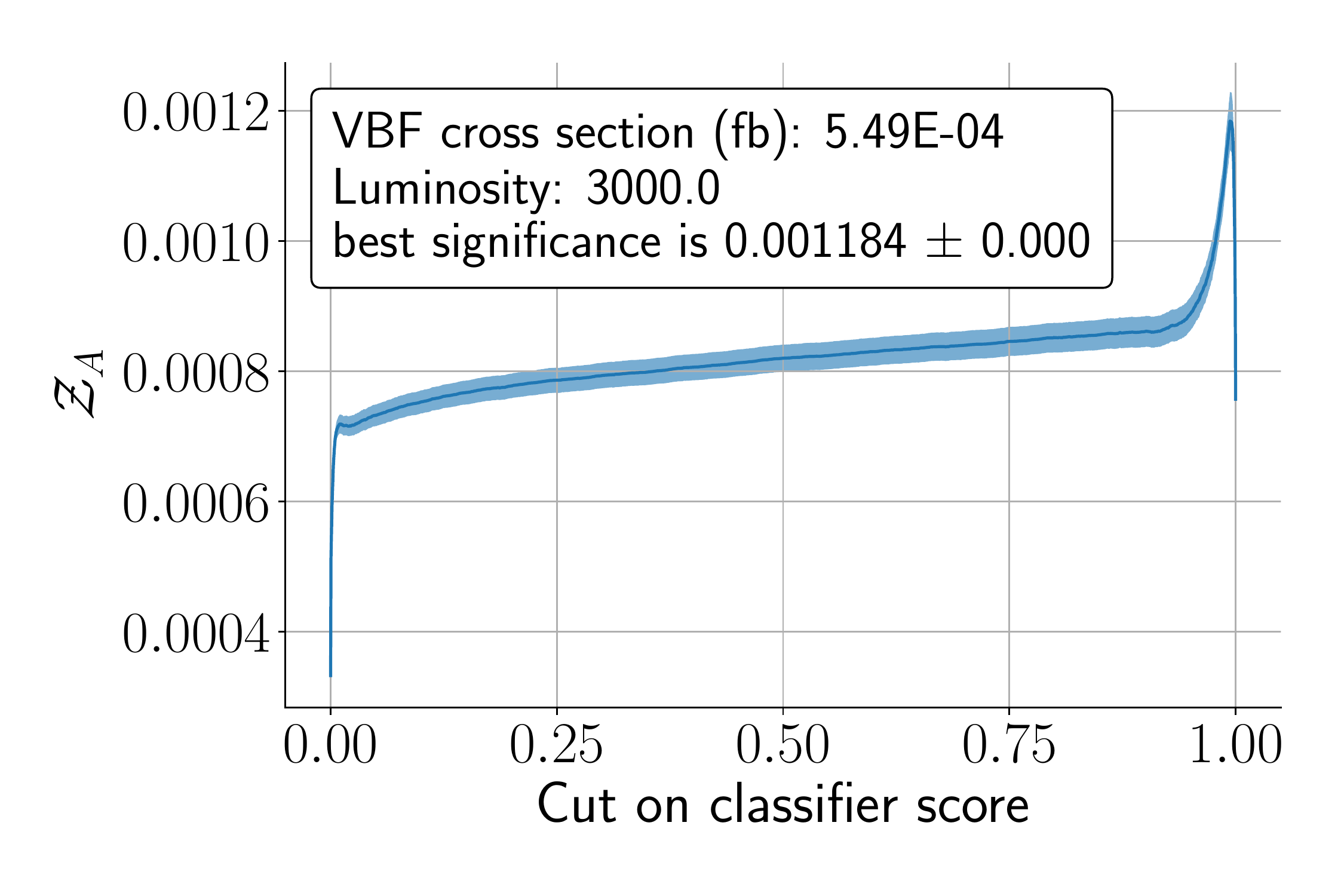} }}\\
	\hspace*{-1.0cm}
	\subfloat[VLBSM topologies]{{\includegraphics[width=0.36\textwidth]{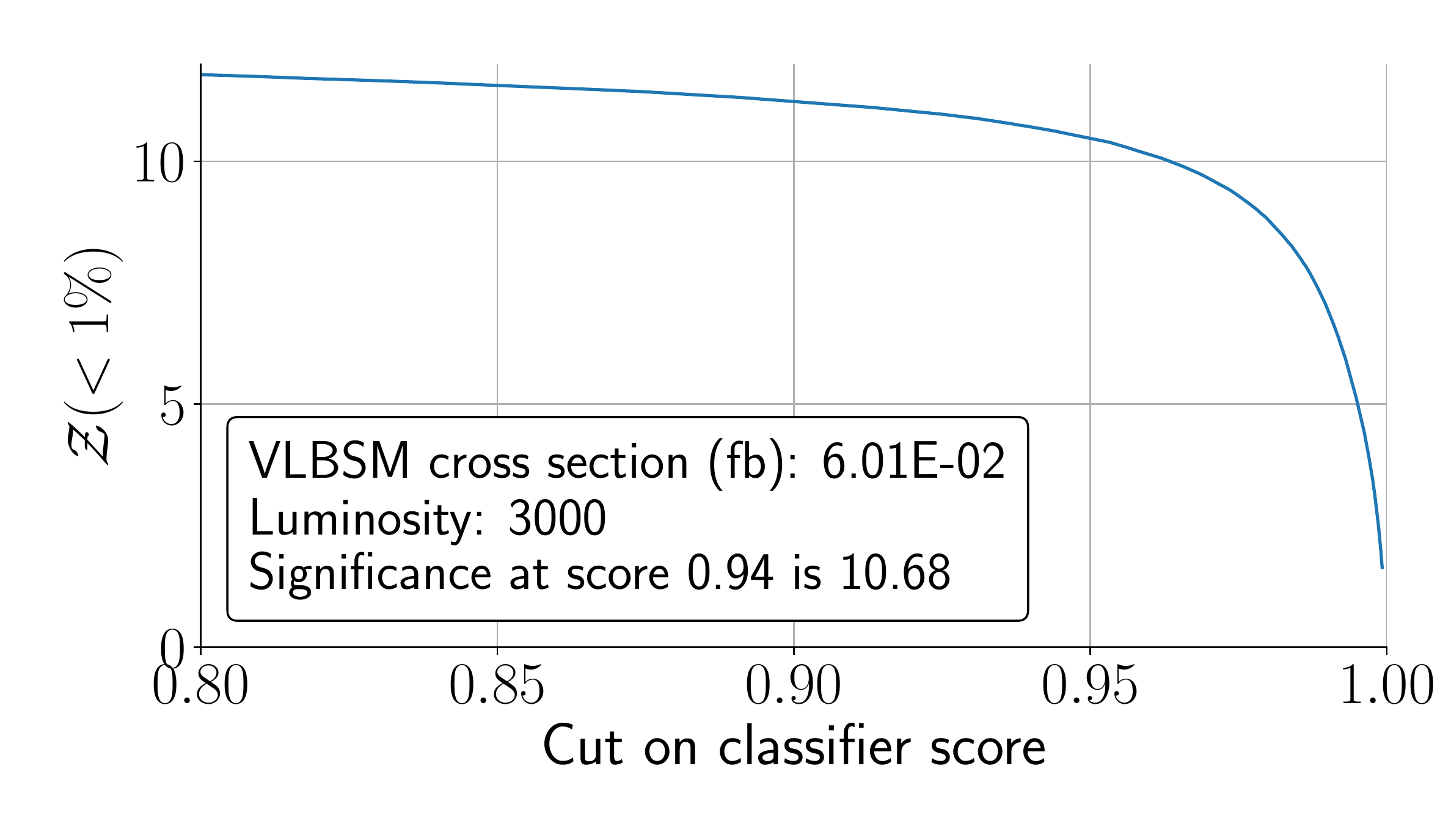} }} 
	\subfloat[VLBSM topologies]{{\includegraphics[width=0.36\textwidth]{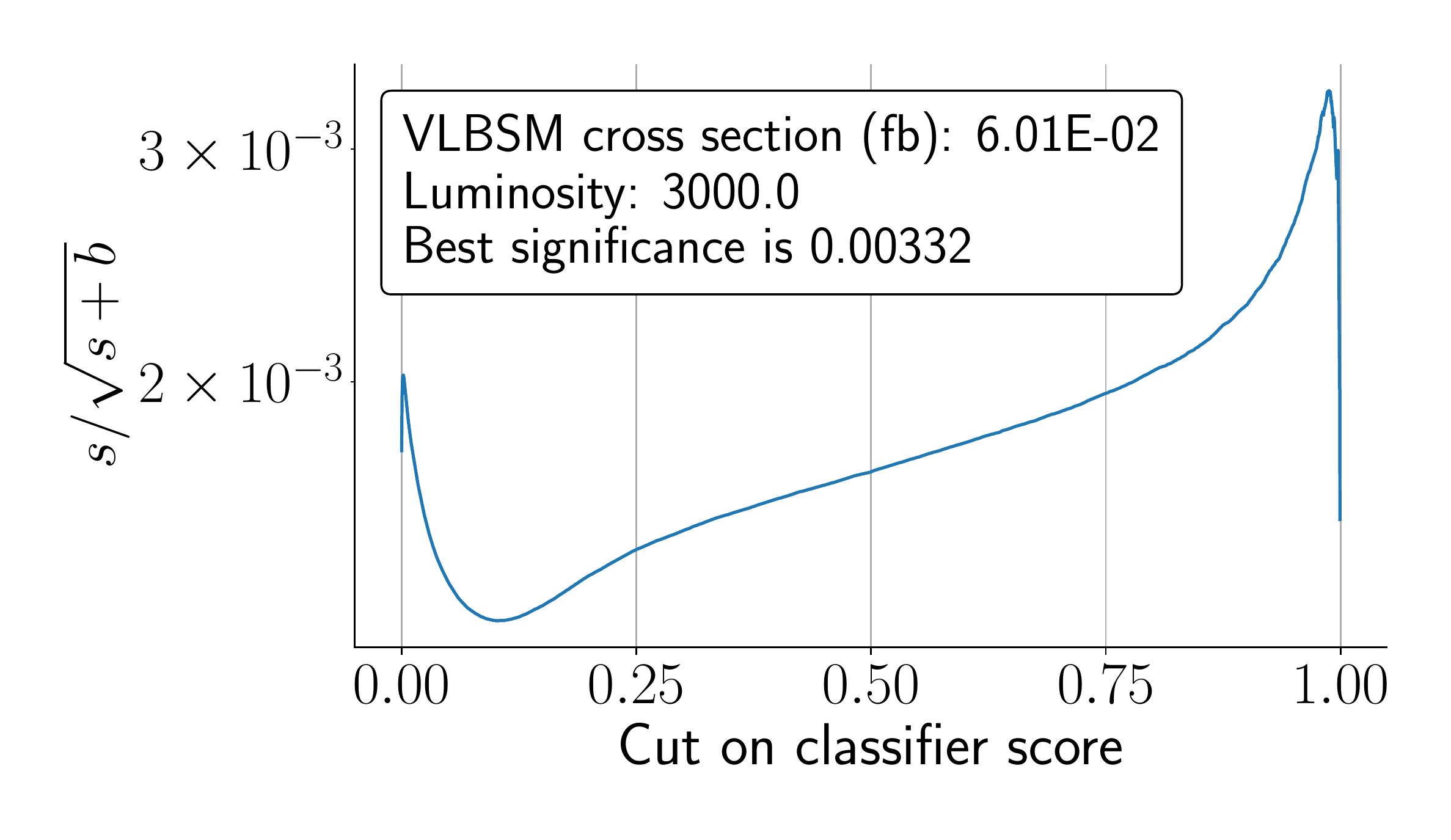} }} 
	\subfloat[VLBSM topologies]{{\includegraphics[width=0.36\textwidth]{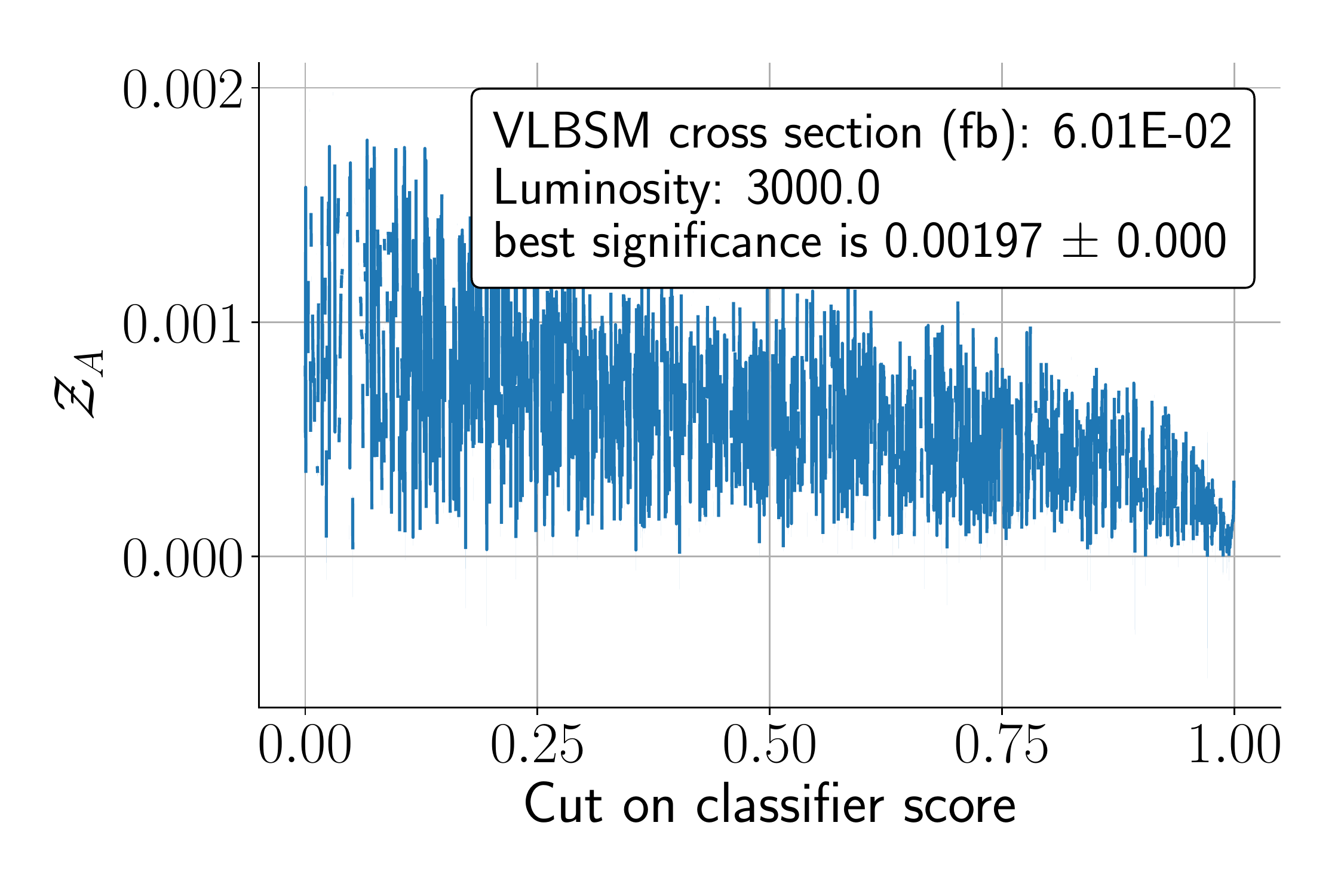} }}\\
	\caption{Significance as a function of the NN scores for each signal topology for a light VLL with mass $m_{e_4} = 677~\mathrm{GeV}$ and an integrated luminosity of $\mathcal{L} = 3000$ $\mathrm{fb}^{-1}$. For plots (a), (d) and (g) we showcase the adapted Asimov significance where backgrounds are known with an error of up to 1$\%$, for (b), (e) and (h) the naive significance $s/\sqrt{s+b}$ and for (c), (f) and (i) the Asimov significance. Plots are computed following an implementation of an evolutive algorithm that maximizes the accuracy metric.
	\label{fig:ACC-Sig-plots}}
\end{figure*}

\begin{figure*}[]
    \hspace{-1.0cm}
	\subfloat[ZA topologies]{{\includegraphics[width=0.36\textwidth]{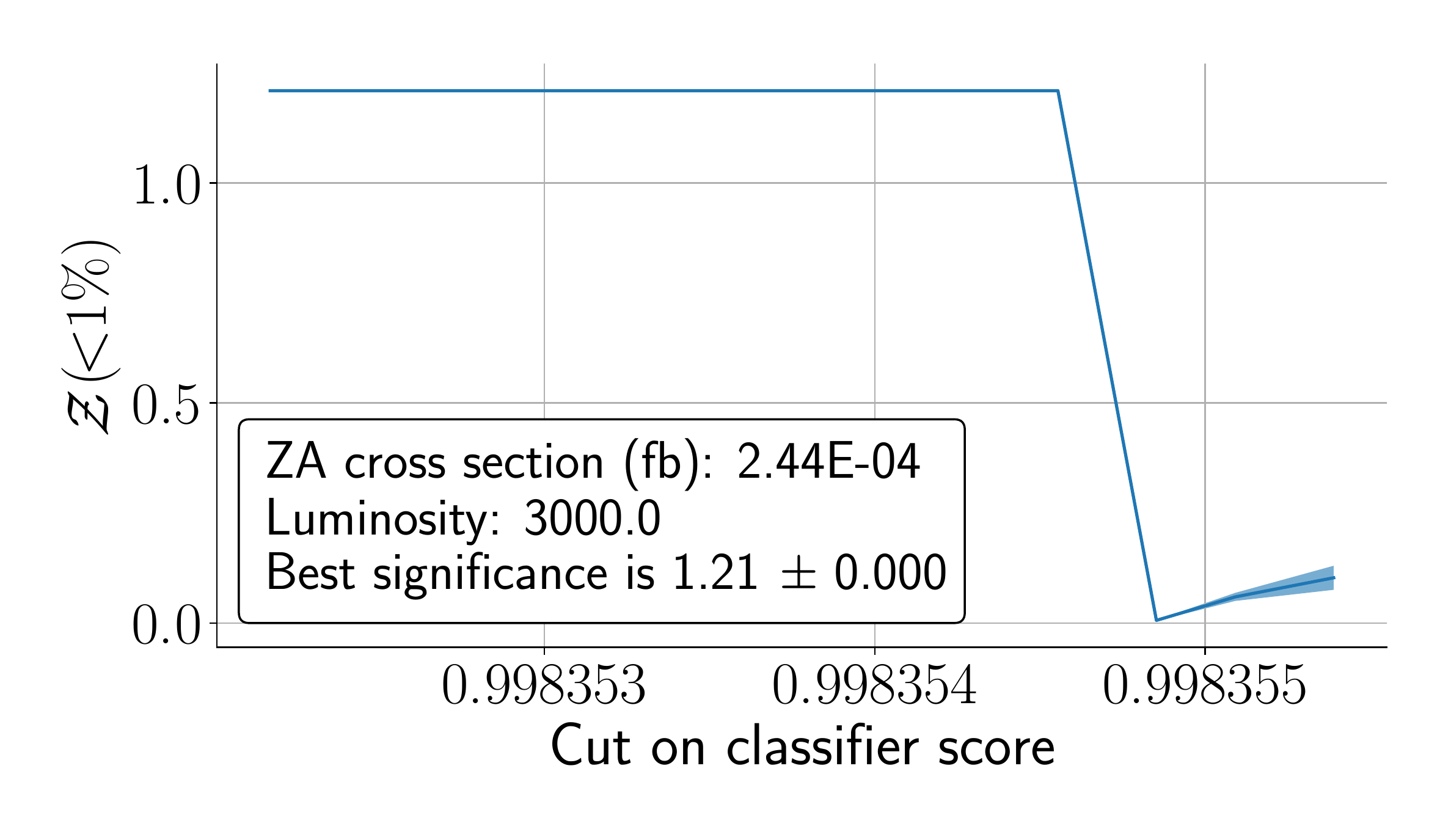} }} 
	\subfloat[ZA topologies]{{\includegraphics[width=0.36\textwidth]{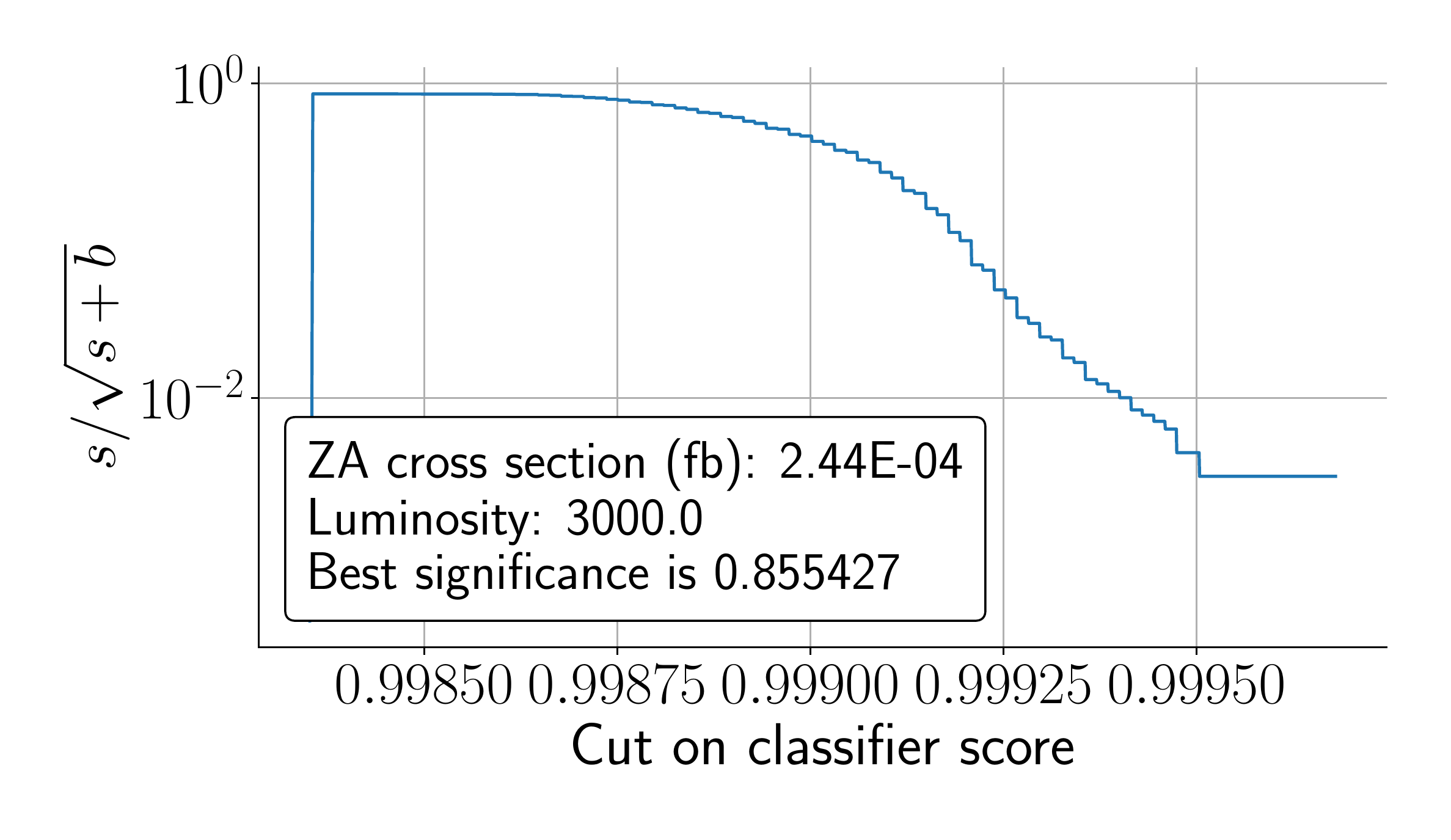} }} 
	\subfloat[ZA topologies]{{\includegraphics[width=0.36\textwidth]{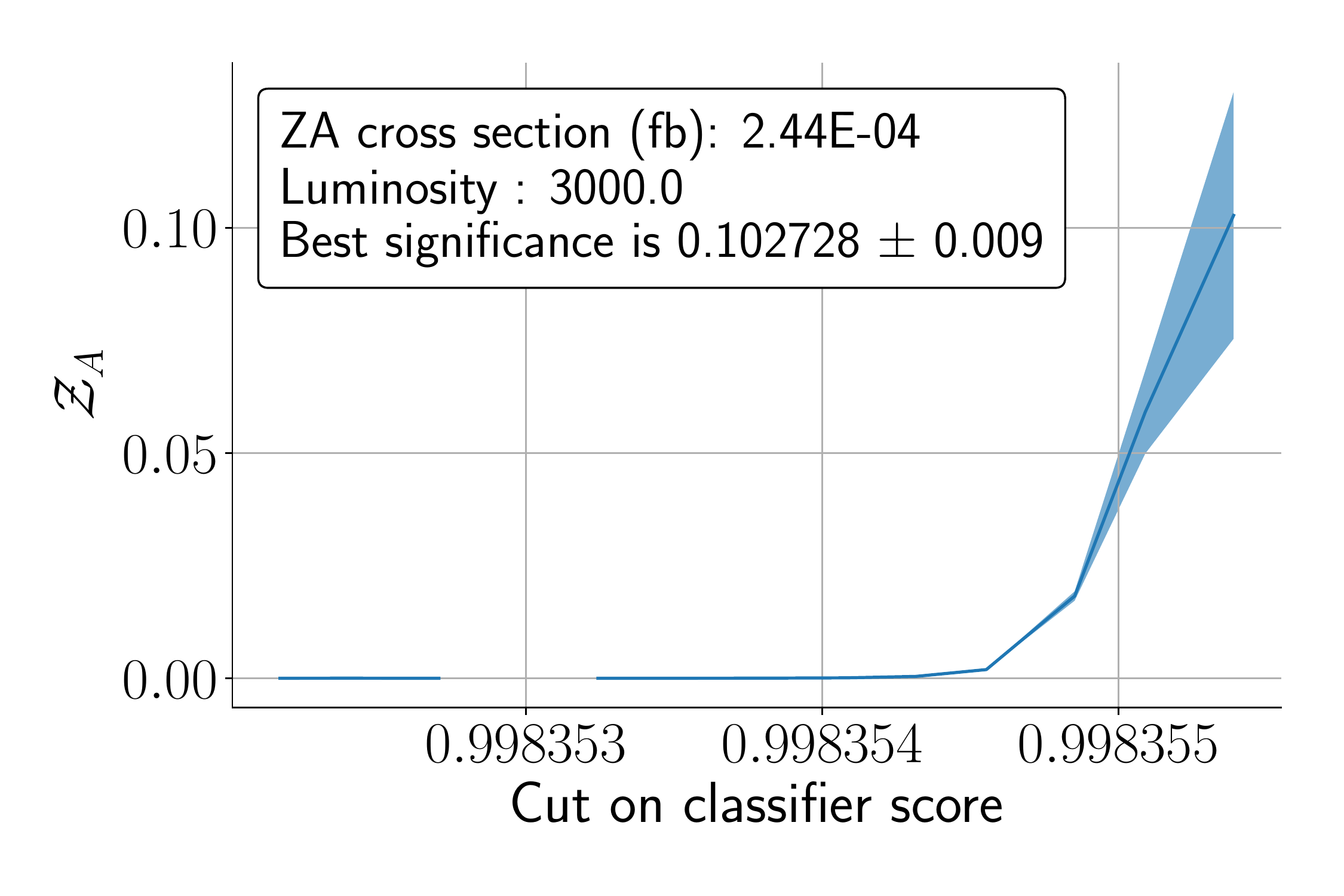} }}\\
    \hspace*{-1.0cm}
	\subfloat[VBF topologies]{{\includegraphics[width=0.36\textwidth]{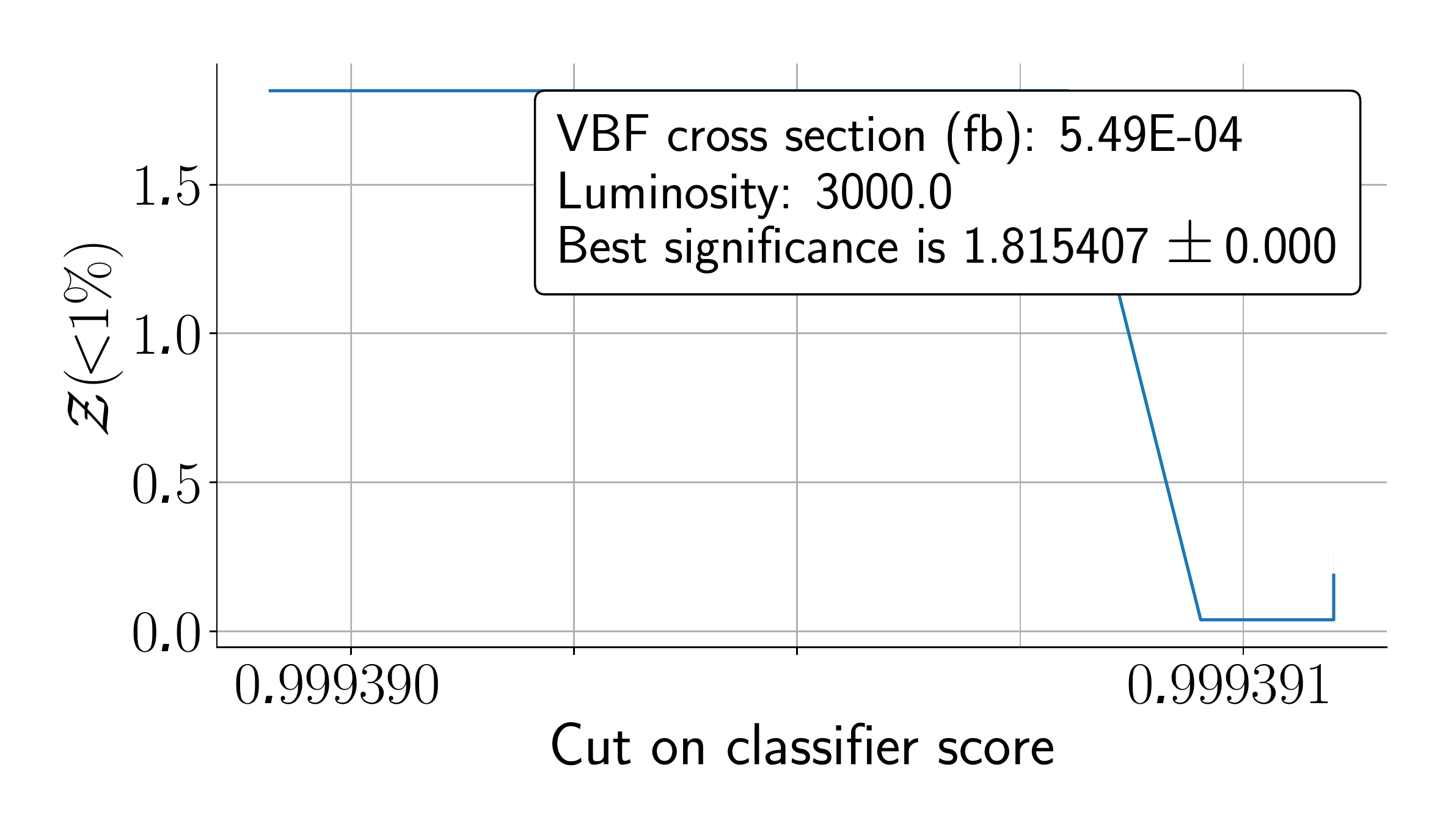} }} 
	\subfloat[VBF topologies]{{\includegraphics[width=0.36\textwidth]{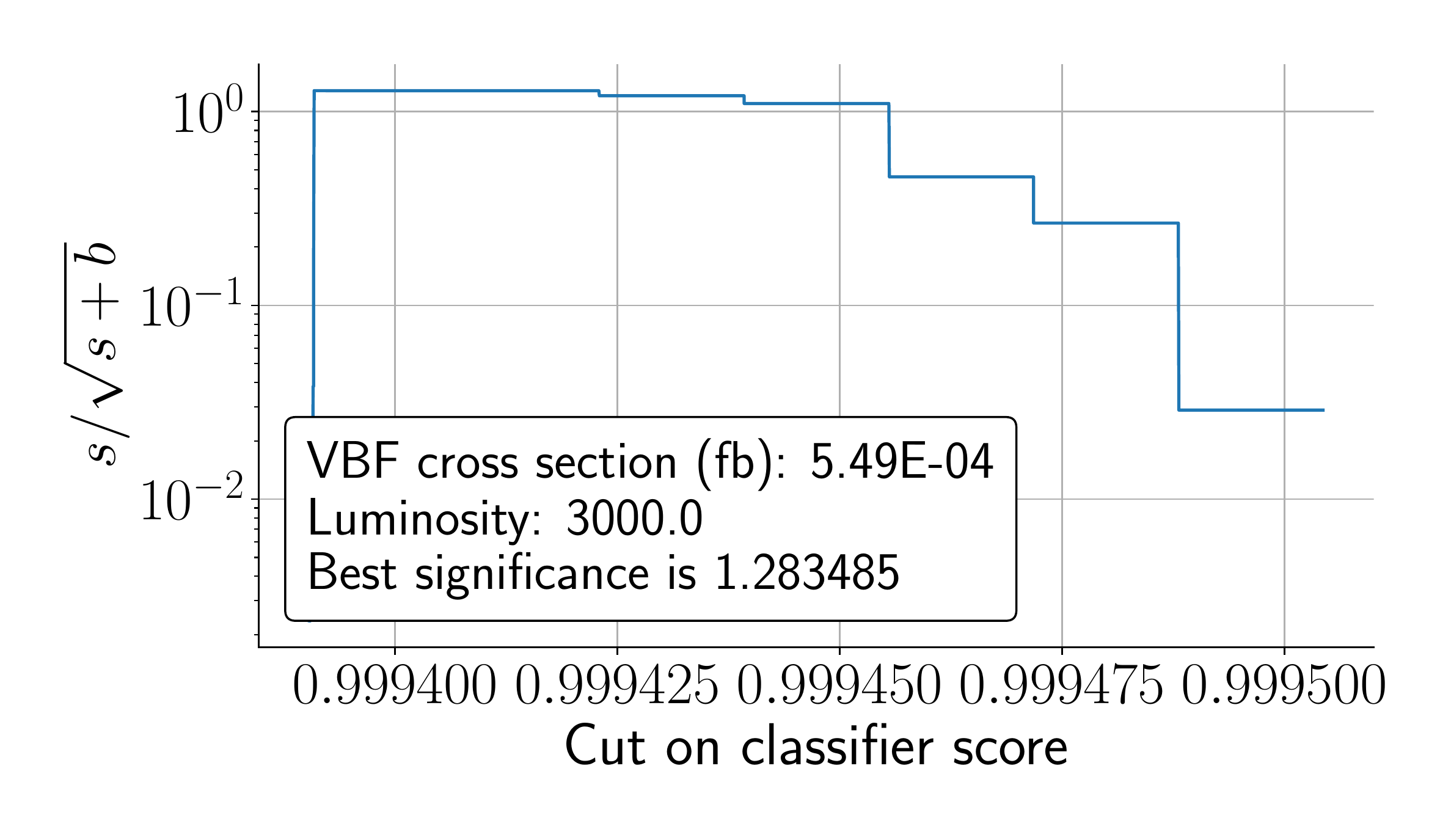} }} 
	\subfloat[VBF topologies]{{\includegraphics[width=0.36\textwidth]{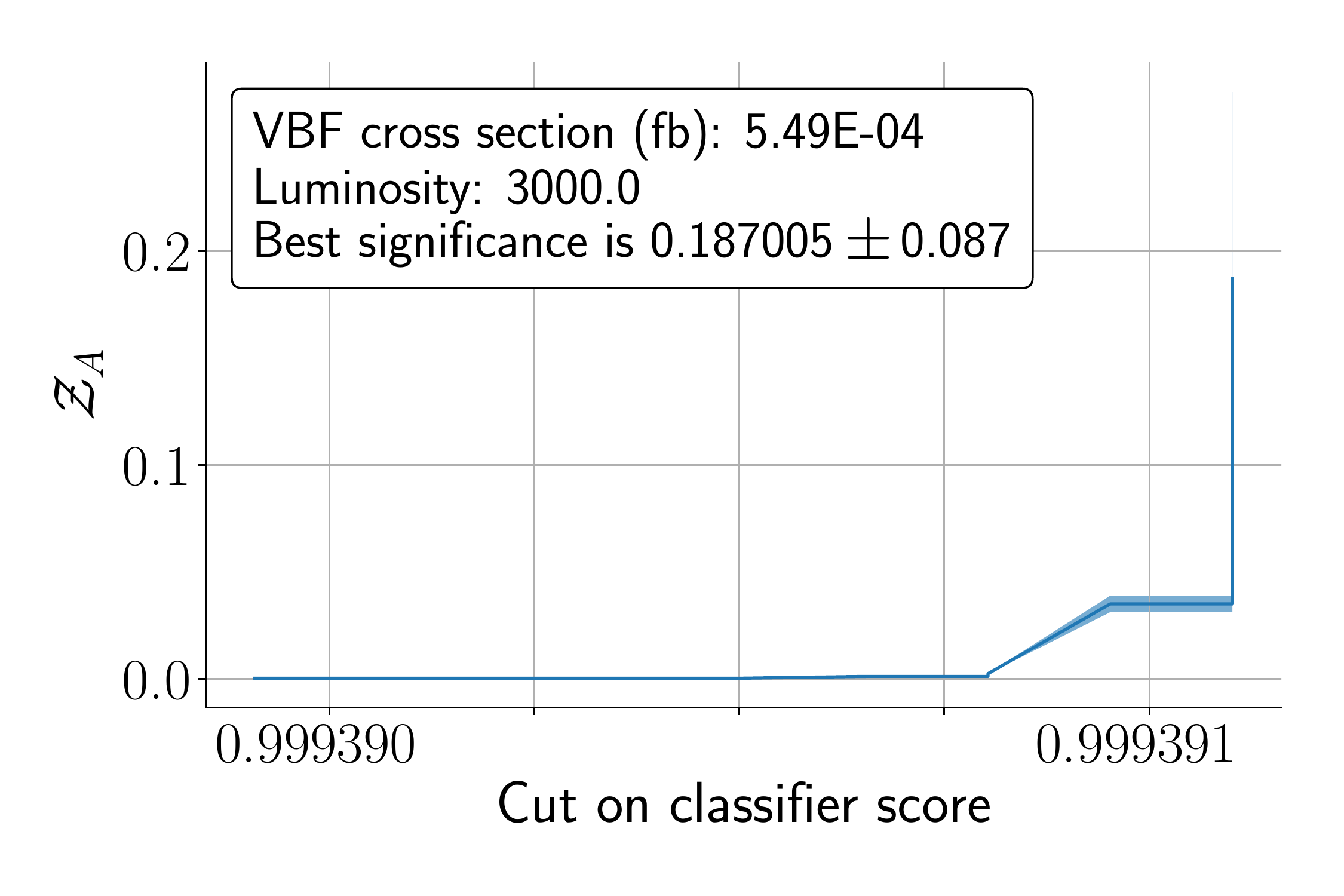} }}\\
	\hspace*{-1.0cm}
	\subfloat[VLBSM topologies]{{\includegraphics[width=0.36\textwidth]{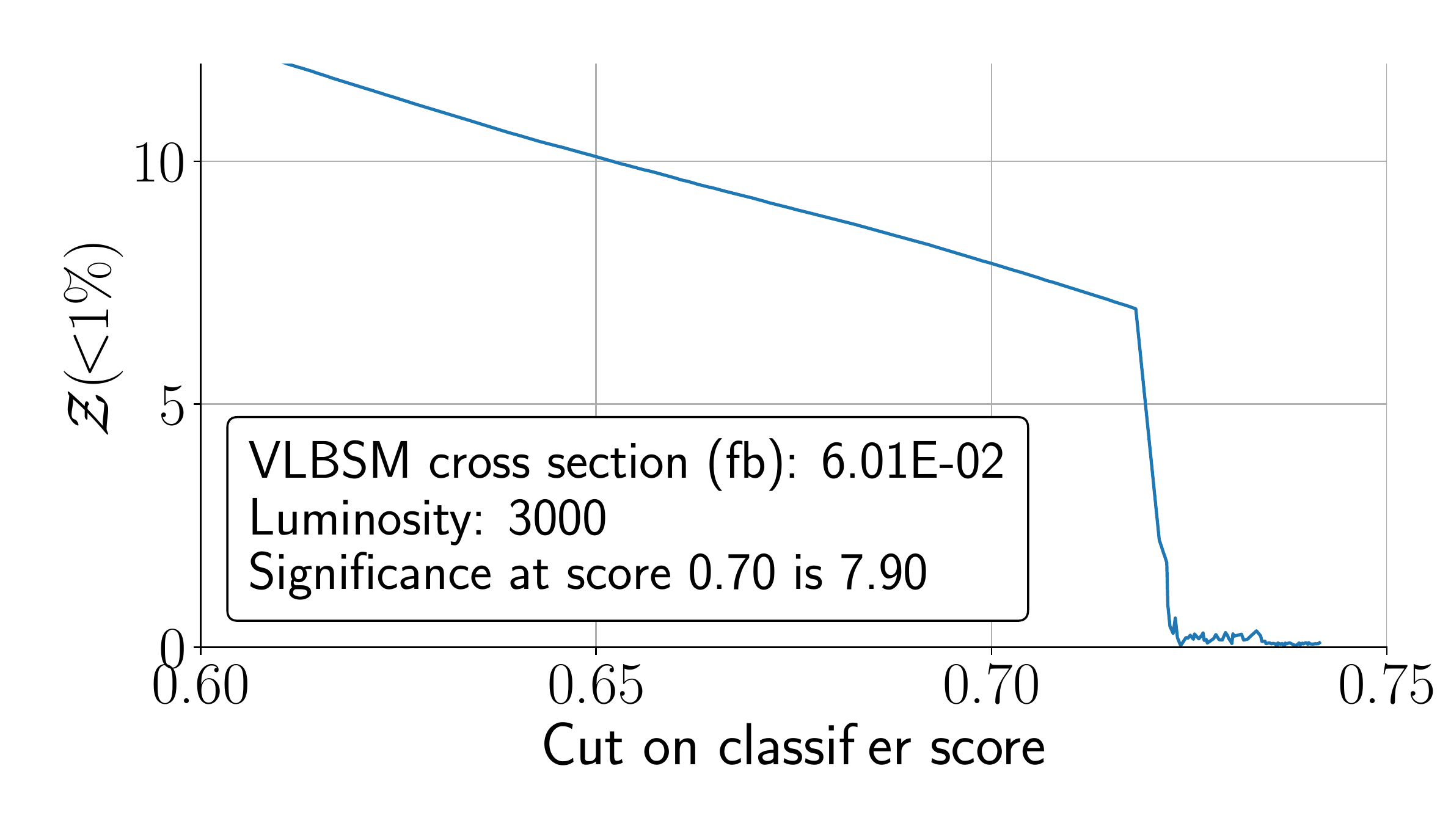} }} 
	\subfloat[VLBSM topologies]{{\includegraphics[width=0.36\textwidth]{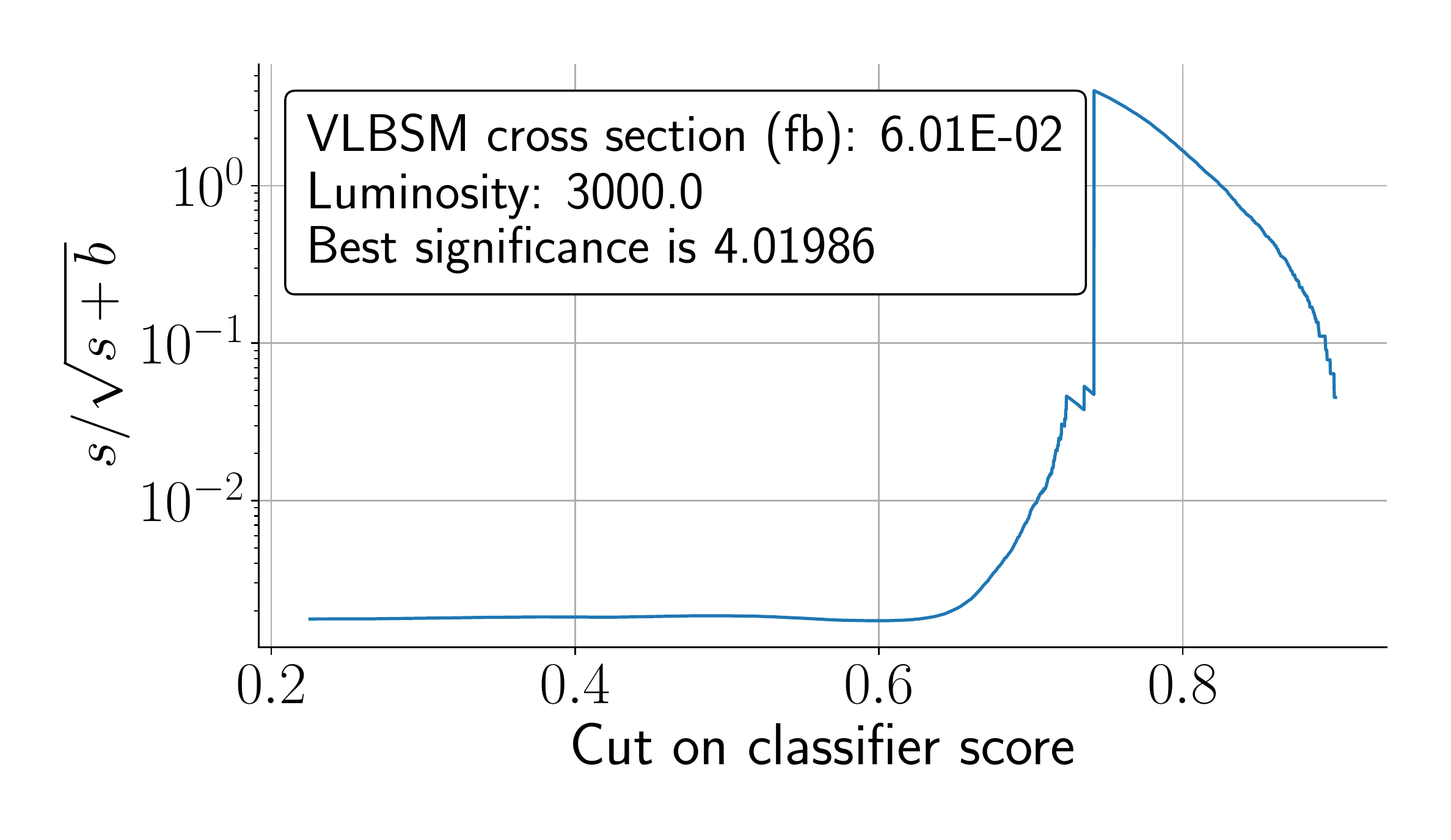} }} 
	\subfloat[VLBSM topologies]{{\includegraphics[width=0.36\textwidth]{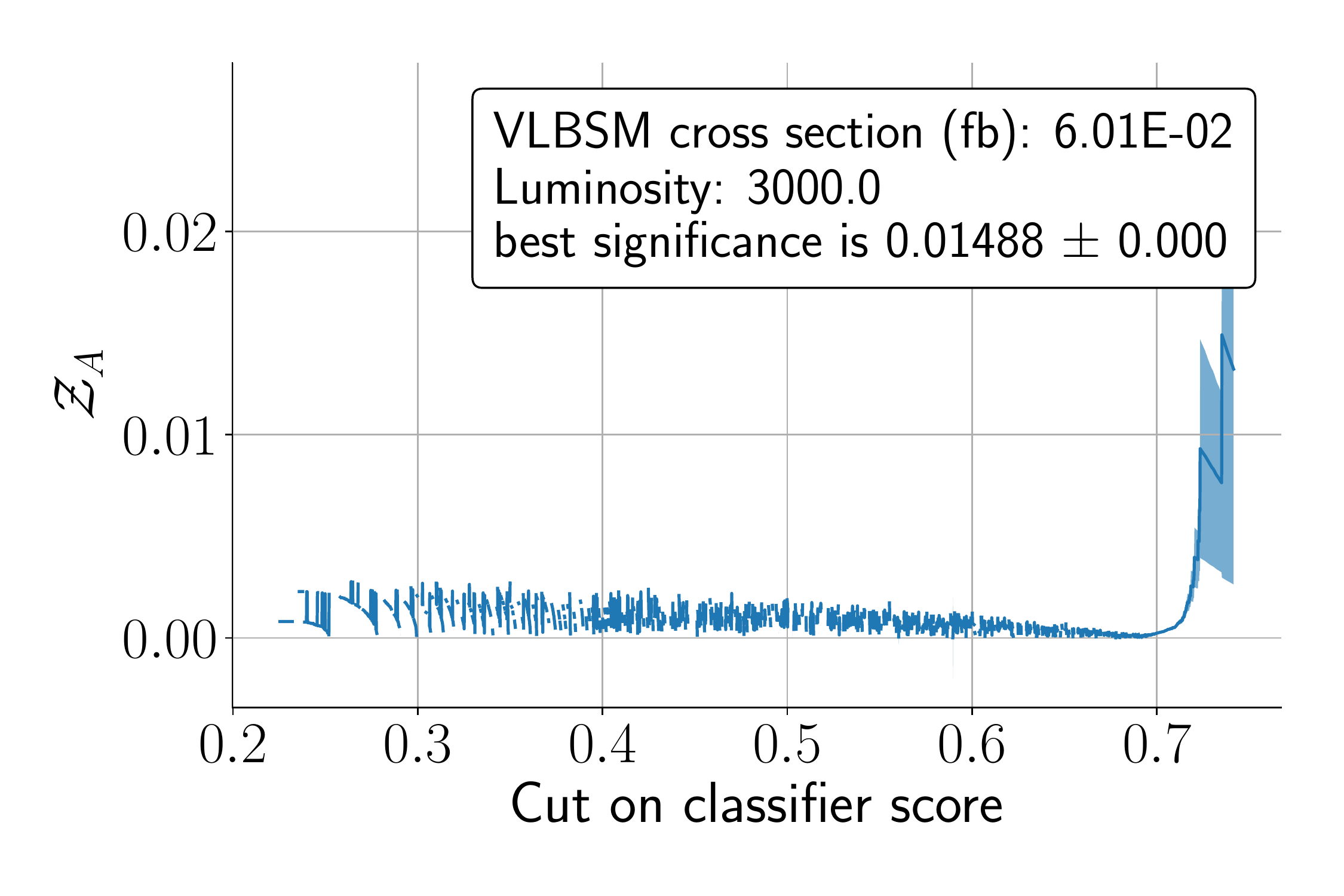} }}\\
	\caption{he same as in Fig.~\ref{fig:ACC-Sig-plots} but for an evolutive algorithm that maximizes the Asimov significance.
	\label{fig:Asimov-Sig-plots}}
\end{figure*}

The significance of signal events is typically regarded by experimental physicists as the most meaningful measure to either claim a discovery or that a given NP candidate is excluded. Therefore, we compute the significance for all channels proposed so far showing our results in Figs.~\ref{fig:ACC-Sig-plots} and \ref{fig:Asimov-Sig-plots}. While in the form the accuracy metric is maximized, in the latter it is the Asimov significance that is maximized. For the sake of rigour and completeness of our analysis we compute the significance for three distinct statistics: 
\begin{enumerate}
    \item First, we consider what we denote as \textit{naive significance}. This is calculated solely by counting the number of background $b$ and signal $s$ events according to the well known formula $s/\sqrt{s+b}$.
    \item The second metric to consider is the \textit{plain} Asimov significance $\mathcal{Z}_A$ as given in Eq.~\eqref{eq:Asimov_sig}. This is typically the most conservative measure in our analysis and it assumes that the background is known with $1\%$ uncertainty.   
    \item Finally, we consider the Asimov significance in the case of backgrounds known with an uncertainty much smaller than $1~\%$ referring to it as $\mathcal{Z}(<1\%)$. In particular, we choose background uncertainty of $10^{-3}\%$ in our studies. This measure typically gives the best results but requires that all physics backgrounds are under control by the experiment. We expect this to be realizable by the ATLAS and CMS communities upon accumulated knowledge and experience over time. Note that in the limit of vanishing background uncertainty we recover the naive significance formula.
\end{enumerate}
We then study these three significance measures in terms of the NN scores specializing to the case of the High-Luminosity (HL) LHC runs, i.e.~$\mathcal{L} = 3000$ $\text{fb}^{-1}$. Considering the results from both scenarios, we note that, as a general rule for all channels we get $\mathcal{Z}(<1\%) > s/\sqrt{s+b} > \mathcal{Z}_A$. For the case of the VLBSM channel, note that we have not selected the highest significance that our algorithm has found. The reason for this is that we have regions of the NN parameter space where only signal is present. For example, in Fig.~\ref{fig:ACC-Sig-plots}(g), we observe that for scores of $\sim 0.80$ we obtain a significance greater of $10\sigma$. However, for a realistic evaluation, we have asked the NN to guarantee the existence of both signal and background events. This is not the case of the ZA and VBF channels where the largest significance is found in a region where both signal and background is always present. 

Under the assumption that all signal topologies represent independent events, we can define the combined significance as the sum of all three contributions, i.e.
\begin{equation}
    \sigma_C = \sigma_{\text{ZA}} + \sigma_{\text{VBF}} + \sigma_{\text{VLBSM}} \,.
    \label{eq:combinedSig}
\end{equation}
For the results in Fig.~\ref{fig:ACC-Sig-plots} we see that if we privilege an evolutionary algorithm that looks for a better accuracy we get
\begin{itemize}
    \item $\mathcal{Z}(<1\%)$: $\sigma_C = 13.71\sigma$,
    \item $s/\sqrt{s+b}$: $\sigma_C = 0.55\sigma$,
    \item $\mathcal{Z}_A$: $\sigma_C = 0.04\sigma$.
\end{itemize}
while if we choose to maximize the Asimov significance, the results depicted in Fig.~\ref{fig:Asimov-Sig-plots} are translated into
\begin{itemize}
    \item $\mathcal{Z}(<1\%)$: $\sigma_C = 10.93\sigma$,
    \item $s/\sqrt{s+b}$: $\sigma_C = 6.16\sigma$,
    \item $\mathcal{Z}_A$: $\sigma_C = 0.33\sigma$.
\end{itemize}
Note that for both metrics we surpass the $5\sigma$ threshold if the $\mathcal{Z}(<1\%)$ measure is considered. However, note that this statistics works under the assumption that all backgrounds are very well under control. Nicely, for the case of the Asimov metric, we obtain for the naive significance $s/\sqrt{s+b} = 6.16\sigma$ providing us a stronger argument towards the possibility of probing VLLs with masses around $700~\mathrm{GeV}$ perhaps even before the end of the HL-LHC runs. We also observe that an evolutive algorithm engineered to maximize the Asimov significance offers overall better results. We should mention here that the combined (or even individual) significance grows with luminosity. Therefore, and based on the results so far discussed, it provides a compelling argument in favour of high-luminosity machines in the longer term.

The results presented so far for a single point are already rather interesting not only in the context of the model formulation we discuss here but also for any other model with VLLs and sterile neutrinos. For completeness and better scrutiny we will study the impact of varying the mass of both the two lightest VLLs as well as of the lightest BSM neutrino. In particular, we are interested in understanding under which circumstances one can reach or surpass a signal significance of 5 standard deviations in order to motivate direct VLL searches for this class of models at the LHC. To do this we repeat the numerical procedure explained above considering the following cases:
\begin{enumerate}
    \item varying the lightest VLL mass, $m_{e_4}$, between $200~\mathrm{GeV}$ and $1.25~\mathrm{TeV}$ while keeping $m_{\nu_4}$ the same as in Eq.~\eqref{eq:choice};
    \item keeping the $m_{e_4}$ fixed as in Eq.~\eqref{eq:choice} and varying $m_{\nu_4}$ between $100~\mathrm{keV}$ and $100~\mathrm{MeV}$;
    \item Varying the masses of the two lightest VLLs such that $m_{e_4} < m_{e_5}$;
    \item the same as in 1 but with varying luminosity.
\end{enumerate}

\begin{figure*}[]
	\centering
	\subfloat{{\includegraphics[width=0.49\textwidth]{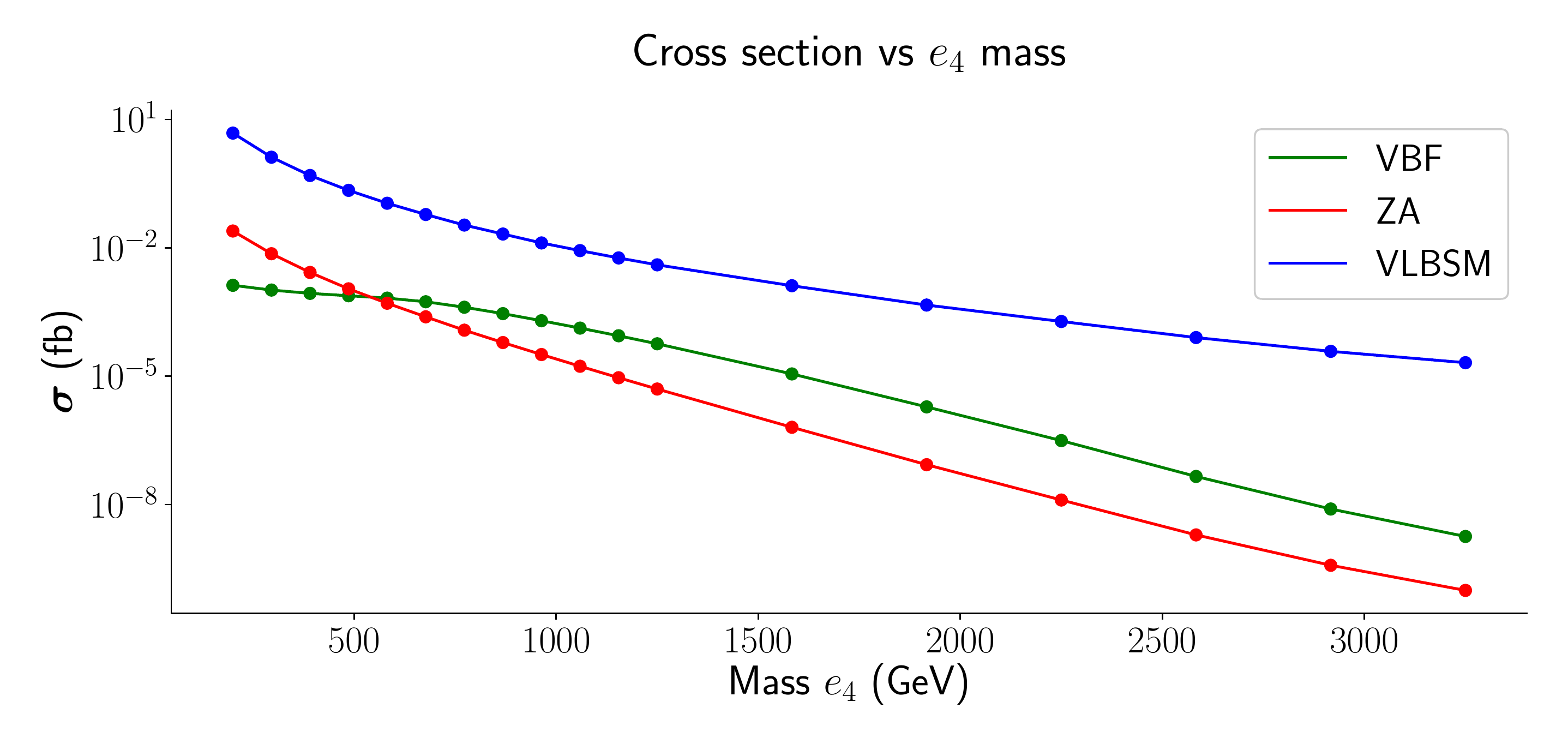} }} 
	\subfloat{{\includegraphics[width=0.49\textwidth]{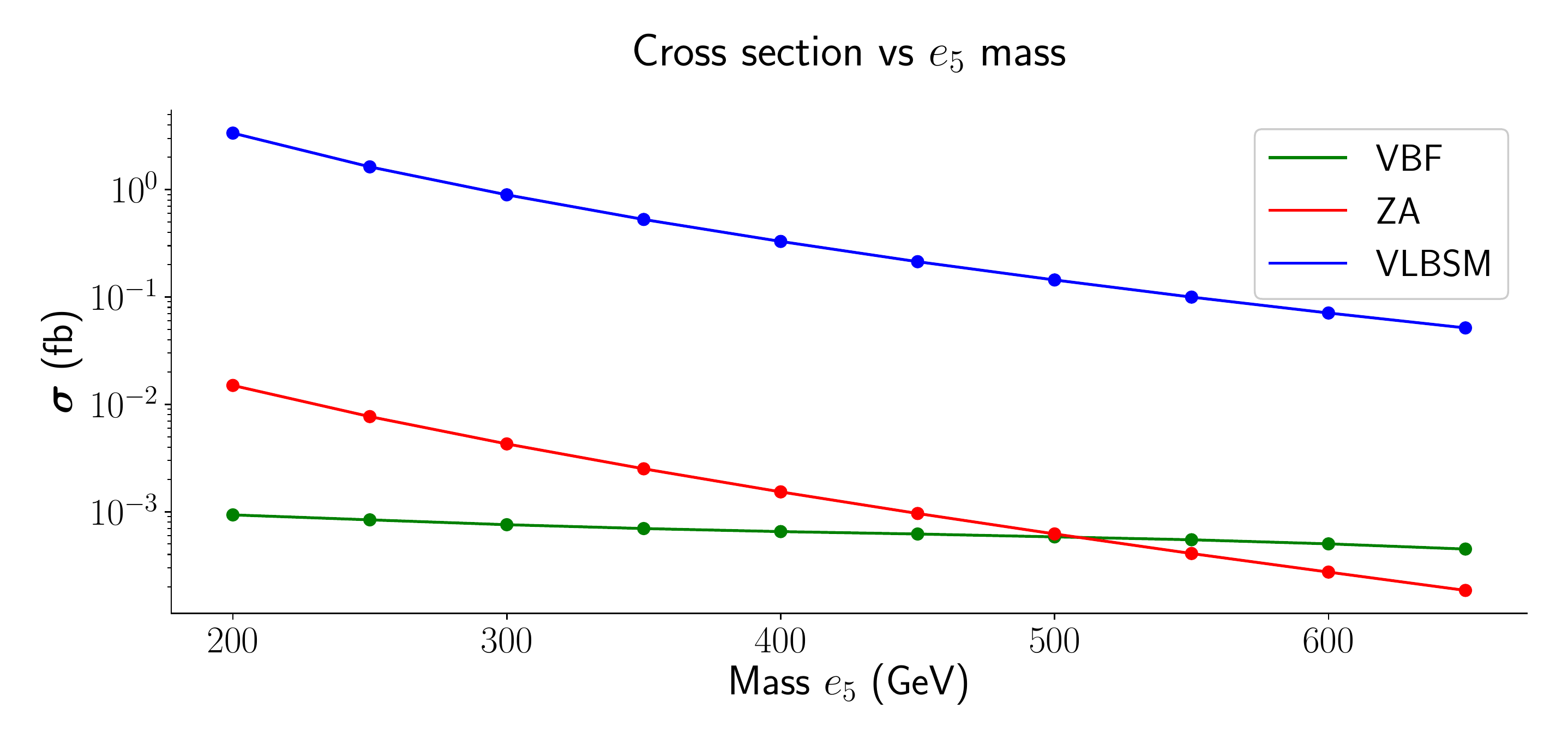} }}
	\caption{Total production and decay cross section (in femtobarn) as a function of $e_4$ (left) and $e_5$ masses (right). While on the left panel $m_{e_5} = 3.2~\mathrm{TeV}$, on the right one we have $m_{e_4} = 200~\mathrm{GeV}$.}
	\label{fig:CrossSections-e4-e5}
\end{figure*}

For a fixed luminosity of $\mathcal{L} = 3000~\mathrm{fb}^{-1}$ we study the signal significance calculated for the VLL masses $m_{e_4} = 200, 486, 868~\textrm{and}~1250~\mathrm{GeV}$. First, we consider an evolutive algorithm that maximizes the accuracy metric showing our results in Tab.~\ref{tab:Evolve_Acc_table}.
\begin{table}[h!]
\resizebox{\textwidth}{!}{\begin{tabular}{c||ccc||ccc||ccc||}
\multirow{2}{*}{Mass of $e_4$} & \multicolumn{3}{c||}{$s/\sqrt{s+b}$}                                                             & \multicolumn{3}{c||}{$\mathcal{Z}(<1\%)$}                               & \multicolumn{3}{c||}{$\mathcal{Z}_{A}$}                                                                    \\ \cline{2-10} 
                      & ZA                          & VBF                                       & VLBSM                & ZA                           & VBF                          & VLBSM   & ZA                                     & VBF                                       & VLBSM                \\ \hline
$200$ GeV  & \multicolumn{1}{c|}{$4.01$} & \multicolumn{1}{c|}{$9.4\times 10^{-3}$}    & $0.31$               & \multicolumn{1}{c|}{$12.18$} & \multicolumn{1}{c|}{$2.83$}  & $12.95$ & \multicolumn{1}{c|}{$2.05$}            & \multicolumn{1}{c|}{$2.47\times 10^{-3}$}  & $1.41$               \\
$486$ GeV  & \multicolumn{1}{c|}{$0.95$} & \multicolumn{1}{c|}{$1.51$}              & $6.66\times 10^{-3}$ & \multicolumn{1}{c|}{$2.59$}  & \multicolumn{1}{c|}{$2.13$}  & $7.83$  & \multicolumn{1}{c|}{$0.12$}            & \multicolumn{1}{c|}{$4.6\times 10^{-4}$} & $2.15\times 10^{-4}$ \\
$677$  GeV & \multicolumn{1}{c|}{$0.53$} & \multicolumn{1}{c|}{$3.15\times 10^{-3}$} & $3.32\times 10^{-3}$ & \multicolumn{1}{c|}{$1.21$}  & \multicolumn{1}{c|}{$1.82$} & $10.68$ & \multicolumn{1}{c|}{$0.040$}           & \multicolumn{1}{c|}{$1.18\times 10^{-3}$} & $1.97\times 10^{-3}$ \\
$868$ GeV   & \multicolumn{1}{c|}{$0.26$} & \multicolumn{1}{c|}{$0.93$}              & $6.18\times 10^{-4}$ & \multicolumn{1}{c|}{$0.52$}  & \multicolumn{1}{c|}{$1.32$} & $6.60$  & \multicolumn{1}{c|}{$0.01$}            & \multicolumn{1}{c|}{$0.30$}              & $2.47\times 10^{-4}$ \\
$1250$ GeV & \multicolumn{1}{c|}{$0.05$} & \multicolumn{1}{c|}{$4.37\times 10^{-4}$}  & $1.20\times 10^{-4}$ & \multicolumn{1}{c|}{$0.17$}  & \multicolumn{1}{c|}{$0.59$} & $4.90$  & \multicolumn{1}{c|}{$4.28\times 10^{-4}$} & \multicolumn{1}{c|}{$2.05\times 10^{-4}$}  & $2.65\times 10^{-3}$
\end{tabular}}
\caption{Signal significance for an evolutive algorithm that maximizes accuracy metric. All significances are computed for $\mathcal{L} = 3000$ fb$^{-1}$.}\label{tab:Evolve_Acc_table}
\end{table}
We have repeated the same procedure for $m_{e_4} = 200~\textrm{and}~486~\mathrm{GeV}$ considering an evolutive algorithm that maximizes the Asimov significance. Our results can be found in Tab.~\ref{tab:Evolve_Asimov_table}.
\begin{table}[h!]
\centering
\resizebox{\textwidth}{!}{\begin{tabular}{c||ccc||ccc||ccc||}
\multirow{2}{*}{Mass of $e_4$} & \multicolumn{3}{c||}{$s/\sqrt{s+b}$}                                               & \multicolumn{3}{c||}{$\mathcal{Z}(<1\%)$}                              & \multicolumn{3}{c||}{$\mathcal{Z}_{A}$}                                            \\ \cline{2-10} 
                               & ZA                          & VBF                         & VLBSM                & ZA                           & VBF                         & VLBSM   & ZA                          & VBF                          & VLBSM                \\ \hline
$200$ GeV                      & \multicolumn{1}{c|}{$6.10$} & \multicolumn{1}{c|}{$2.00$} & $12.65$              & \multicolumn{1}{c|}{$12.18$} & \multicolumn{1}{c|}{$2.83$} & $12.70$ & \multicolumn{1}{c|}{$4.44$} & \multicolumn{1}{c|}{$0.145$} & $4.28$               \\
$486$ GeV                      & \multicolumn{1}{c|}{$1.77$} & \multicolumn{1}{c|}{$1.50$} & $11.26$ & \multicolumn{1}{c|}{$2.60$} & \multicolumn{1}{c|}{$2.13$} & $8.62$  & \multicolumn{1}{c|}{$0.30$} & \multicolumn{1}{c|}{$0.53$}  & $0.20$ \\
$677$ GeV                      & \multicolumn{1}{c|}{$0.86$} & \multicolumn{1}{c|}{$1.28$} & $4.02$               & \multicolumn{1}{c|}{$1.21$}  & \multicolumn{1}{c|}{$1.82$} & $7.90$  & \multicolumn{1}{c|}{$0.11$} & \multicolumn{1}{c|}{$0.187$} & $0.015$              
\end{tabular}}
\caption{Signal significance for an evolutive algorithm that maximizes Asimov significance metric. All significances are computed for $\mathcal{L} = 3000$ fb$^{-1}$.}\label{tab:Evolve_Asimov_table}
\end{table}
From the aforementioned tables we notice that for heavy VLL scenarios with $m_{e_4} = 1.25~\mathrm{TeV}$ the combined significance for all event signals drops to values near zero indicating that such channels can be rather challenging for direct searches at the LHC. However, if the background is well under control, the $\mathcal{Z}(<1\%)$ statistics offers a combined significance of $5.66\sigma$. Note that the larger component of the combined significance results from the VLBSM channel with $4.9\sigma$, which means that a potential observation of heavy VLLs with masses in the TeV range can only be possible if the $pp \to \ell \nu_\ell + (0 j, j, jj)$ backgrounds are known and with a high precision. The fast decrease in significance for larger masses is a consequence of the, also fast, decrease in cross-section with increasing mass, as shown in Fig.~\ref{fig:CrossSections-e4-e5}, left panel.
A small significance is also obtained for the heavier VLLs, $e_5$ and $e_6$, whose masses lie beyond $\mathcal{O}(3~\mathrm{TeV})$ and likely out of the reach of the HL-LHC. 

However, recall that all three signal events represent independent variables, which means that we can evaluate a global significance as the sum of the individual ones from each process. This implies that we can consider additional event signals that would boost this global significance. In particular, this entails that including channels with jets from $W$ decays to quarks can be relevant due to a larger expected number of events. In fact, the $W$ decay width is larger for light jets with a branching ratio (BR) of approximately $67.4\%$, rather than for leptons, whose BR is $10.86\%$ \cite{Tanabashi:2018oca}. For the present case, and when the accuracy metric maximisation is concerned, the combined significance at $m_{e_4} = 200~\mathrm{GeV}$ is $27.16\sigma$ for $\mathcal{Z}(<1\%)$, $3.46\sigma$ for $\mathcal{Z}_A$ and $4.33\sigma$ for the naive significance. If one instead maximizes the Asimov metric, the same benchmark point ($m_{e_4} = 200~\mathrm{GeV}$) yields $27.71\sigma$ for $\mathcal{Z}(<1\%)$, $8.57\sigma$ for $\mathcal{Z}_A$ and $20.75\sigma$ for the naive significance. These results, and in particular those for the naive significance, indicate that a light VLL characteristic of our model is expected to have a strong presence for a high-luminosity run at the LHC and can be probed well before the end of the LHC programme.

In general, we note that for an ever increasing mass of $e_4$, the overall significance reduces, as we have already discussed above. We also see that the rate of decrease is faster for the ZA channel rather than for the VBF and VLBSM ones. Note that VBF signals yield a maximum significance in $\mathcal{Z}(<1\%)$ of $2.83\sigma$ for $m_{e_4} = 200$ GeV and for both the Asimov and accuracy metrics maximization. On the other hand, for the ZA and VLBSM channels with maximized accuracy, the same VLL mass yields a significance of $12.18\sigma$ for ZA events and $12.95\sigma$ for the VLBSM channel (see Tab.~\ref{tab:Evolve_Acc_table}). However, for results obtained upon maximization of the Asimov metric the naive significance can be as large as $6.10\sigma$ for ZA events and $12.65$ for VLBSM ones (see Tab.~\ref{tab:Evolve_Asimov_table}).
\begin{figure*}[]
	\centering
	\subfloat[]{{\includegraphics[width=0.55\textwidth]{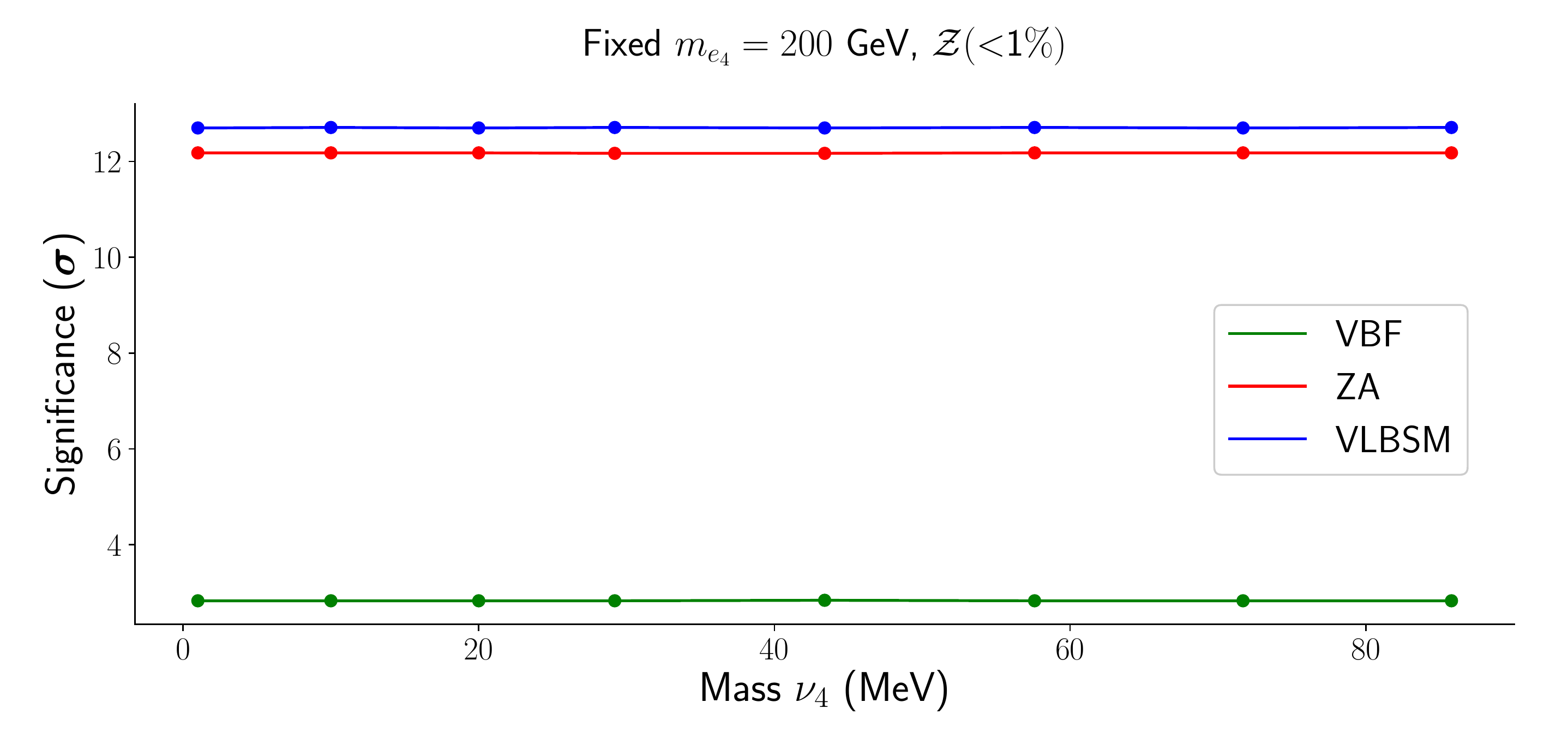} }} 
	\subfloat[]{{\includegraphics[width=0.55\textwidth]{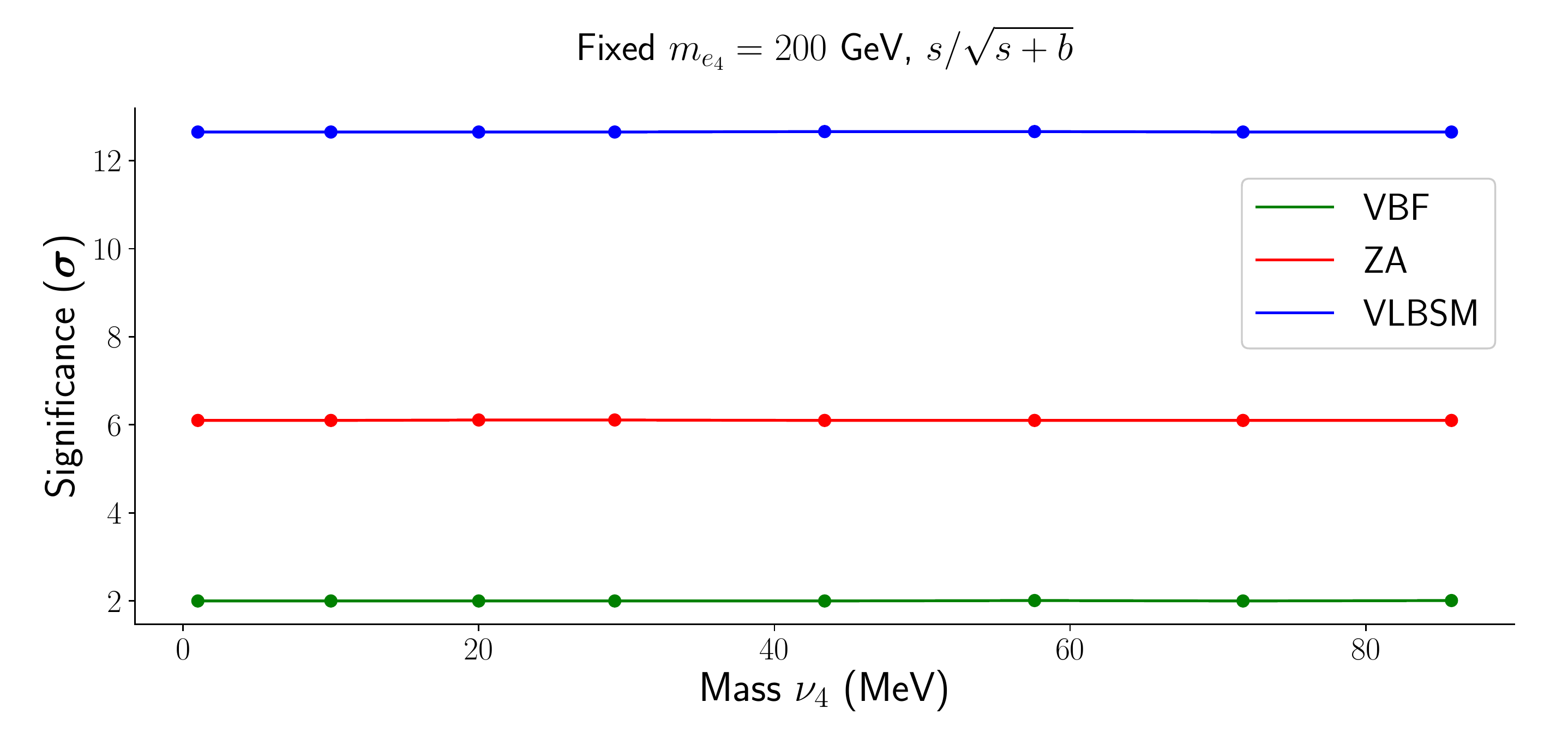} }} \\
	\subfloat[]{{\includegraphics[width=0.55\textwidth]{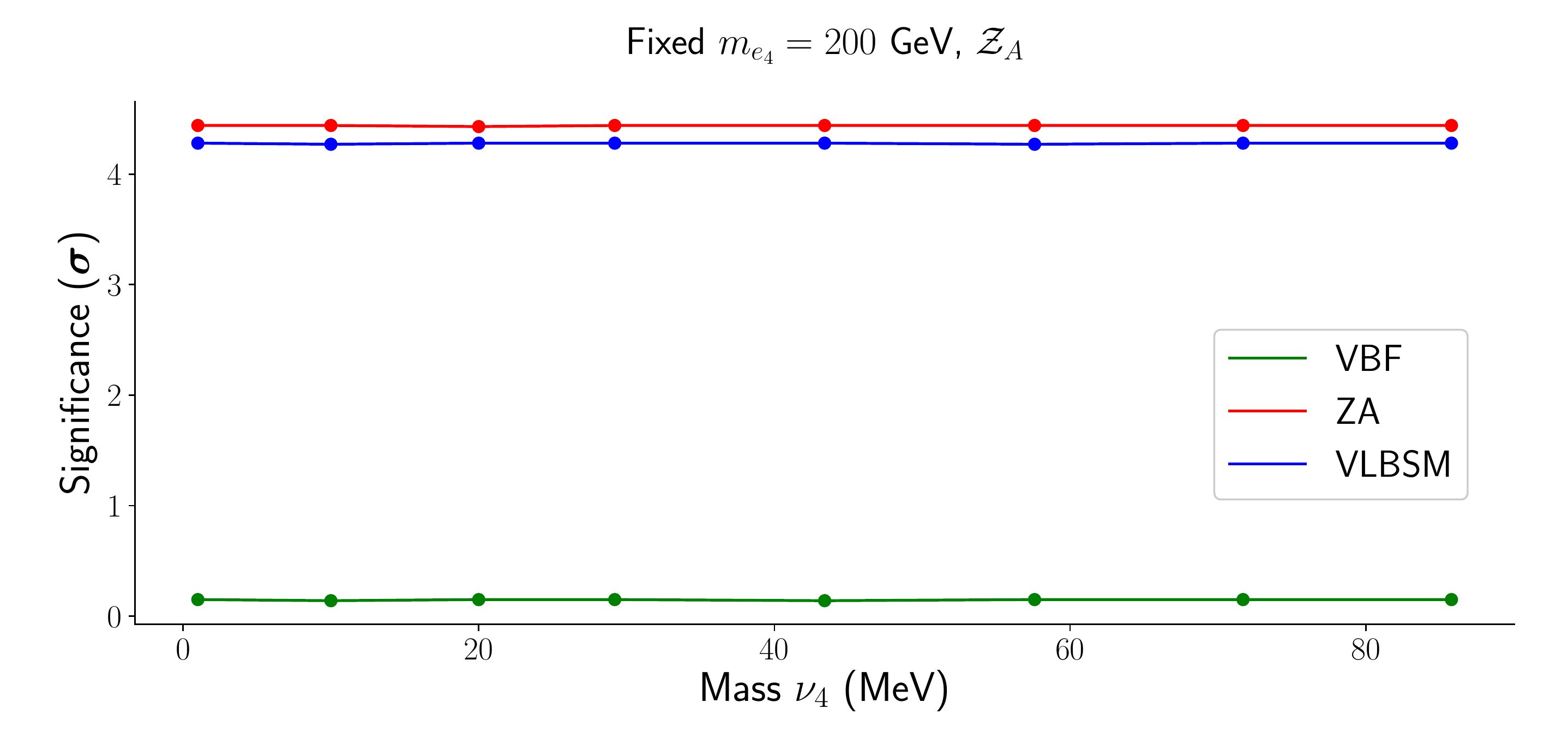} }}
	\caption{Significance as a function of the neutrino's mass for a luminosity of $\mathcal{L} = 3000$ fb$^{-1}$. The significance is computed following an evolutive algorithm that maximises the Asimov metric.}
	\label{fig:Neutrino-scan-plots}
\end{figure*}

For completeness of information, we show in Fig.~\ref{fig:Neutrino-scan-plots} that fixing $m_{e_4} = 200~\mathrm{GeV}$ and for $m_{\nu_4} < m_{\nu_5}$, the effect of varying the sterile neutrino mass, $m_{\nu_4}$, is negligible
for any value of the lightest BSM neutrino mass in the range $100~\mathrm{keV}$ to $100~\mathrm{MeV}$. Note that, while our analysis is generic enough, due to a combination between the seesaw nature of neutrino masses and the radiative origin of Yukawa interactions in the low-scale SHUT model, these mass scales for $\nu_4$ are realistic, compatible with the model' structure and should be seriously considered.

\begin{figure}[]
	\centering
	\subfloat[]{{\includegraphics[width=0.55\textwidth]{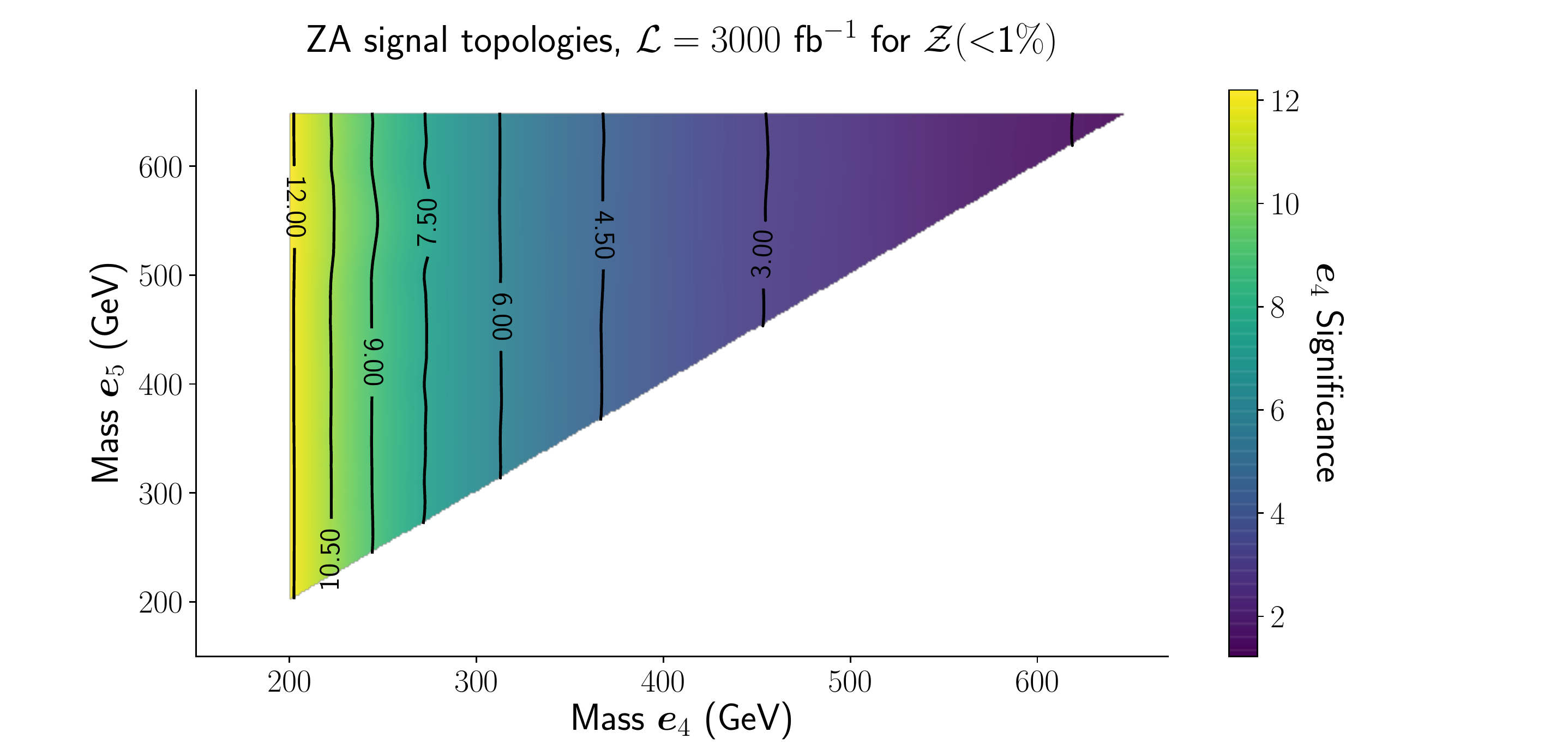} }} 
	\subfloat[]{{\includegraphics[width=0.55\textwidth]{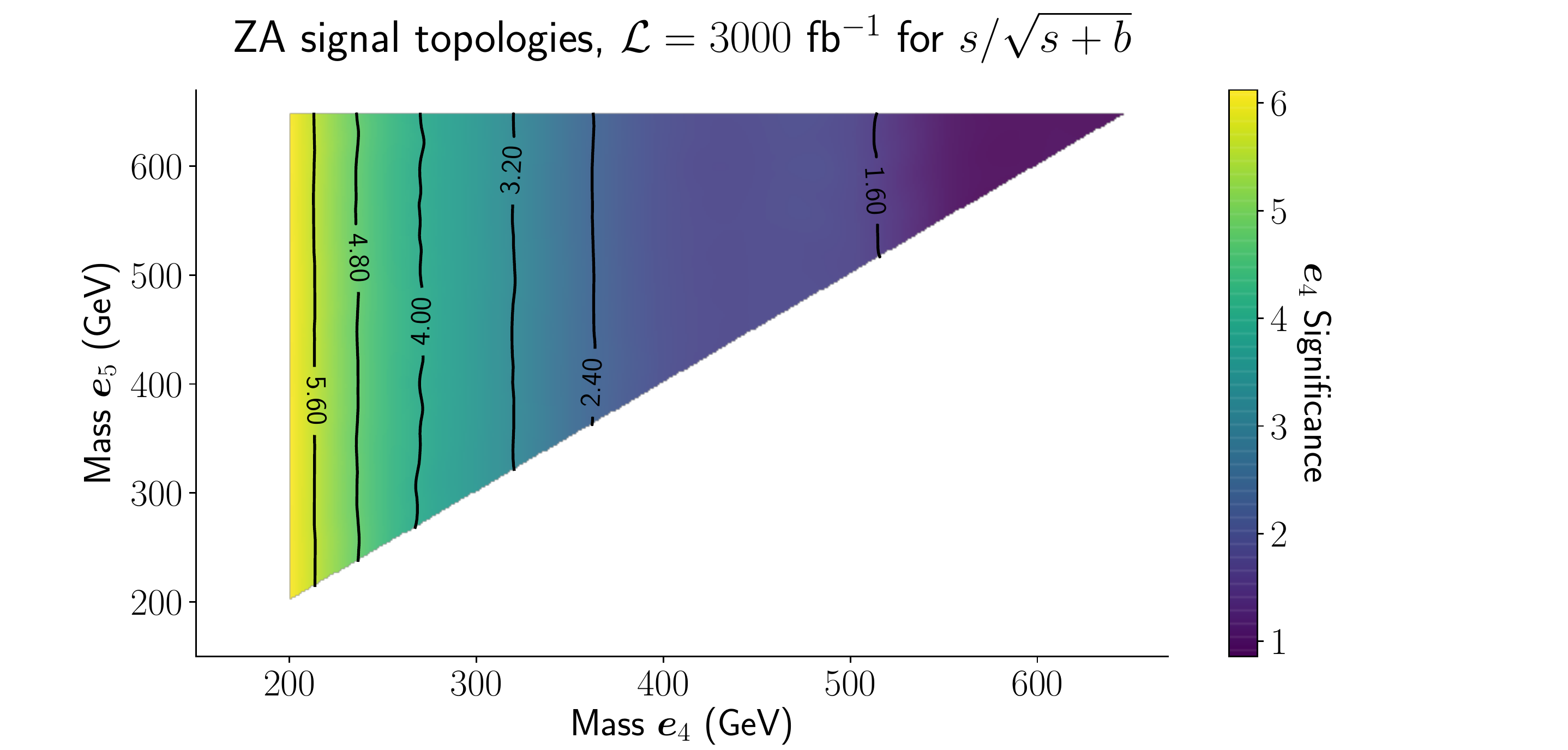} }} \\
	\subfloat[]{{\includegraphics[width=0.55\textwidth]{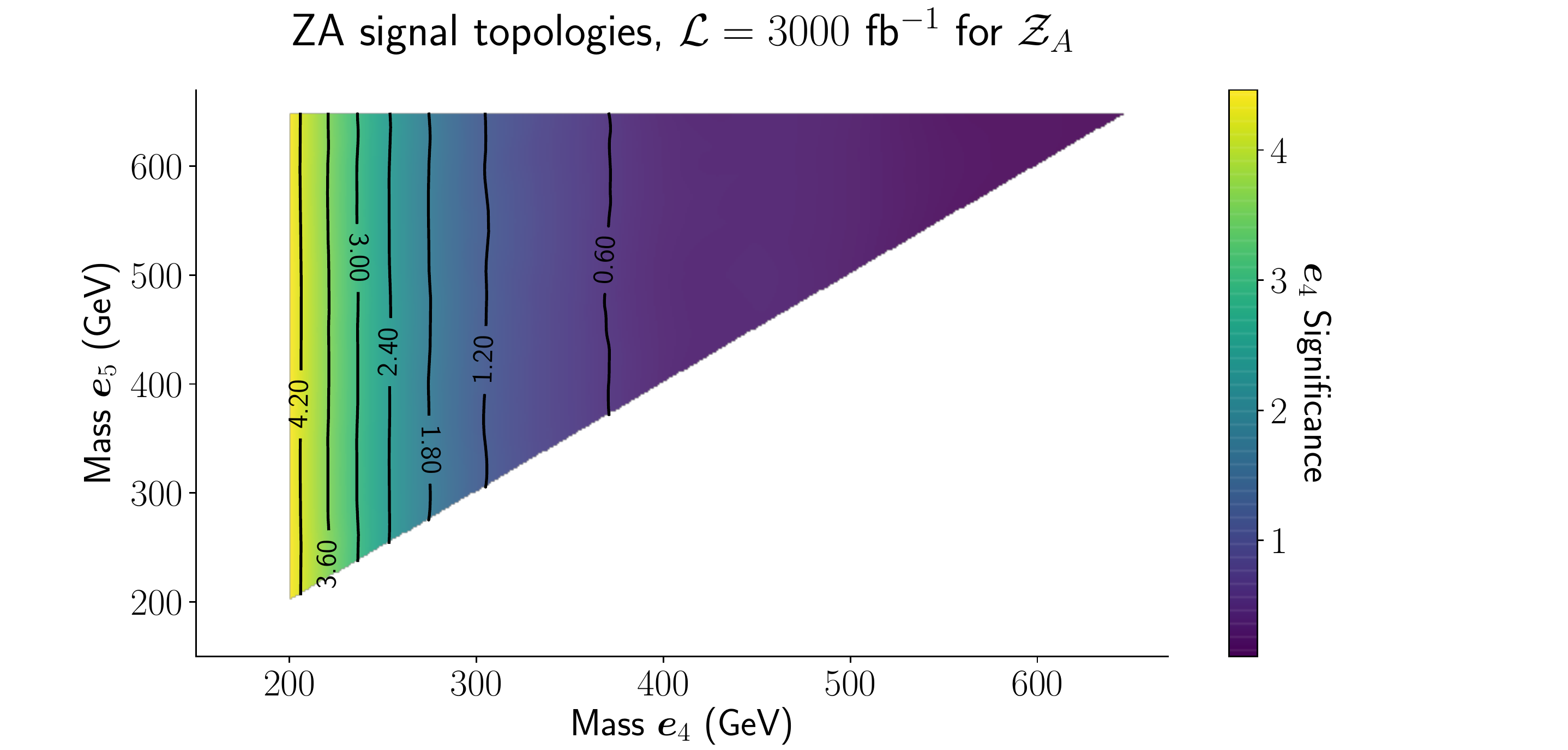} }}
	\caption{Significance contour plots for (a) $\mathcal{Z}(<1\%)$, (b) $s/\sqrt{s+b}$ and (c) $\mathcal{Z}_A$. The colour scale represents a ZA event signal for $e_4$ at a luminosity of $\mathcal{L} = 3000$ $\text{fb}^{-1}$. The significance is computed with an evolutive algorithm that maximises the Asimov metric.}
	\label{fig:E4-E5-sig-Contur-plots}
\end{figure}

Provided that lighter VLLs represent a more interesting case for forthcoming explorations at the LHC we will essentially focus our attention in the mass range $[200,700]~\mathrm{GeV}$ for both $e_4$ and $e_5$, in such a way that $m_{e_4} < m_{e_5}$. This choice is based on our discussion in Sec.~\ref{subsubsec:Benchmarks_masses} just below Eq.~\eqref{eq:VLL_mass_taylor}, where two light VLLs below $1~\mathrm{TeV}$ order with a heavy one at around $5~\mathrm{TeV}$ is a viable scenario. First, we fix the mass of the $e_4$ to be $m_{e_4} = 200$ GeV, as this represents the case where we obtain the greatest significance, as shown in Tabs.~\ref{tab:Evolve_Acc_table} and \ref{tab:Evolve_Asimov_table}. We also consider the BSM neutrino to be in the hundreds of keV order. The results of these scans can be seen in Fig.~\ref{fig:E4-E5-sig-Contur-plots} where we show $e_4$ significance contours in terms of $m_{e_4}$ and $m_{e_5}$.
One can immediately see that a varying $e_5$ mass has a very marginal impact on the $e_4$ significance, indicating that it is independent of $m_{e_5}$. This can be understood from the $e_5 \to W \nu_i$ decay branching fractions shown in Tab.~\ref{tab:BR-data}.
\begin{table}[h!]
    \centering
	\begin{tabular}{c|c|c|c|}
		& \begin{tabular}[c]{@{}c@{}}$m_{e_4} = 200$ GeV\\ $m_{e_5} = 200$ GeV\end{tabular} & \begin{tabular}[c]{@{}c@{}}$m_{e_4} = 200$ GeV\\ $m_{e_5} = 250$ GeV\end{tabular} & \begin{tabular}[c]{@{}c@{}}$m_{e_4} = 200$ GeV\\ $m_{e_5} = 300$ GeV\end{tabular} \\ \hline
		$e_5 \rightarrow W\nu_4$ & 0.5082513                                                                               & 0.5069336                                                                               & 0.5063337                                                                               \\
		$e_5 \rightarrow W\nu_5$ & 0.3111182                                                                               & 0.3119451                                                                               & 0.3123246                                                                               \\
		$e_5 \rightarrow W\nu_6$ & 0.1806305                                                                               & 0.1811213                                                                               & 0.1813417                                                                              
	\end{tabular}
	\caption{\label{tab:BR-data} Branching fractions of $e_5$ decaying into $W$ and BSM neutrinos for three different $e_5$ masses and fixed $m_{e_4} = 200$ GeV}
\end{table}
In fact, we observe that the overall BRs do not suffer significant alterations with the varying $e_5$ mass, and as such, the computed $e_4$ production and decay cross-section remains very much the same. Therefore we do not expect visible changes in the significance and thus, the impact on the significance is understandably small.

\begin{figure}[]
	\centering
	\includegraphics[width=0.80\textwidth]{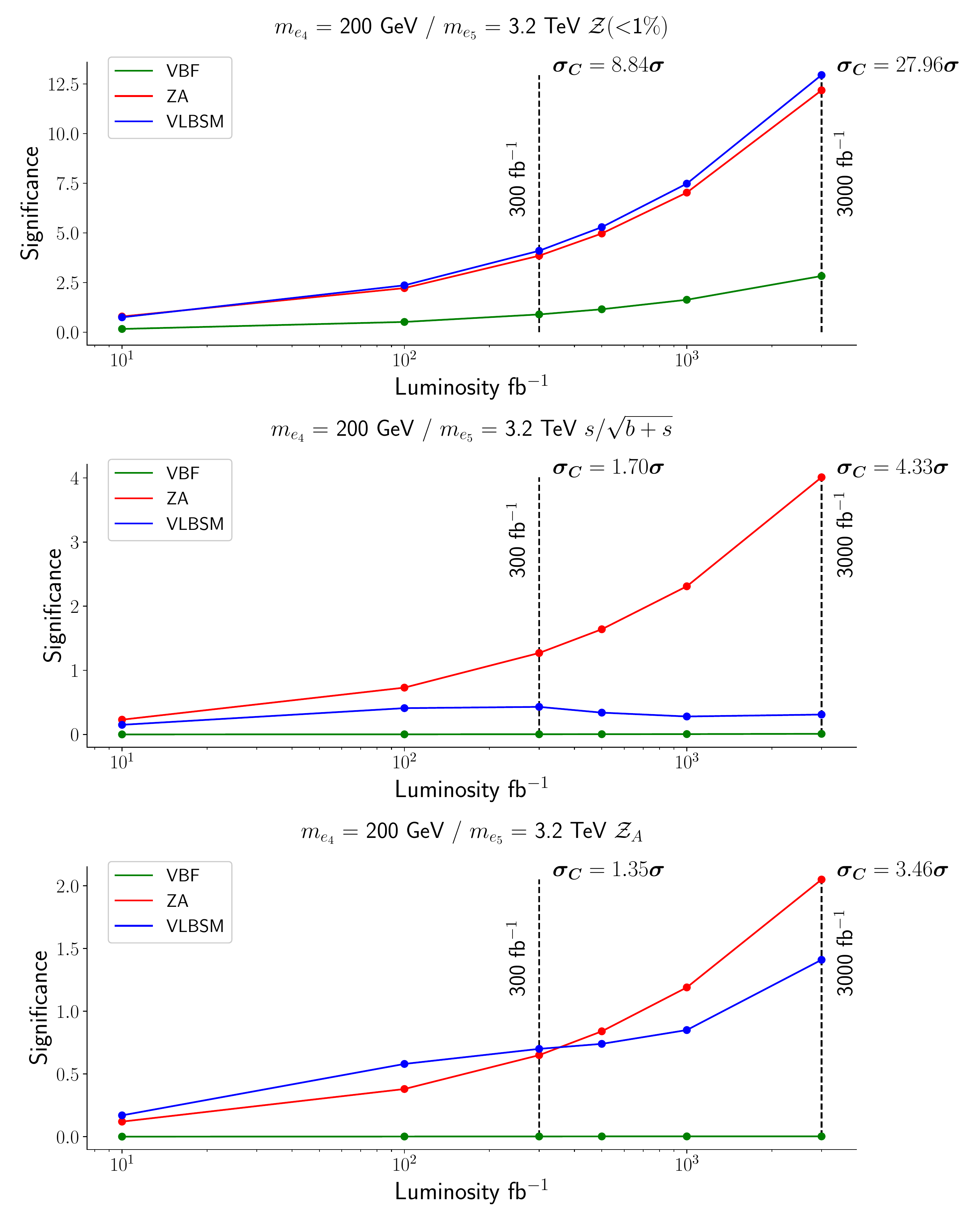}
	\caption{Significance as a function of the luminosity for different statistics and topologies that results from an evolutive algorithm that maximises accuracy. Here, $x$ axis is in logarithmic scale for all plots and in the bottom plot the $y$ axis is also in logarithmic scale. Green plots correspond to VBF signal events, red plots correspond to ZA signals and blue plots are representative of VLBSM topologies. The top plot corresponds to the Asimov significance where backgrounds are exactly known with a systematics of 1\%, the middle plot is representative of the naive significance ($\sigma =s/\sqrt{s+b}$) and the bottom plot corresponds to the Asimov significance where backgrounds are not exactly known. The combined significance shown for $\mathcal{L} = 300$ fb$^{-1}$ and $\mathcal{L} = 300$ fb$^{-1}$ is defined as $\sigma_C = \sigma_{\text{VBF}} + \sigma_{\text{ZA}} + \sigma_{\text{VLBSM}}$. The lightest VLL has mass of 200 GeV, while the second lightest has mass of 3.2 TeV.}
	\label{fig:Lum-plots-ACC-EVO}
\end{figure}

\begin{figure}[]
	\centering
	\includegraphics[width=0.80\textwidth]{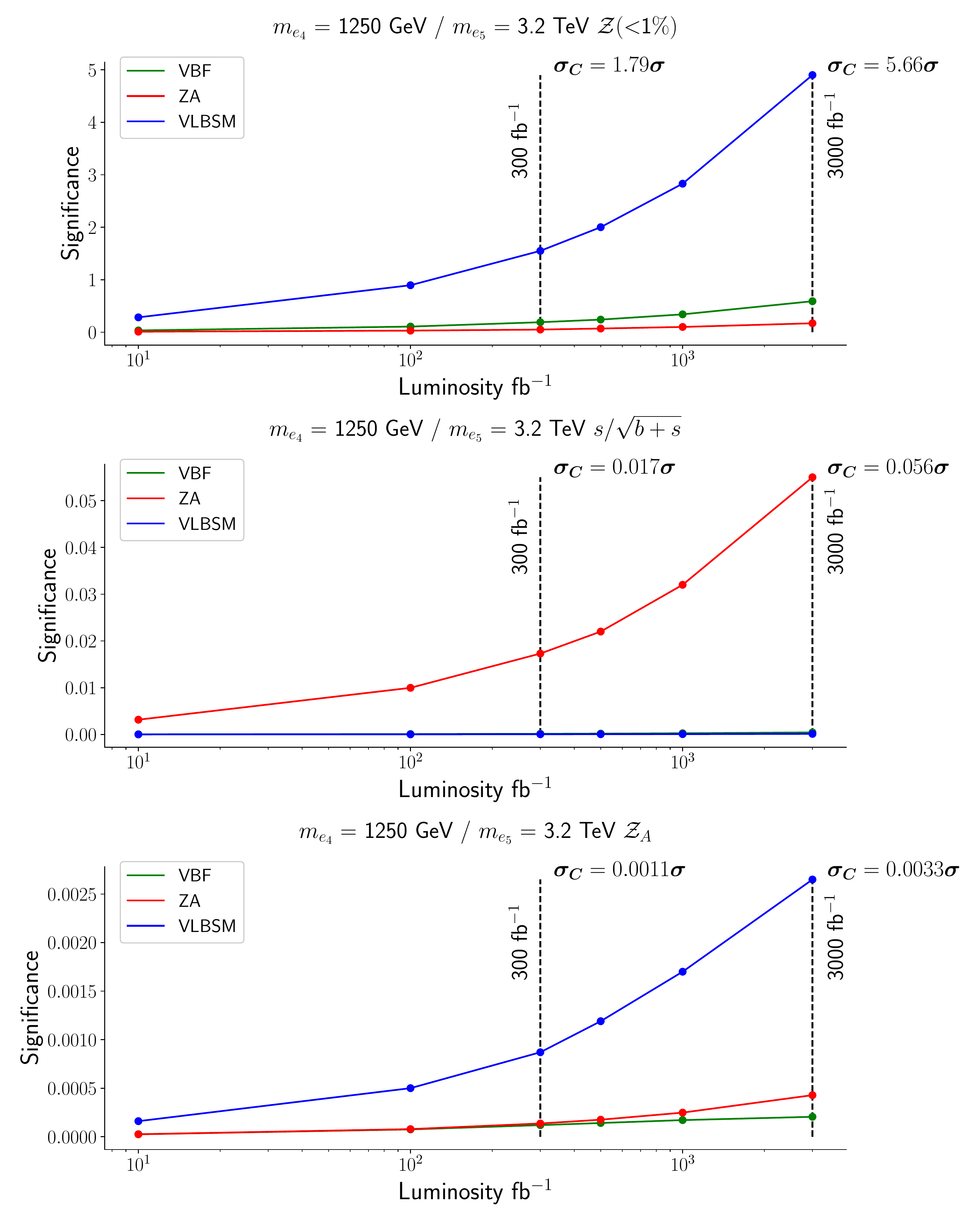}
	\caption{The same as in Fig.~\ref{fig:Lum-plots-ACC-EVO} but for a lightest VLL mass $m = 1250$ GeV.}
	\label{fig:Lum-plots-ACC-EVO-1250}
\end{figure}

\begin{figure}[]
	\centering
	\includegraphics[width=0.80\textwidth]{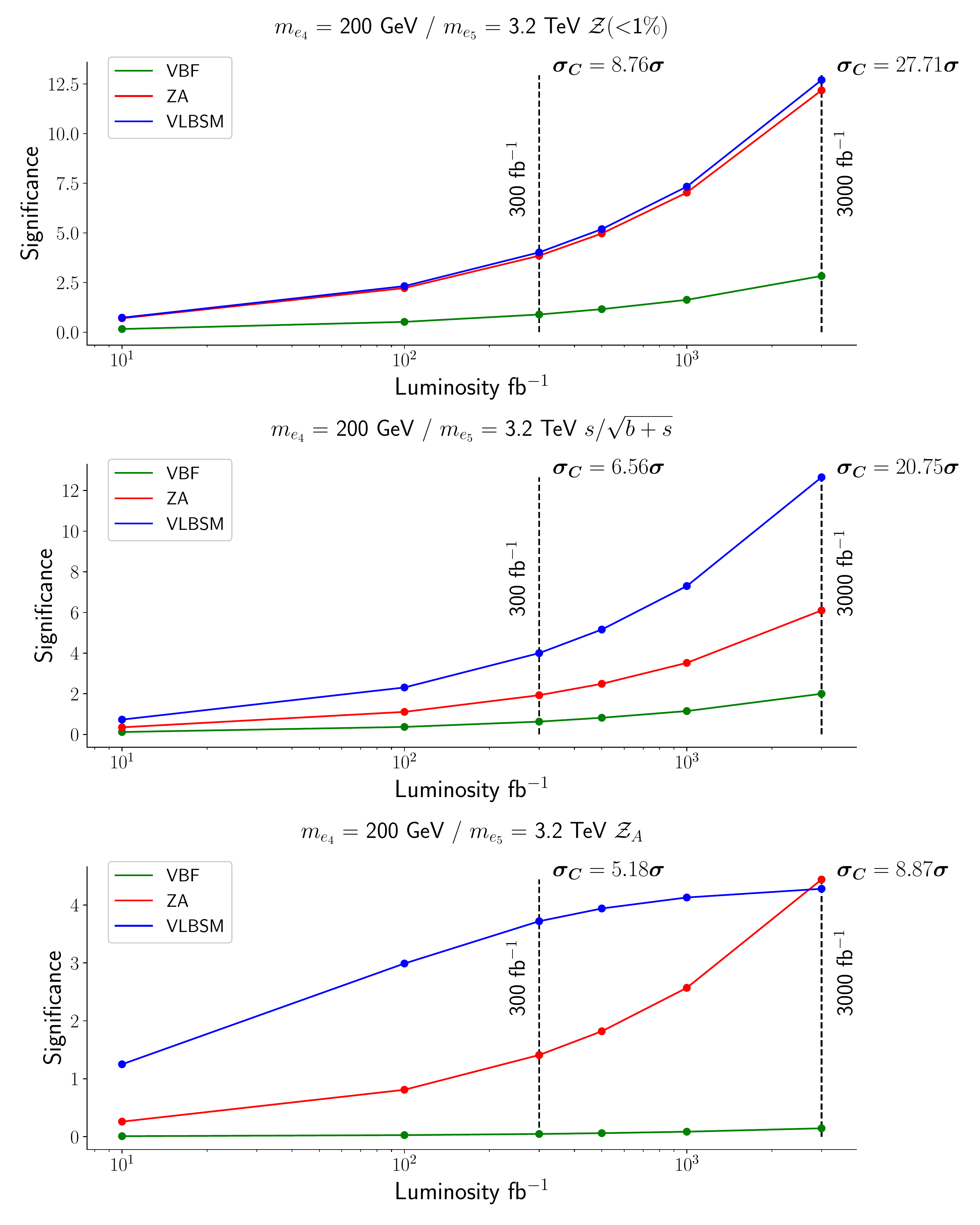}
	\caption{The same as in Fig.~\ref{fig:Lum-plots-ACC-EVO} but for an evolutive algorithm that maximizes the Asimov significance.}
	\label{fig:Lum-plots-ASIMOV-EVO}
\end{figure}

\begin{figure}[]
	\centering
	\includegraphics[width=0.80\textwidth]{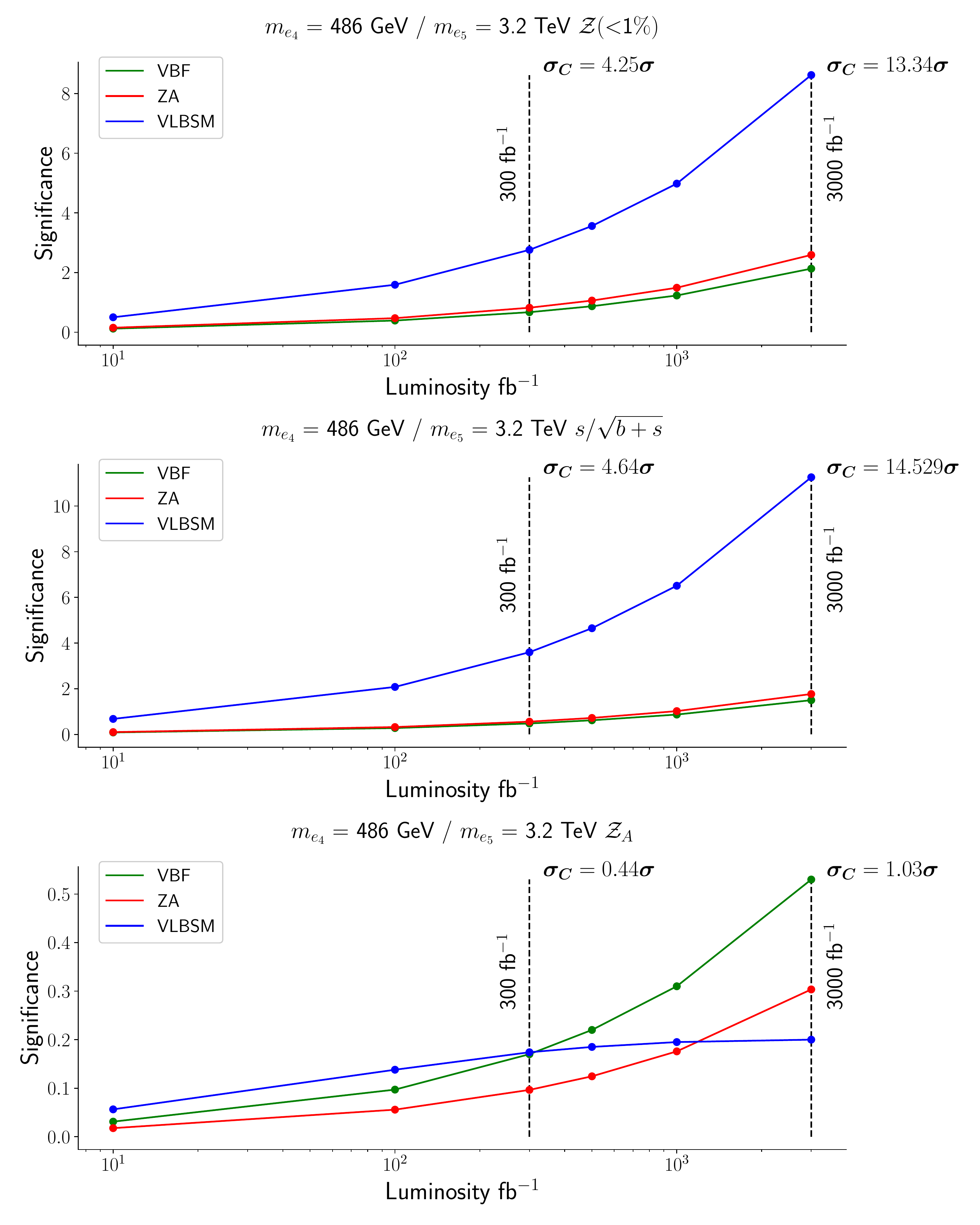}
	\caption{The same as in Fig.~\ref{fig:Lum-plots-ASIMOV-EVO} but for a lightest VLL mass $m = 486~\mathrm{GeV}$.}
	\label{fig:Lum-plots-ASIMOV-EVO-486}
\end{figure}

\begin{figure}[]
	\centering
	\includegraphics[width=0.80\textwidth]{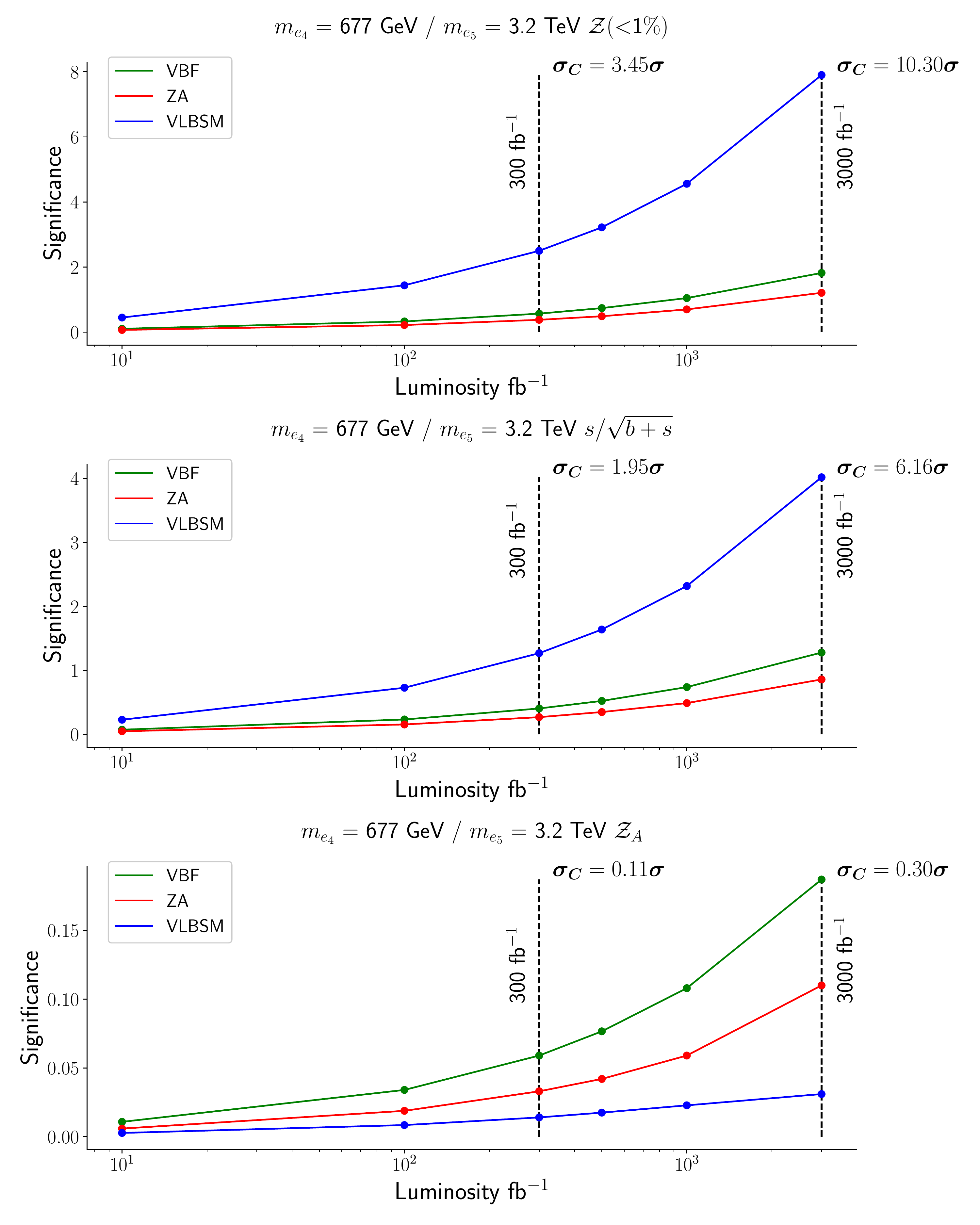}
	\caption{The same as in Fig.~\ref{fig:Lum-plots-ASIMOV-EVO} but for a lightest VLL mass $m = 677~\mathrm{GeV}$.}
	\label{fig:Lum-plots-ASIMOV-EVO-677}
\end{figure}

So far all results have been computed for a luminosity of $\mathcal{L} = 3000~\mathrm{fb}^{-1}$. However, the LHC is only scheduled to run at such luminosities around 2026-2030 \cite{Monica}. Therefore it is equally relevant to study how the significance changes for lower luminiosities, in particular, for $300$ $\mathrm{fb}^{-1}$, which is planned to be delivered in the Run III, scheduled to start during 2021. With this in mind, we show in Figs.~\ref{fig:Lum-plots-ACC-EVO} and \ref{fig:Lum-plots-ACC-EVO-1250} the dependency of the significance over the projected luminosities for a maximized accuracy. Results for the case where the Asimov significance is maximized are shown in Figs.~\ref{fig:Lum-plots-ASIMOV-EVO}, \ref{fig:Lum-plots-ASIMOV-EVO-486} and \ref{fig:Lum-plots-ASIMOV-EVO-677}.

Looking first at Figs.~\ref{fig:Lum-plots-ACC-EVO} and \ref{fig:Lum-plots-ACC-EVO-1250}, for $\mathcal{L} = 300$ $\mathrm{fb}^{-1}$, we note that a signal significance at or beyond $5\sigma$ level is only achievable for a light VLL of 200 GeV. In particular, the combined significance is $8.84\sigma$ for $\mathcal{Z}(<1\%)$. This means that, if all backgrounds introduced in Sec.~\ref{section:Numerics} are known with high precision, it will be possible to either discover or exclude a VLL with mass $200~\mathrm{GeV}$, possibly even before the end of Run III. 

For the case of heavy VLLs, the $\mathcal{Z}(<1\%)$ significance is no larger than $1.78\sigma$ for the case of $m_{e_4} = 1.25~\mathrm{TeV}$, where VLBSM signals provide the larger contribution. With these results in mind, we would only expect to observe such heavy states in a high luminosity run. In fact, since we are simulating proton-proton collisions at $\sqrt{s} = 14$ TeV, at such high masses, pair production of VLLs is increasingly unlikely, meaning that we only expect such heavy states to become visible in either high-luminosity runs or at higher energy colliders.

These results confirm that light states are favored to be probed at Run III of the LHC. This is especially noticeable when we use an evolutive algorithm that maximizes the Asimov significance. In plots \ref{fig:Lum-plots-ASIMOV-EVO}, \ref{fig:Lum-plots-ASIMOV-EVO-486} and \ref{fig:Lum-plots-ASIMOV-EVO-677}, we obtain, for a VLL of 200 GeV, significances well above $5\sigma$, at $\mathcal{L} = 300~\mathrm{fb}
^{-1}$ in all three statistics. Therefore, a $200~\mathrm{GeV}$ VLL characteristic of our model can already be probed by the LHC Run III. We also note that for a VLL of 486 GeV, we are already able to obtain a significance of $4.25\sigma$ for $\mathcal{Z}(<1\%)$ and $4.64\sigma$ for $s/\sqrt{s+b}$. While the latter two do not pass a $5\sigma$ baseline, they already represent significant deviations from pure SM processes. We argue that the addition of new signals, such as the ones with jets as mentioned before, should offer the necessary boost to achieve $5\sigma$. This type of argumentation can also apply, for example, for a VLL of 677 GeV where we read a combined significance of $3.45\sigma$.
\cleardoublepage
\section{Conclusions}\label{sec:Conclusions}

In this work we have studied the collider phenomenology inherent to $\SU{2}{L}$-doublet VLLs at the LHC, relevant for Run III and beyond. The properties of such exotic leptons were based on a framework built upon unification principles where the strong and EW interactions are ultimately unified with a local family symmetry. One of the major goals of the model under consideration is to offer a potential solution to the flavour problem, including a description for neutrino masses. It features a low-scale theory where new TeV-scale VLLs and VLQs are a natural consequence of the unification of Higgs and matter in common representations, sharing the same gauge and flavour quantum numbers.

We have performed Monte Carlo simulations relying on DL techniques with the aim of determining the statistical significance of an hypothetical VLL discovery at future LHC runs. In this work, simple neural networks were considered, following the implementation of an evolutive algorithm that maximizes either the accuracy metric or the Asimov significance. For the first scenario, we are able to distinguish background events from signal events with an accuracy between 98\% to 100\%, depending in the signal topology in question, while the second scenario provides an increase in the Asimov significance at the cost of lower overall accuracy (32\% to 70\%).

We have proposed three distinct signatures for VLLs in our model which can be searched for at the LHC in the ZA, VBF and VLBSM channels with purely leptonic final states. Three distinct statistical significances were subject of our analysis, namely, the Asimov significance $\mathcal{Z}_A$, an adapted version of the Asimov significance $\mathcal{Z}(<1\%)$ and the well known \textit{naive} significance $s/\sqrt{s+b}$. A combined result of $27.71\sigma$ for $\mathcal{Z}(<1\%)$, $20.75\sigma$ for $s/\sqrt{s+b}$ and finally $8.87\sigma$ for $\mathcal{Z}_A$ was obtained for a luminosity of $3000~\mathrm{fb}^{-1}$, a center of mass beam energy $\sqrt{s}=14~\mathrm{TeV}$, a VLL mass $m_{e_4} = 200~\mathrm{GeV}$ and a maximized Asimov metric significance. Under the same conditions, but for an accuracy metric search we have obtained a combined significance of $27.96\sigma$ for $\mathcal{Z}(<1\%)$, $4.33\sigma$ for $s/\sqrt{s+b}$ and $3.46\sigma$ for $\mathcal{Z}_A$. In this mass range, the ZA and VLBSM channels provide the dominant contributions. Note that we have also considered relatively light BSM neutrinos with a mass of the order of $\mathcal{O}(100~\mathrm{keV})$. Furthermore, we have shown that varying the mass of the lightest BSM neutrino up to $\mathcal{O}(100~\mathrm{MeV})$ has a residual effect on the significance. As expected, the luminosity also has a noticeable impact on the significance. In particular, a value at or above $5\sigma$ can be achieved for $\mathcal{L} = 300~\text{fb}^{-1}$, meaning that a $200~\mathrm{GeV}$ VLL as predicted in our model can already be probed at the LHC Run-III. However, for larger VLL masses, and in particular for $m_{e_4} = 1.25$ TeV, we have observed that a combined $5\sigma$ significance in the fully leptonic channels can only be achieved for $\mathcal{L} = 3000~\mathrm{fb}^{-1}$, that is, at the high-luminosity LHC, and with a combined significance of $5.66\sigma$ if the backgrounds are known with a high precision.

A scenario with two light VLLs, both below $650~\mathrm{GeV}$, was also studied, where an identical significance for an $e_4$ discovery was achieved. This follows from a negligible effect played by different $m_{e_5}$ masses on the decay branching fractions and thus on the signal cross-section.

For all studied cases one has observed that the significance quickly drops if the VLL masses lie beyond $1~\mathrm{TeV}$. One of the first steps beyond the work presented here is to add jets to the final states from $W$ decays into light quarks. Since such channels offer larger branching ratios it is likely possible that the significance for higher masses can be increased. Adding these channels would possibly increase the significance of the region around 1 TeV, but also provide a decisive boost for the $m_{e_4} = 486 \mathrm{GeV}$ scenario, where significances for Run III luminosities over $4\sigma$ are already obtained. As such, we conclude that VLLs in the mass range $\sim[200,500]~\mathrm{GeV}$ are under the capabilities of being either discovered or excluded before the end of Run III.

Based on the model under consideration, the observation of VLLs at the reach of forthcoming LHC runs can offer a crucial probe to falsify our model and obtain hints about the high scale dynamics. For example, if the New Physics scale above the EW one, which was defined by the $p$, $f$ and $\omega$ VEVs, is of the order $100~\mathrm{TeV}$, the $e_4$ mass as given in \eqref{eq:VLL_mass_taylor} would imply that the radiatively generated Yukawa couplings $\kappa_{5,8}$ need to be approximately $\mathcal{O}(10^{-2.7})$. However, one should comment here that an even stronger link to the high-scale can be obtained through the study of VLQs due to the tree-level nature of their masses. In fact, while for VLLs there is still some degree of uncertainty steaming from a non-trivial functional dependency of the $\kappa_{i}$ parameters with masses and couplings, leading contributions to the two lightest VLQ masses are well understood and are proportional to the Yukawa coupling $\mathcal{Y}_{2} \sim \mathcal{O}(10^{-2})$. In turn, this would allow us to fix the $\omega \sim f$ and $p$ scales establishing a direct link to the scale where larger symmetries are broken. Having said this, performing collider phenomenology studies for VLQs using similar methods to those employed in this work is one of our key priorities for the near future. A combined study of both VLQ and VLL sectors can offer us a rather complete information about the $\omega$, $f$ and $p$ scales, the sizes of the radiatively generated Yukawa couplings and the physics involved at such high energies. 

Other important studies to perform concern Higgs and flavour physics which, due to the presence of three $\SU{2}{L}$ scalar doublets, is highly relevant. In fact, we have chosen a basis where the SM lepton sector has zero mixing with other BSM fermions. However, a more complete approach should consider small deviations from this limit up to flavour physics constraints. Last but not least, the presence of keV-MeV scale neutrinos can potentially offer a DM candidate if, in a basis with a more generic neutrino mixing, it is stable enough. In the longer term, with all such phenomenological studies, we can determine more precisely what are the viable regions of the parameter space that will help us in performing a direct matching between the low-scale and the high-scale regimes of the theory.

\section*{Acknowledgments}

The authors want to thank António Onofre for rather thorough and insightful discussions about the subjects addressed in this manuscript. APM and FFF are supported by the Center for Research and Development in Mathematics and Applications (CIDMA) through the Portuguese Foundation for Science and Technology (FCT - Fundação para a Ciência e a Tecnologia), references UIDB/04106/2020 and UIDP/04106/2020. APM, FFF and JG are supported by the project PTDC/FIS-PAR/31000/2017. APM is also supported by the projects CERN/FIS-PAR/0027/2019, CERN/FISPAR/0002/2017 and by national funds (OE), through FCT, I.P., in the scope of the framework contract foreseen in the numbers 4, 5 and 6 of the article 23, of the Decree-Law 57/2016, of August 29, changed by Law 57/2017, of July 19. R.P.~is supported in part by the Swedish Research Council grants, contract numbers
621-2013-4287 and 2016-05996, as well as by the European Research Council (ERC) under 
the European Union's Horizon 2020 research and innovation programme (grant agreement No 668679).

\appendix
\section{Low-scale SHUT model Lagrangian and Feynman rules}\label{app:Feynman Rules}

In this appendix the tree-level Lagrangian and Feynman rules between the EW gauge bosons and leptons are presented. To simplify some notation, projection operators are defined as
\begin{equation}\label{eq:notation}
\begin{aligned}
P_\text{R} = \frac{1+ \gamma^5}{2}, \quad P_\text{L} = \frac{1-\gamma^5}{2}
\end{aligned}
\end{equation}

The Yukawa and fermion bilinear interactions are presented in Sec.~\ref{subsec:Low-energy} in Eqs.~\eqref{eq:YukawaTerms} and \eqref{eq:Bilinear}. Here we write the remaining Lagrangian terms of the low-scale SHUT model by considering all renormalizable, Lorentz and gauge invariant operators. We start by writing out all kinetic terms for the fermions,
\begin{equation}{\label{eq:KinFer}}
\begin{aligned}
& \mathcal{L}_{\text{kin,f}} = i\qty(\bar{Q}_L)^i \slashed{D}\qty(Q_L)_i + i\qty(\bar{L})^i \slashed{D}\qty(L)_i + i\qty(\bar{d}_R)^i \slashed{D}\qty(d_R)_i + i\qty(\bar{u}_R)^i \slashed{D}\qty(u_R)_i + i\qty(\bar{e}_R)^i \slashed{D}\qty(e_R)_i + \\
& + i\qty(\bar{E}_L)^i \slashed{D}\qty(E_L)_i + i\qty(\bar{E}_R)^i \slashed{D}\qty(E_R)_i + i\qty(\bar{D}_L)^i \slashed{D}\qty(D_L)_i + i\qty(\bar{D}_R)^i \slashed{D}\qty(D_R)_i + i\qty(\bar{\nu}_R)^i \slashed{D}\qty(\nu_R)_i,
\end{aligned}
\end{equation}
where repeated index $i$ represents summation over the different generations. The covariant derivative is defined as\footnote{
	We have a quite strong abuse of language here. In fact, the covariant derivate for the term $i\qty(\bar{Q}_L)^i \slashed{D}\qty(Q_L)_i$ is different than, for example, $i\qty(\bar{E}_L)^i \slashed{D}\qty(E_L)_i$. That is because quarks couple to gluons, while the leptons do not, so the covariant derivative for leptons \textbf{does not} have the last term of \ref{eq:CovDeriv}. One should interpret the definition in this fashion, that is, if it interacts, it exists, if it does not, it does not exist.}
\begin{equation}\label{eq:CovDeriv}
D_\mu = \partial_\mu - ig\frac{Y}{2}B_\mu - ig_w\frac{\sigma_a}{2}A^a_{\mu} - ig_s\frac{\lambda_a}{2}G^a_{\mu}.
\end{equation}

The kinetic terms for the bosonic sector reads
\begin{equation}\label{eq:BosFer}
\mathcal{L}_{\text{bos,f}} = -\frac{1}{4}B\indices{^\mu^\nu}B\indices{_\mu_\nu} - \frac{1}{4}A_b^{\mu\nu}A^b_{\mu\nu} - \frac{1}{4}G_c^{\mu\nu}G^c_{\mu\nu} + \frac{1}{2}\qty(D_\mu\phi_a)\qty(D^\mu\phi^a)^\dagger,
\end{equation}
where $b$ and $c$ represent $\text{SU(2)}_\text{L}$ and $\text{SU(3)}_\text{C}$ adjoint indices, while $a$ denotes scalar generations. The scale potential is that of a generic 3HDM model and reads as:
\begin{equation}\label{eq:Potential}
V\qty(\phi,\phi^\dagger)= \qty(m_i)^2\abs{\phi^i}^2 + \qty(m_{ij}^2\phi^i\qty(\phi^j)^\dagger + \text{H.c}.) + \lambda_{ijkl}\qty(\phi^i\qty(\phi^j)^\dagger\phi^k\qty(\phi^l)^\dagger + \text{H.c}.).
\end{equation}
The full Lagrangian density for the effective low-energy 3HDM is the sum of all previous sectors and reads as
\begin{equation}\label{eq:3HDM}
\mathcal{L}_{\text{3HDM}} = \mathcal{L}_{\text{kin,f}} + \mathcal{L}_{\text{bos,f}} + \mathcal{L}_{\text{y}} + \mathcal{L}_{\text{bil}} - V\qty(\phi,\phi^\dagger)\,.
\end{equation}

Expanding and rotating \eqref{eq:3HDM} to the mass basis one arrives to the following Feynman rules:
\begin{itemize}

\item \textbf{Lepton and Gauge bosons interactions}
\begin{itemize}

\item Charged Leptons - Photon vertex
\begin{center} 
\includegraphics[width=0.30\textwidth]{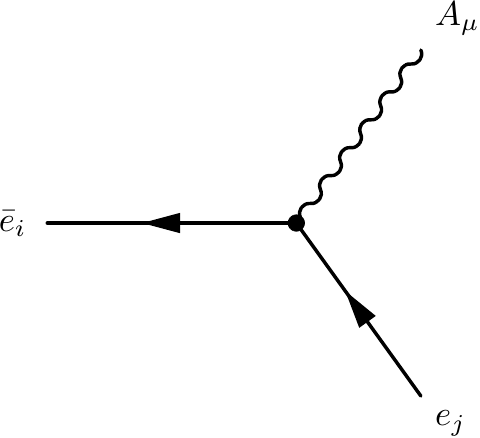}
\end{center}  
\begin{equation}
\begin{aligned} 
 &\frac{i}{2} \delta_{i j} \Big(g_1 \cos\theta_W   + g_2 \sin\theta_W  \Big)\Big(\gamma^{\mu}\cdot P_\text{L}\Big) + \,\frac{i}{2} \Big(2 g_1 \cos\theta_W  \sum_{a=1}^{3}U^{e,*}_{R,{i a}} U_{R,{j a}}^{e} \\ 
  &  + \Big(g_1 \cos\theta_W   + g_2 \sin\theta_W  \Big)\sum_{a=1}^{3}U^{e,*}_{R,{i 3 + a}} U_{R,{j 3 + a}}^{e}  \Big)\Big(\gamma^{\mu}\cdot P_\text{R}\Big)\end{aligned}
  \end{equation}

    \item Charged Leptons - Z boson vertex
    \begin{center}
\includegraphics[width=0.30\textwidth]{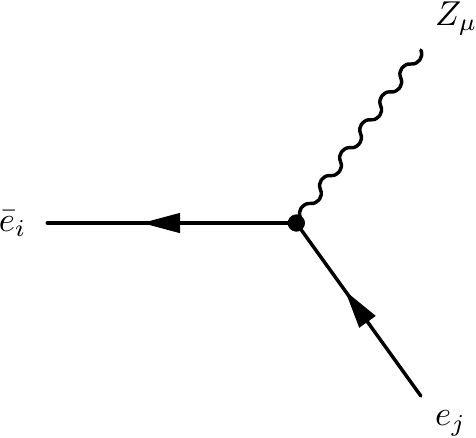}
\end{center}  
\begin{equation}\begin{aligned} 
 &\frac{i}{2} \delta_{i j} \Big(- g_1 \sin\theta_W   + g_2 \cos\theta_W  \Big)\Big(\gamma^{\mu}\cdot P_\text{L}\Big) + \,-\frac{i}{2} \Big(2 g_1 \sin\theta_W  \sum_{a=1}^{3}U^{e,*}_{R,{i a}} U_{R,{j a}}^{e}  \\ 
  & + \Big(g_1 \sin\theta_W   - g_2 \cos\theta_W  \Big)\sum_{a=1}^{3}U^{e,*}_{R,{i 3 + a}} U_{R,{j 3 + a}}^{e}  \Big)\Big(\gamma^{\mu}\cdot P_\text{R}\Big)\end{aligned}\end{equation}
     \item Neutrinos - Z boson vertex
\begin{center} 
\includegraphics[width=0.30\textwidth]{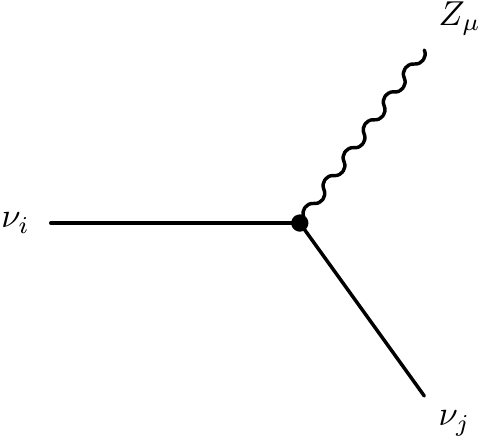}
\end{center}  
\begin{equation}\begin{aligned} 
 &\frac{i}{2} \Big(g_1 \sin\theta_W   + g_2 \cos\theta_W  \Big)\Big(- \sum_{a=1}^{6}U^*_{{\nu},{j a}} U_{\nu,{i a}}   + \sum_{a=1}^{3}U^*_{{\nu},{j 6 + a}} U_{\nu,{i 6 + a}} \Big)\Big(\gamma^{\mu}\cdot P_\text{L}\Big)\\ 
  & + \,-\frac{i}{2} \Big(g_1 \sin\theta_W   + g_2 \cos\theta_W  \Big)\Big(- \sum_{a=1}^{6}U^*_{{\nu},{i a}} U_{\nu,{j a}}   + \sum_{a=1}^{3}U^*_{{\nu},{i 6 + a}} U_{\nu,{j 6 + a}} \Big)\Big(\gamma^{\mu}\cdot P_\text{R}\Big)\end{aligned}\end{equation}

\item Neutrinos - Charged Leptons - $W^{-}$ boson vertex

\begin{center} 
\includegraphics[width=0.30\textwidth]{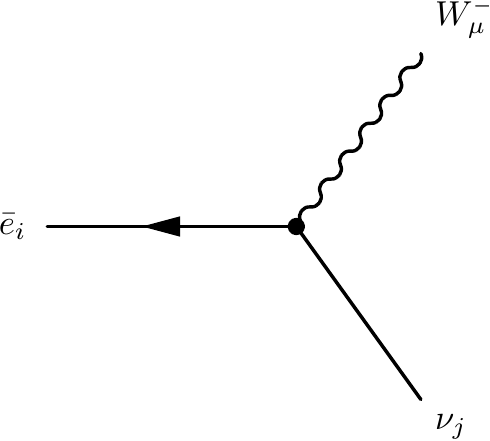}
\end{center}  
\begin{equation}\begin{aligned} 
 &-i \frac{1}{\sqrt{2}} g_2 \sum_{a=1}^{6}U^*_{{\nu},{j a}} U_{L,{i a}}^{e}  \Big(\gamma^{\mu}\cdot P_\text{L} \Big)\\ 
  & + \,-i \frac{1}{\sqrt{2}} g_2 \sum_{a=1}^{3}U^{e,*}_{R,{i 3 + a}} U_{\nu,{j 6 + a}}  \Big(\gamma^{\mu}\cdot P_\text{R}\Big)\end{aligned}\end{equation}

  \item Neutrinos - Charged Leptons - $W^{+}$ boson vertex  
  
  \begin{center} 
\includegraphics[width=0.30\textwidth]{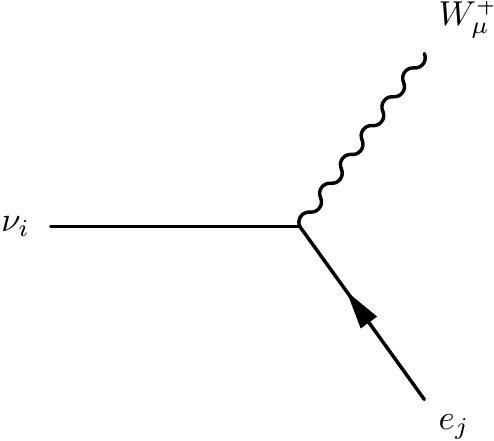}
\end{center}  
\begin{equation}\begin{aligned} 
 &-i \frac{1}{\sqrt{2}} g_2 \sum_{a=1}^{6}U^{e,*}_{L,{j a}} U_{\nu,{i a}}  \Big(\gamma^{\mu}\cdot P_\text{L}\Big)\\ 
  & + \,-i \frac{1}{\sqrt{2}} g_2 \sum_{a=1}^{3}U^*_{{\nu},{i 6 + a}} U_{R,{j 3 + a}}^{e}  \Big(\gamma^{\mu}\cdot P_\text{R}\Big)\end{aligned}\end{equation} 
    
\end{itemize} 

\end{itemize}

\section{Kinematic and angular variable for all topologies with detector effects}\label{app:Kin-Ang-vars}

\begin{figure*}[htb!]
	\centering
	\includegraphics[width=\textwidth]{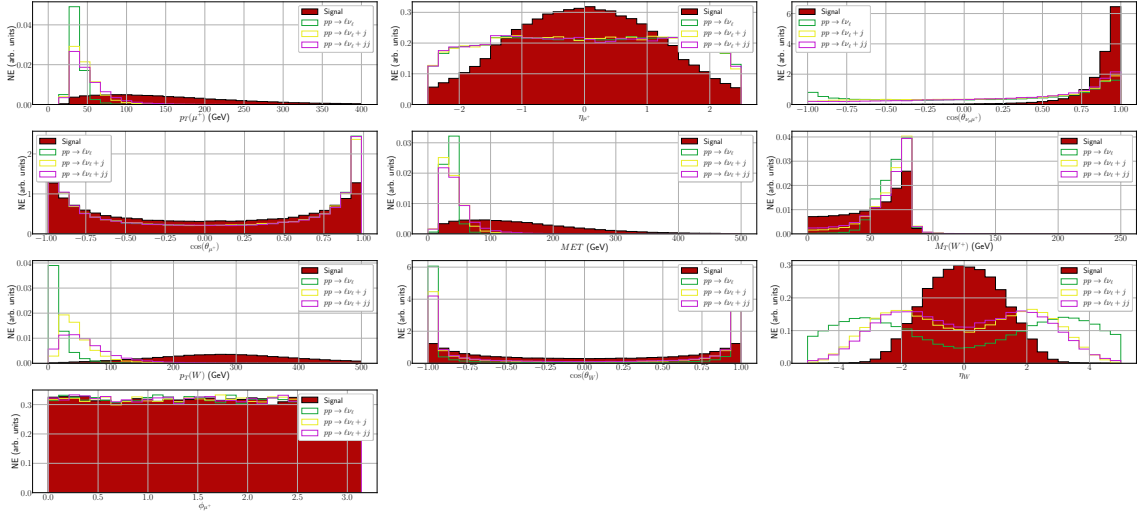}
	\caption{Dimensionless (angular) and dimension-full (kinematic) observables at lab reference frame for VLBSM channel (solid red), with $pp \rightarrow \ell\nu_\ell$ (green line), $pp \rightarrow \ell \nu_\ell + j$ (yellow line) and $pp \rightarrow \ell\nu_\ell + jj$ (purple line) backgrounds where it is considered 30 bins for all histograms. From top left to bottom right, we have distributions for transverse momentum $\mu^+$, pseudo-rapidity $\mu^+$, $\cos(\theta_\mu^{+})$, $\cos(\theta_{\nu_\mu^{+}\mu^{+}})$, MET, transverse mass for $W$, transverse momentum for $W$, $\cos(\theta_W)$, pseudo-rapidity for $W$ and azimuthal angle for $\mu^{+}$}
	\label{fig:VLBSM-vars}
\end{figure*}

\begin{figure*}[ht!]
	\vspace*{-3cm}
	\centering
	\includegraphics[width=\textwidth]{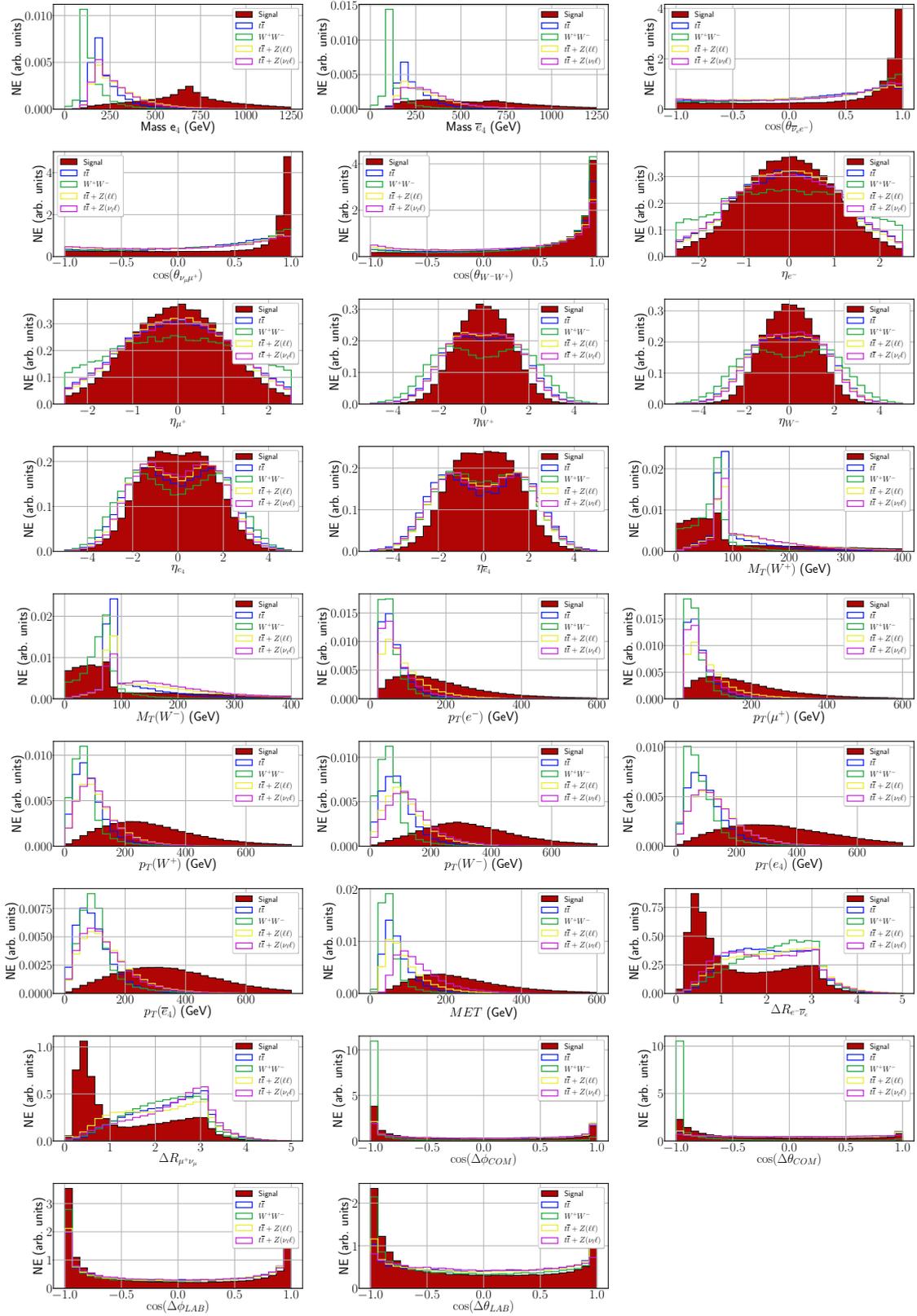}
	\caption{Dimensionless (angular) and dimension-full (kinematic) observables at lab reference frame for ZA channel (solid red), with $t\bar{t}$ (blue line), $W^+W^-$ (green line), $t\bar{t} + Z (\ell^+\ell^-)$ (yellow line) and $t\bar{t} + Z (\nu_\ell\ell)$ (purple line) backgrounds where it is considered 30 bins for all histograms. From top left to bottom right, we have distributions for $e_4$ and $\bar{e}_4$ mass, $\cos(\theta_{\bar{\nu}_e e^-})$, $\cos(\theta_{\nu_\mu \mu^+})$, $\cos(\theta_{W^+ W^-})$, pseudo-rapidity for $e^-$, $\mu^+$, $W^+$, $W^-$, $e_4$ and $\bar{e}_4$, transverse mass for $W^+$ and $W^-$, transverse momentum for $e^-$, $\mu^+$, $W^+$, $W^-$, $e_4$ and $\bar{e}_4$, missing energy MET, $\Delta R_{e^-\bar{\nu}_e}$, $\Delta R_{\mu^+\nu_\mu}$, $\cos(\Delta\phi)$ and $\cos(\Delta\theta)$ in the lab and $e_4/\bar{e}_4$ CM frames.}
	\label{fig:ZA-vars}
\end{figure*}

\begin{figure*}[ht!]
    \vspace*{-3cm}
	\centering
	\includegraphics[width=\textwidth]{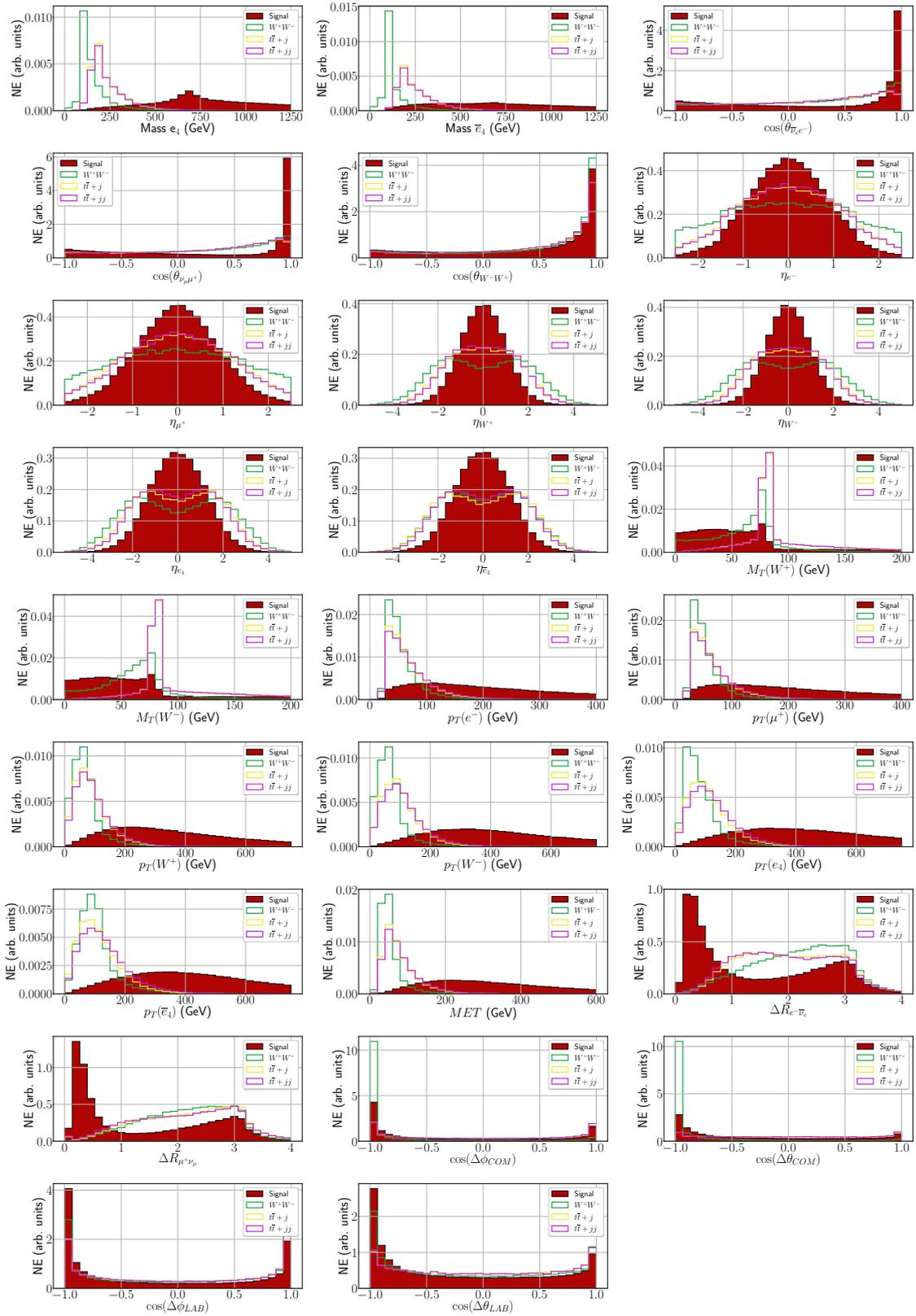}
	\caption{Dimensionless (angular) and dimension-full (kinematic) observables at lab reference frame for VBF channel (solid red), with $W^+W^-$ (green line), $t\bar{t} + j$ (yellow line) and $t\bar{t} + jj$ (purple line) backgrounds where it is considered 30 bins for all histograms. From top to bottom left to bottom right, we have distributions for $e_4$ and $\bar{e}_4$ mass, $\cos(\theta_{\bar{\nu}_e e^-})$, $\cos(\theta_{\nu_\mu \mu^+})$, $\cos(\theta_{W^+ W^-})$, pseudo-rapidity for $e^-$, $\mu^+$, $W^+$, $W^-$, $e_4$ and $\bar{e}_4$, transverse mass for $W^+$ and $W^-$, transverse momentum for $e^-$, $\mu^+$, $W^+$, $W^-$, $e_4$ and $\bar{e}_4$, missing energy MET, $\Delta R_{e^-\bar{\nu}_e}$, $\Delta R_{\mu^+\nu_\mu}$, $\cos(\Delta\phi)$ and $\cos(\Delta\theta)$ in the lab and $e_4/\bar{e}_4$ CM frame.}
	\label{fig:VBF-vars}
\end{figure*}

\begin{figure*}[ht!]
	\centering
	\includegraphics[width=\textwidth]{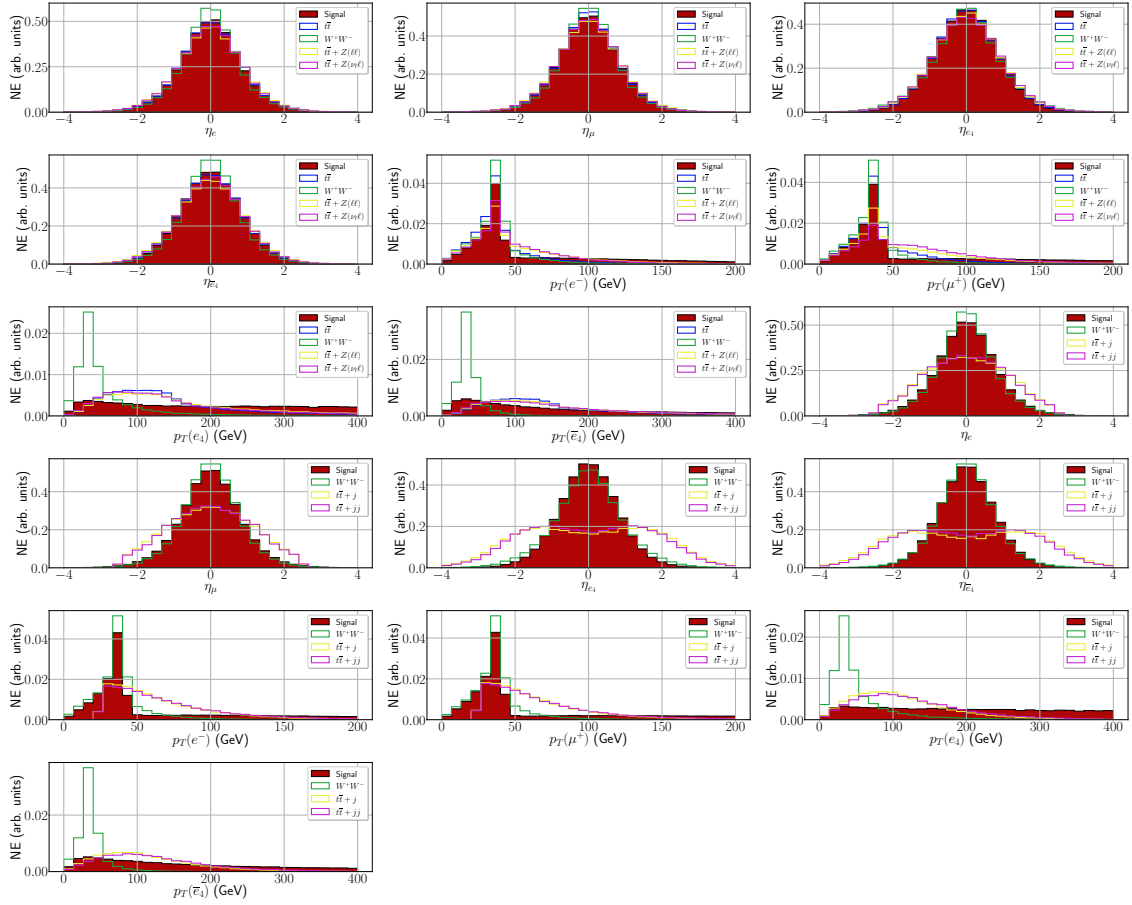}
	\caption{Dimensionless (angular) and dimension-full (kinematic) observables at the $W^\pm$ reference frame for VBF and ZA channels (solid red) where it is considered 30 bins for all histograms. VBF channel correspond to plots with 3 distinct backgrounds, while ZA has 4. The same backgrounds from previous plots also applies here. From top left to bottom right, for both channels, we have distributions for pseudo-rapidity for $e^-$, $\mu^+$, $e_4$ and $\bar{e}_4$ and transverse momentum for $e^-$, $\mu^+$, $e_4$ and $\bar{e}_4$.}
	\label{fig:Boosted-vars}
\end{figure*}

\cleardoublepage
\section{Neural Network models for different masses}\label{app:Models-Mass}

\begin{table*}[ht!]    
	\centering
\resizebox{0.91\textwidth}{!}{\begin{tabular}{|c|c|}
		\hline
    	Mass	& ZA \\
		\hline
		\hline
		\midrule
		\makecell{200 GeV}  & \makecell{\underline{Layers} : 1 input + 2 hidden + 1 output. \\Hidden and input layers with 2048 neurons each, \\ output layer with 5 neurons \\
		\underline{Regularizer} : L2 (for layers 1,2,3) and   none (for layer 4)\\
		\underline{Activation function} : tanh (for layers 1,2,3) and  sigmoid (for layer 4)  \\
		\underline{Initializer} : VarianceScaling, with normal distribution \\
		(for layers 1,2,3) in fan\_in mode \\ and uniform distribution (layer 4) in fan\_avg mode \\
		}  \\
		\hline
		\makecell{486 GeV} & \makecell{\underline{Layers} : 1 input + 4 hidden + 1 output. \\Hidden and input layers with 2048 neurons each, \\ output layer with 5 neurons \\
		\underline{Regularizer} : L2 (for layers 1 to 5) and   none (for layer 6)\\
		\underline{Activation function} : elu (for layers 1 until 5) and sigmoid (for layer 6)  \\
		} \\
		\hline
		\makecell{677 GeV}  & \makecell{\underline{Layers} : 1 input + 1 hidden + 1 output. \\Hidden and input layers with 2048 neurons each, \\ output layer with 5 neurons \\
		\underline{Regularizer} : L2 (for layers 1 and 2) and   none (for layer 3)\\
		\underline{Activation function} : tanh (for layers 1 and 2) and  sigmoid (for layer 3)  \\
		\underline{Initializer} : VarianceScaling, with normal distribution \\
		(for layers 1 and 2) in fan\_in mode and \\ uniform distribution (layer 3) in fan\_avg mode \\
		} \\
		\hline
		\makecell{868 GeV}  & \makecell{\underline{Layers} : 1 input + 4 hidden + 1 output. \\Hidden and input layers with 2048 neurons each, \\ output layer with 5 neurons \\
		\underline{Regularizer} : L2 (for layers 1 to 5) and   none (for layer 6)\\
		\underline{Activation function} : elu (for layers 1 to 5) and  sigmoid (for layer 6)  \\
		\underline{Initializer} : RandomNormal (for layers 1 until 5)  and VarianceScaling \\ with uniform distribution (layer 6) in fan\_avg mode \\
		} \\
		\hline
		\makecell{1250 GeV}  & \makecell{\underline{Layers} : 1 input + 1 hidden + 1 output. \\Hidden and input layers with 512 neurons each, \\ output layer with 5 neurons \\
		\underline{Regularizer} : L2 (for layers 1,2) and  none (for layer 3)\\
		\underline{Activation function} : tanh (for layers 1,2) and sigmoid (for layer 3)  \\
		\underline{Initializer} : VarianceScaling, with uniform distribution 
		(for layers 1,2) \\ in fan\_in mode and uniform distribution (layer 3) in fan\_avg mode \\
		} \\
		\hline
	\end{tabular}}
	\caption{Neural networks architectures employed for each mass of the lightest VLL for ZA channel. The architecture is determined following the implementation of an evolutive algorithm that maximizes the accuracy metric.}
	\label{tab:table-ZA-EVO}
\end{table*}

\begin{table*}[ht!]    
	\centering
	\resizebox{0.91\textwidth}{!}{\begin{tabular}{|c|c|}
		\hline
    	Mass	& VBF \\
		\hline
		\hline
		\midrule
		\makecell{200 GeV}  & \makecell{\underline{Layers} : 1 input + 3 hidden + 1 output. \\Hidden and input layers with 512 neurons each, \\ output layer with 3 neurons \\
		\underline{Regularizer} : L2 (for layers 1 to 4) and   none (for layer 5)\\
		\underline{Activation function} : elu (for layers 1 to 4) and  sigmoid (for layer 5)  \\
		\underline{Initializer} : VarianceScaling, with uniform distribution \\
		(for layers 1 to 4) in fan\_in mode \\ and uniform distribution (layer 5) in fan\_avg mode \\
		}  \\
		\hline
		\makecell{486 GeV} & \makecell{\underline{Layers} : 1 input + 4 hidden + 1 output. \\Hidden and input layers with 512 neurons each, \\ output layer with 3 neurons \\
		\underline{Regularizer} : L2 (for layers 1 to 5) and   none (for layer 6)\\
		\underline{Activation function} : relu (for layers 1 until 5) and sigmoid (for layer 6)  \\
		\underline{Initializer} : VarianceScaling, with normal distribution \\ (for layers 1 to 5) in fan\_in mode and \\ uniform distribution (layer 6) in fan\_avg mode \\
		} \\
		\hline
		\makecell{677 GeV}  & \makecell{\underline{Layers} : 1 input + 3 hidden + 1 output. \\Hidden and input layers with 2048 neurons each, \\ output layer with 3 neurons \\
		\underline{Regularizer} : L2 (for layers 1 and 4) and   none (for layer 5)\\
		\underline{Activation function} : sigmoid (for layers 1 to 5)  \\
		\underline{Initializer} : VarianceScaling, with normal distribution \\
		(for layers 1 and 4) in fan\_in mode and \\ uniform distribution (layer 5) in fan\_avg mode \\
		} \\
		\hline
		\makecell{868 GeV}  & \makecell{\underline{Layers} : 1 input + 2 hidden + 1 output. \\Hidden and input layers with 2048 neurons each, \\ output layer with 3 neurons \\
		\underline{Regularizer} : L2 (for layers 1 to 3) and   none (for layer 4)\\
		\underline{Activation function} : sigmoid (for layers 1 to 4)\\
		\underline{Initializer} : VarianceScaling, with uniform distribution \\
		(for layers 1 and 3) in fan\_in mode and \\ uniform distribution (layer 3) in fan\_avg mode \\
		} \\
		\hline
		\makecell{1250 GeV}  & \makecell{\underline{Layers} : 1 input + 4 hidden + 1 output. \\Hidden and input layers with 2048 neurons each, \\ output layer with 3 neurons \\
		\underline{Regularizer} : L2 (for layers 1 to 5) and  none (for layer 6)\\
		\underline{Activation function} : tanh (for layers 1 to 5) and sigmoid (for layer 6)  \\
		\underline{Initializer} : VarianceScaling, with normal distribution 
		(for layers 1 to 5) \\ in fan\_in mode and uniform distribution (layer 3) in fan\_avg mode \\
		} \\
		\hline
	\end{tabular}}
	\caption{Neural networks architectures employed for each mass of the lightest VLL for VBF channel. The architecture is determined following the implementation of an evolutive algorithm that maximizes the accuracy metric.}
	\label{tab:table-VBF-EVO}
\end{table*}

\begin{table*}[ht!]    
	\centering
	\resizebox{0.91\textwidth}{!}{\begin{tabular}{|c|c|}
		\hline
    	Mass	& VLBSM \\
		\hline
		\hline
		\midrule
		\makecell{200 GeV}  & \makecell{\underline{Layers} : 1 input + 3 hidden + 1 output. \\Hidden and input layers with 2048 neurons each, \\ output layer with 3 neurons \\
		\underline{Regularizer} : L2 (for layers 1 to 4) and   none (for layer 5)\\
		\underline{Activation function} : Sigmoid (for layers 1 to 5) \\
		\underline{Initializer} : VarianceScaling, with normal distribution \\
		(for layers 1 to 4) in fan\_in mode \\ and uniform distribution (layer 5) in fan\_avg mode \\
		}  \\
		\hline
		\makecell{486 GeV} & \makecell{\underline{Layers} : 1 input + 3 hidden + 1 output. \\Hidden and input layers with 1024 neurons each, \\ output layer with 3 neurons \\
		\underline{Regularizer} : L2 (for layers 1 to 4) and   none (for layer 5)\\
		\underline{Activation function} : relu (for layers 1 until 4) and sigmoid (for layer 5)  \\
		\underline{Initializer} : VarianceScaling, with normal distribution \\ (layer 1) in fan\_in mode and \\ uniform distribution (for layers 2 to 5) in fan\_avg mode \\
		} \\
		\hline
		\makecell{677 GeV}  & \makecell{\underline{Layers} : 1 input + 4 hidden + 1 output. \\Hidden and input layers with 256 neurons each, \\ output layer with 3 neurons \\
		\underline{Regularizer} : L2 (for layers 1 and 5) and   none (for layer 6)\\
		\underline{Activation function} : relu (for layers 1 to 5) and Sigmoid (layer 6) \\
		\underline{Initializer} : VarianceScaling, with normal distribution \\
		(for layers 1 and 5) in fan\_in mode and \\ uniform distribution (layer 6) in fan\_avg mode \\
		} \\
		\hline
		\makecell{868 GeV}  & \makecell{\underline{Layers} : 1 input + 3 hidden + 1 output. \\Hidden and input layers with 256 neurons each, \\ output layer with 3 neurons \\
		\underline{Regularizer} : L2 (for layers 1 to 4) and   none (for layer 5)\\
		\underline{Activation function} : tanh (for layers 1 to 4) and sigmoid (layer 5)\\
		\underline{Initializer} : RandomNormal (for layers 1 to 4) and VarianceScaling \\ with uniform distribution (layer 5) in fan\_avg mode \\
		} \\
		\hline
		\makecell{1250 GeV}  & \makecell{\underline{Layers} : 1 input + 1 output. \\Input layers with 256 neurons, \\ output layer with 3 neurons \\
		\underline{Regularizer} : L2 (layer 1) and  none (layer 2)\\
		\underline{Activation function} : elu (layer 1) and sigmoid (layer 2)  \\
		\underline{Initializer} : VarianceScaling, with uniform distribution 
		(layer 1) \\ in fan\_in mode and uniform distribution (layer 2) in fan\_avg mode \\
		} \\
		\hline
	\end{tabular}}
	\caption{Neural networks architectures employed for each mass of the lightest VLL for VLBSM. The architecture is determined following the implementation of an evolutive algorithm that maximises the accuracy metric.}
	\label{tab:table-VLBSM-EVO}
\end{table*}

\begin{table*}[ht!]    
	\centering
	\begin{tabular}{|c|c|}
		\hline
    	Mass	& ZA \\
		\hline
		\hline
		\midrule
		\makecell{200 GeV}  & \makecell{\underline{Layers} : 1 input + 1 output. \\Input layer with 256 neurons, \\ output layer with 5 neurons \\
		\underline{Regularizer} : L2 (for layer 1) and  none (for layer 2)\\
		\underline{Activation function} : Sigmoid (for layers 1 and 2) \\
		\underline{Initializer} : VarianceScaling, with normal distribution \\
		(layer 1) in fan\_in mode \\ and uniform distribution (layer 2) in fan\_avg mode \\
		}  \\
		\hline
		\makecell{486 GeV} & \makecell{\underline{Layers} : 1 input + 3 hidden + 1 output. \\Hidden and input layers with 512 neurons each, \\ output layer with 5 neurons \\
		\underline{Regularizer} : L2 (for layers 1 to 4) and none (for layer 5)\\
		\underline{Activation function} : sigmoid (for layers 1 until 5)\\
		\underline{Initializer} : VarianceScaling, with normal distribution \\ (for layers 1 to 4) in fan\_in mode and \\ uniform distribution (for layer 5) in fan\_avg mode \\
		} \\
		\hline
		\makecell{677 GeV}  & \makecell{\underline{Layers} : 1 input + 3 hidden + 1 output. \\Hidden and input layers with 256 neurons each, \\ output layer with 5 neurons \\
		\underline{Regularizer} : L2 (for layers 1 and 4) and   none (for layer 5)\\
		\underline{Activation function} : sigmoid (for layers 1 to 5) \\
		\underline{Initializer} : VarianceScaling, with normal distribution \\
		(for layers 1 and 4) in fan\_in mode and \\ uniform distribution (layer 5) in fan\_avg mode \\
		} \\
		\hline
	\end{tabular}
	\caption{Neural networks architectures employed for each mass of the lightest VLL for ZA. The architecture is determined following the implementation of an evolutive algorithm that maximises the Asimov significance.}
	\label{tab:table-ZA-Asimov-EVO}
\end{table*}

\begin{table*}[ht!]    
	\centering
	\begin{tabular}{|c|c|}
		\hline
    	Mass	& VBF \\
		\hline
		\hline
		\midrule
		\makecell{200 GeV}  & \makecell{\underline{Layers} : 1 input + 4 hidden + 1 output. \\Hidden and input layers with 256 neurons each, \\ output layer with 3 neurons \\
		\underline{Regularizer} : L2 (for layers 1 to 5) and none (for layer 6)\\
		\underline{Activation function} : tanh (for layers 1 to 5) and sigmoid (layer 6)\\
		\underline{Initializer} : VarianceScaling, with normal distribution \\
		(for layers 1 to 5) in fan\_in mode \\ and uniform distribution (layer 6) in fan\_avg mode \\
		}  \\
		\hline
		\makecell{486 GeV} & \makecell{\underline{Layers} : 1 input + 2 hidden + 1 output. \\Hidden and input layers with 256 neurons each, \\ output layer with 3 neurons \\
		\underline{Regularizer} : L2 (for layers 1 to 3) and   none (for layer 4)\\
		\underline{Activation function} : sigmoid (for layers 1 to 4)  \\
		\underline{Initializer} : VarianceScaling, with uniform distribution \\ (for layers 1 to 3) in fan\_in mode and \\ uniform distribution (layer 4) in fan\_avg mode \\
		} \\
		\hline
		\makecell{677 GeV}  & \makecell{\underline{Layers} : 1 input + 4 hidden + 1 output. \\Hidden and input layers with 256 neurons each, \\ output layer with 3 neurons \\
		\underline{Regularizer} : L2 (for layers 1 to 5) and   none (for layer 6)\\
		\underline{Activation function} : sigmoid (for layers 1 to 5) \\
		\underline{Initializer} : VarianceScaling, with uniform distribution \\
		(for layers 1 and 5) in fan\_in mode and \\ uniform distribution (layer 6) in fan\_avg mode \\
		} \\
		\hline
	\end{tabular}
	\caption{Neural networks architectures employed for each mass of the lightest VLL for VBF. The architecture is determined following the implementation of an evolutive algorithm that maximises the Asimov significance.}
	\label{tab:table-VBF-Asimov-EVO}
\end{table*}

\begin{table*}[ht!]    
	\centering
	\resizebox{0.91\textwidth}{!}{\begin{tabular}{|c|c|}
		\hline
    	Mass	& VLBSM \\
		\hline
		\hline
		\midrule
		\makecell{200 GeV}  & \makecell{\underline{Layers} : 1 input + 2 hidden + 1 output. \\Hidden and input layers with 512 neurons each, \\ output layer with 3 neurons \\
		\underline{Regularizer} : L2 (for layers 1 to 3) and none (for layer 4)\\
		\underline{Activation function} : relu (for layers 1 until 3) and sigmoid (layer 4) \\
		\underline{Initializer} : VarianceScaling, with normal distribution \\
		(for layers 1 to 3) in fan\_in mode \\ and uniform distribution (layer 4) in fan\_avg mode \\
		}  \\
		\hline
		\makecell{486 GeV} & \makecell{\underline{Layers} : 1 input + 2 hidden + 1 output. \\Hidden and input layers with 256 neurons each, \\ output layer with 3 neurons \\
		\underline{Regularizer} : L2 (for layers 1 to 3) and none (layer 4)\\
		\underline{Activation function} : sigmoid (for layers 1 to 4)  \\
		\underline{Initializer} : VarianceScaling, with normal distribution \\ (for layers 1 to 3) in fan\_in mode and \\ uniform distribution (layer 4) in fan\_avg mode \\
		} \\
		\hline
		\makecell{677 GeV}  & \makecell{\underline{Layers} : 1 input + 1 output. \\Input layer with 256 neurons, \\ output layer with 3 neurons \\
		\underline{Regularizer} : L2 (layer 1) and none (layer 2)\\
		\underline{Activation function} : relu (layer 1) and sigmoid (layer 2) \\
		\underline{Initializer} : VarianceScaling, with normal distribution \\
		(layer 1) in fan\_in mode and \\ uniform distribution (layer 2) in fan\_avg mode \\
		} \\
		\hline
	\end{tabular}}
	\caption{Neural networks architectures employed for each mass of the lightest VLL for VLBSM. The architecture is determined following the implementation of an evolutive algorithm that maximises the Asimov significance.}
	\label{tab:table-VLBSM-Asimov-EVO}
\end{table*}
 
\cleardoublepage
\section{Hierarchical clustering for the neural networks}\label{app:Dendograms}
\begin{figure*}[h!]
	\centering
	\includegraphics[width=\textwidth]{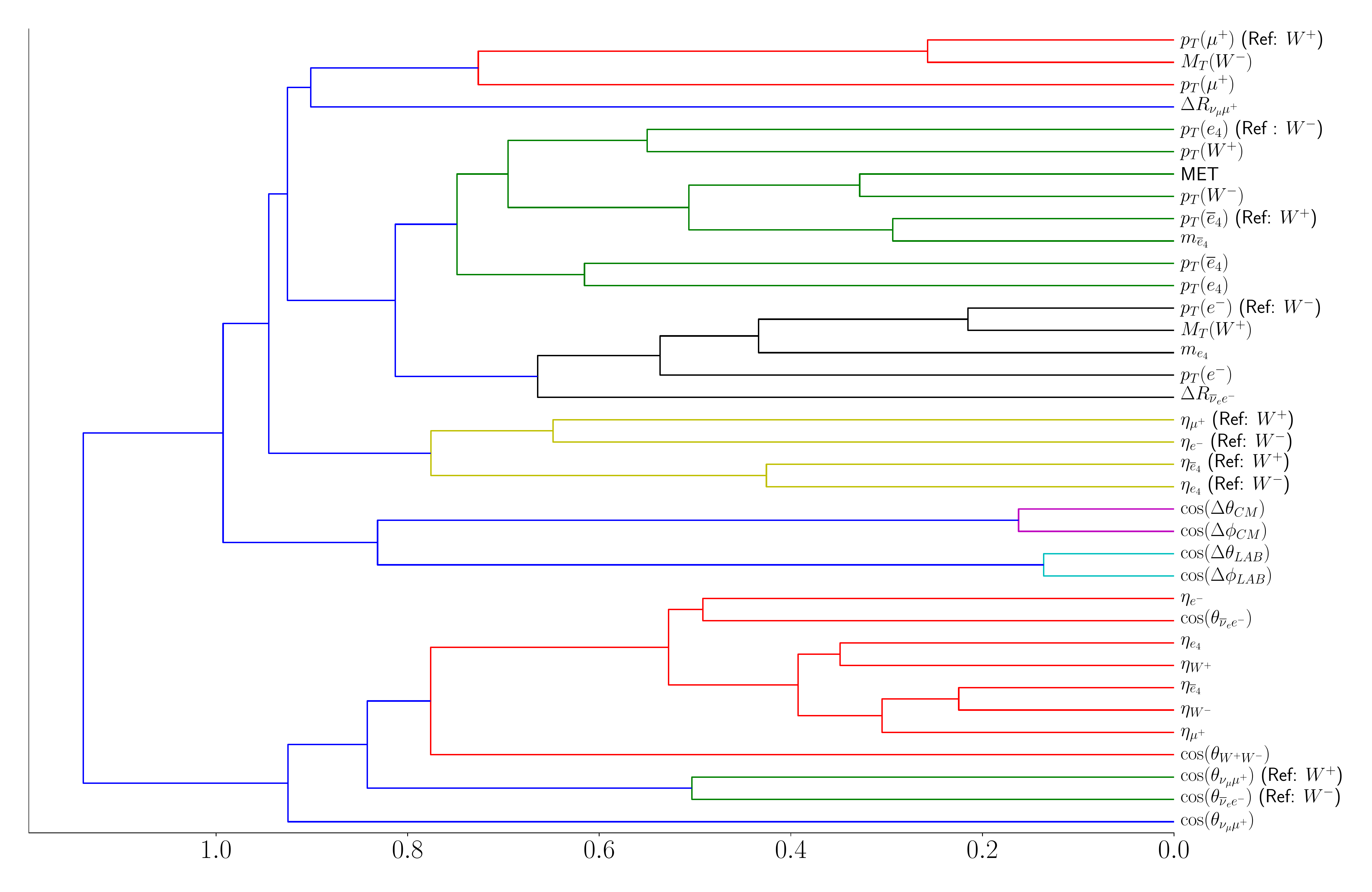}
	\caption{Hierarchical clustering of the various kinematic/angular variables utilized by the neural network model for a light VLL of 200 GeV and a ZA topology signal, under the Asimov metric. This dendrograms show how the NN groups the input features in different clusters. The dendrogram illustrates how each cluster is composed by drawing a U-shaped link between a non-singleton cluster and its children. The length of the two legs of the U-link represents the distance between the child clusters.}
	\label{fig:dend_ZA}
\end{figure*}

\begin{figure*}[ht!]
	\centering
	\includegraphics[width=\textwidth]{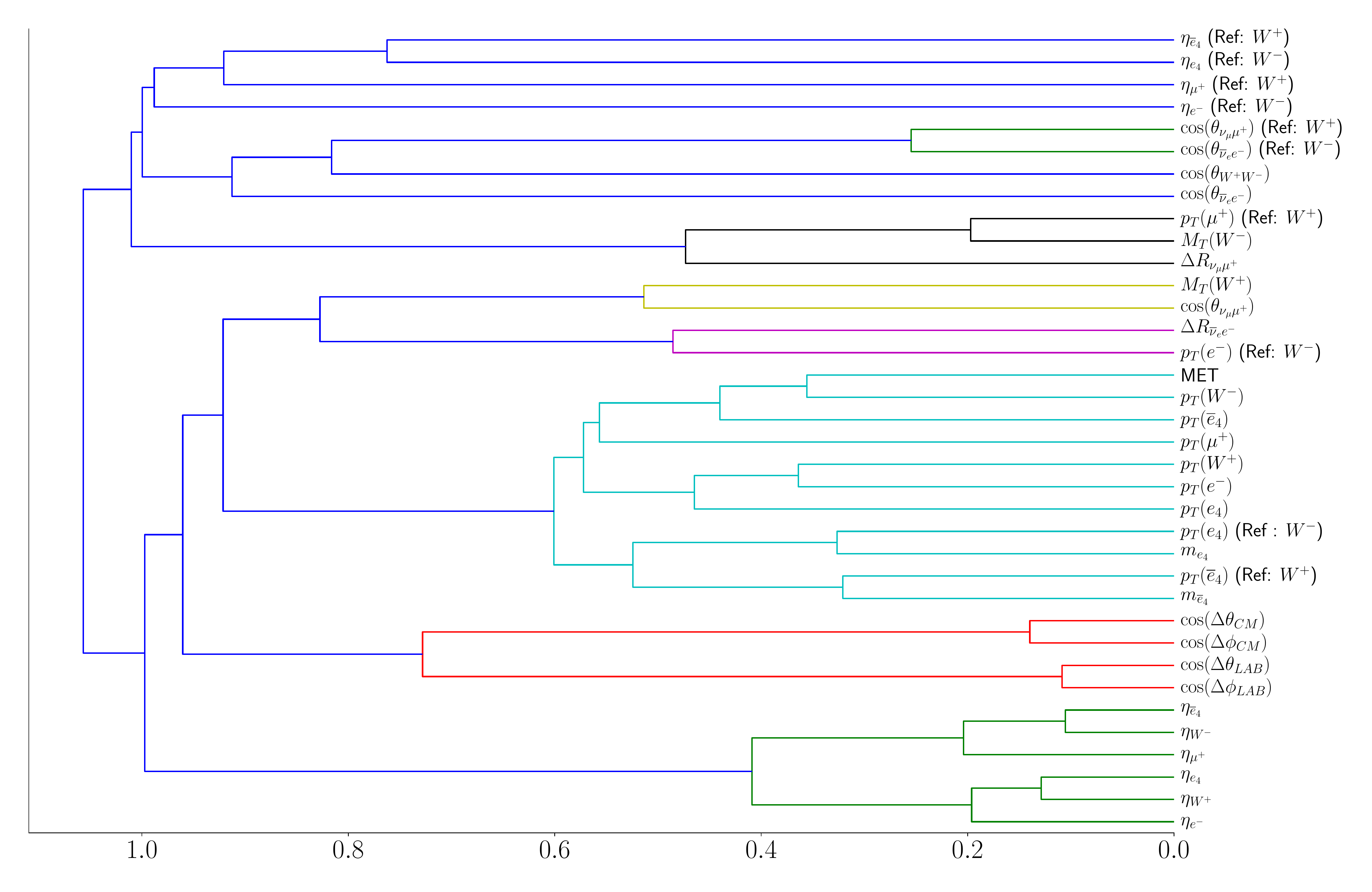}
	\caption{Hierarchical clustering of the various kinematic/angular variables utilized by the neural network model for a light VLL of 200 GeV and VBF topology signal, under the Asimov metric.}
	\label{fig:dend_VBF}
\end{figure*}

\begin{figure*}[ht!]
	\centering
	\includegraphics[width=\textwidth]{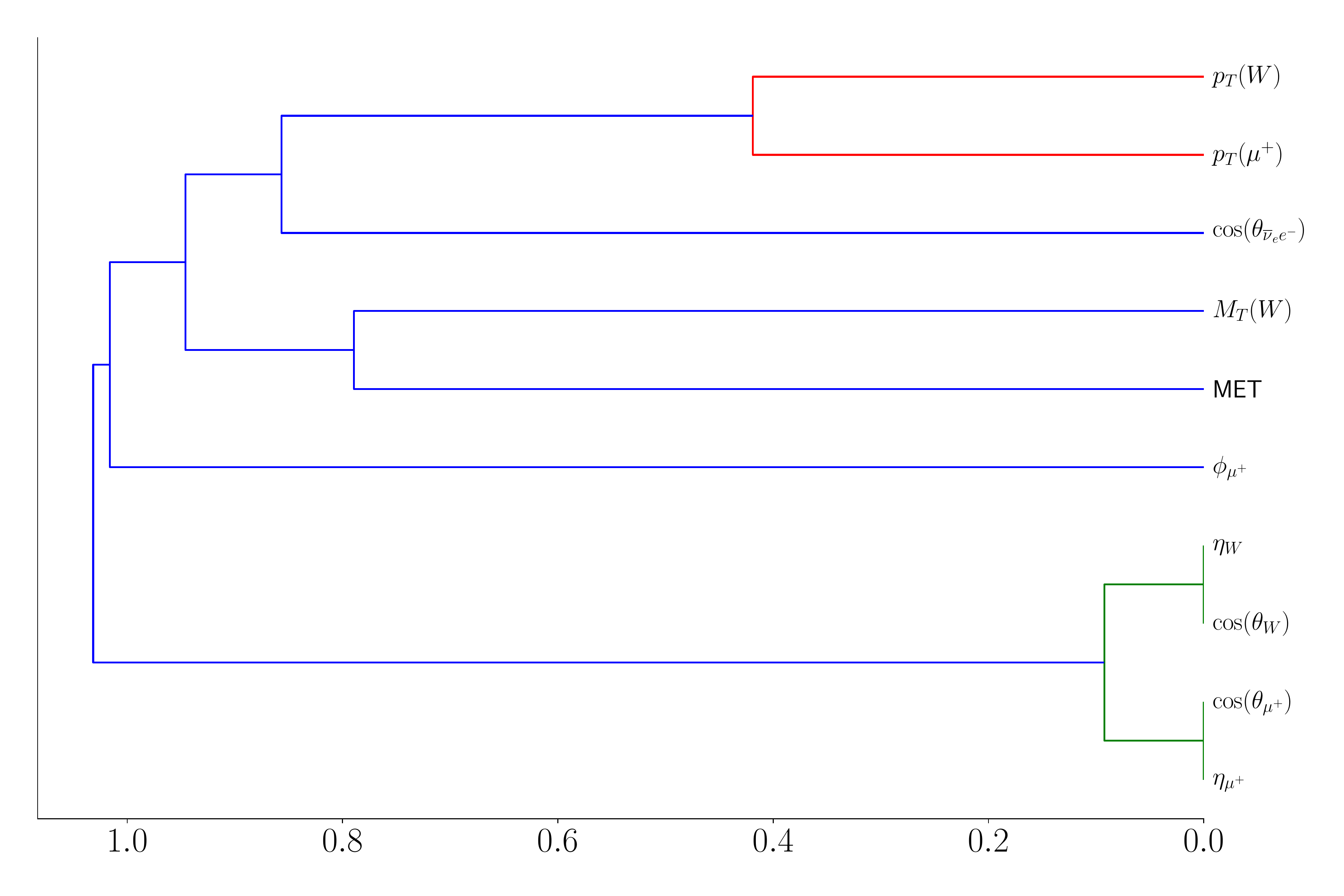}
	\caption{Hierarchical clustering of the various kinematic/angular variables utilized by the neural network model for a light VLL of 200 GeV and VLBSM topology signal, under the Asimov metric.}
	\label{fig:dend_VLBSM}
\end{figure*}

\cleardoublepage
\bibliographystyle{JHEP}
\bibliography{bib}

\end{document}